\documentclass[a4paper,fleqn,usenatbib]{mnras}

\usepackage{newtxtext,newtxmath}

\usepackage[T1]{fontenc}
\usepackage{ae,aecompl}

\usepackage{graphicx}	
\usepackage{amsmath}	
\usepackage{amssymb}	
\usepackage{subfig}     
\usepackage{ctable}

\defcitealias{PT11}{PT11}
\defcitealias{2012A&A...539A..67C}{C12}

\title[A RT model for M33]{A radiative transfer model for the spiral galaxy M33.\thanks{The solutions for the radiation fields are available in electronic form at the CDS via anonymous ftp to cdsarc.u-strasbg.fr or via http://cdsweb.u-strasbg.fr/cgi-bin/
}}

\author[J. J. Thirlwall et al.]{
Jordan J. Thirlwall\hspace{1px}{\textsuperscript{{\href{https://orcid.org/0000-0003-0390-8199}{\includegraphics[scale=0.05]{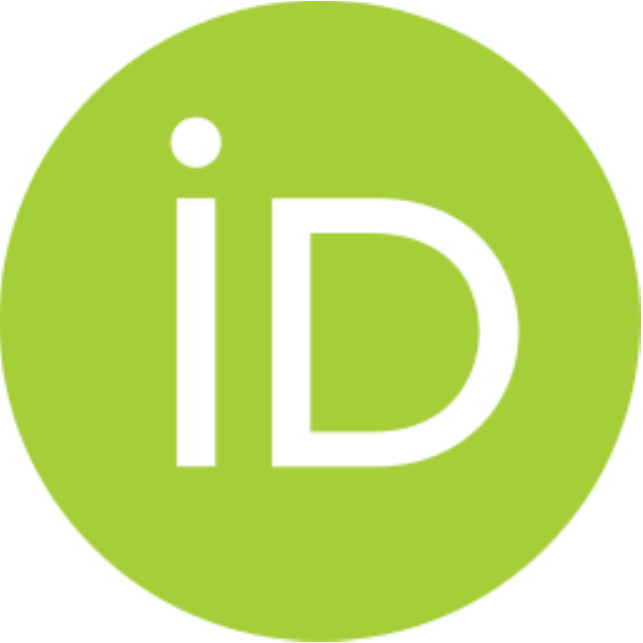}}}},$^{1}$\thanks{jordanjthirlwall@gmail.com}}
Cristina C. Popescu\hspace{1px}{\textsuperscript{{\href{https://orcid.org/0000-0002-7866-702X}{\includegraphics[scale=0.05]{FIGURES/ORCID-iD_icon-vector.pdf}}}}},$^{1,2,3}$\thanks{cpopescu@uclan.ac.uk}
Richard J. Tuffs,$^{3}$
\newauthor Giovanni Natale,$^{1}$
Mark Norris\hspace{1px}{\textsuperscript{{\href{https://orcid.org/0000-0002-7001-805X}{\includegraphics[scale=0.05]{FIGURES/ORCID-iD_icon-vector.pdf}}}}},$^{1}$
Mark Rushton,$^{2}$ 
Meiert Grootes$^{3}$
\newauthor and Ben Carroll$^{1}$
\\
$^{1}$Jeremiah Horrocks Institute, University of Central Lancashire, PR1 2HE Preston, UK \\
$^{2}$The Astronomical Institute of the Romanian Academy, Str. Cutitul de Argint 5, Bucharest, Romani\\
$^{3}$Max Planck Institut f\"{u}r Kernphysik, Saupfercheckweg 1, D-69117 Heidelberg, Germany
}

\date{Accepted XXX. Received YYY; in original form ZZZ}

\pubyear{2020}

\begin{document}
\label{firstpage}
\pagerange{\pageref{firstpage}--\pageref{lastpage}}
\maketitle

\begin{abstract}

We present the first {radiative transfer (RT)} model of a non-edge-on disk galaxy in which the large-scale {geometry} of stars and dust {is} self-consistently derived through fitting of multiwavelength imaging observations from the UV to the submm. To this end we used the axi-symmetric RT model of Popescu et al. and a new methodology for deriving geometrical parameters, and applied this to decode the {spectral energy distribution (SED)} of M33. We successfully account for both the spatial and spectral energy distribution, with residuals {typically within  $7\%$} in the profiles of surface brightness and within {$8\%$} in the spatially-integrated SED. We predict well  the energy balance between absorption and re-emission by dust, with no need to invoke modified grain properties, and we find no submm emission that is in excess of our model predictions. We calculate that 
{$80\pm8\%$} of the dust heating is powered by the young stellar populations.  We identify several morphological components in M33, a nuclear, an inner, a main and an outer disc, showing a monotonic trend in decreasing star-formation surface-density ($\Sigma_{\rm SFR}$) from the nuclear to the outer disc. In relation to surface density of stellar mass, the $\Sigma_{\rm SFR}$ of these components define a steeper relation than the {``}main sequence" of star-forming galaxies, {which we call a ``structurally resolved main sequence". Either environmental or stellar feedback mechanisms could explain the slope of the newly defined sequence.}
We find the star-formation rate to be  
${\rm SFR}=0.28^{+0.02}_{-0.01}{\rm M}_{\odot}{\rm yr}^{-1}$. 

\end{abstract}




\section{Introduction}

Star-forming galaxies contain dust {\citep{1930PASP...42..214T,1963ARA&A...1..267G,1987ARA&A..25..187S,2000ARA&A..38..761G,2001ARA&A..39..249H}}, and, although this is an insignificant component in terms of the mass budget of a galaxy, usually contributing only  $1\%$ of  its interstellar medium (ISM) {\citep{1963ARA&A...1..267G,1984ApJ...285...89D,1996A&A...312..256B,1997ApJ...480..173S}}, it is a major component in term of its effects {\citep{1963ARA&A...1..267G,1995A&ARv...6..271D,2001PASP..113.1449C,2005SSRv..119..313S}}. Because dust is widespread throughout the ISM, it absorbs and scatters the stellar photons produced by different stellar populations, and re-emits the absorbed ultraviolet (UV)/optical light into the infrared (IR) regime. Dust changes both the direction of propagation of stellar photons and their energy: it transforms the otherwise highly isotropic processes related to the propagation of stellar light in galaxies into highly anisotropic processes due to scattering, and continuously re-processes the relatively higher energy photons into lower energy ones, diminishing the direct-light output of galaxies.  Dust thus changes not only the flux of direct stellar light received by the observer {\citep{2004A&A...419..821T,2004ApJ...617.1022P,2007MNRAS.379.1022D, 2007ApJ...658..884C,2007ApJ...659.1159S, 2008ApJ..678.L101,2008ApJ...687..976U, 2008MNRAS.388.1321P, 2009ApJ...691..394M}}, but also the appearance of UV/optical images through attenuation effects {\citep{1994ApJ...432..114B,2001MNRAS.323..130C,2006A&A...456..941M,2012MNRAS.421.1007K,2013A&A...553A..80P, 2013A&A...557A.137P}}. 

In highly obscured regions of galaxies dust completely blocks the light from stars within these regions. In particular young stars in compact star forming clouds are only visible through the dust re-emitted stellar photons {\citep{2019MNRAS.486.4508M}}. On large galactic scales, UV light from the plane of edge-on discs is also highly obscured by dust, and even optical images of edge-on galaxies exhibit strong dark lanes because of dust obstruction. In the Milky Way, an edge-on galaxy seen from our observing point at the Solar position, dust prevents the {detection of UV-optical emission from the disc beyond the immediate vicinity of the Sun.  This makes dust emission} an important tool in constraining the spatial distribution of the young stellar populations throughout the Galaxy \citep{Popescu17}. But even in face-on galaxies that are seen through more translucent lines of sight\footnote {Disk galaxies are very thin objects (height much smaller than the length), and as such the face-on dust optical depth is much smaller than the edge-on optical depth. Because of this, when seen face-on, disc galaxies appear more transparent than their edge-on counterparts.}, dust strongly affects the propagation of stellar photons and distorts the projected images of stars, in particular in the UV range \citep{2004A&A...419..821T,MPT_2006,Gadotti_2010,2013A&A...553A..80P,2013A&A...557A.137P}.

In addition to the strong effect dust has in attenuating stellar light, it can also affect the thermodynamic state of gas inside and outside galaxies \citep{2010AIPC.1240...35P}, and thus the formation of structure in the Universe.  Dust is a primary coolant for gas in the highly opaque cores of star-forming clouds {\citep{1995A&ARv...6..271D}}, and may also play a major role in cooling virialised components of the intergalactic medium (IGM) {\citep{2008A&A...490..547G,2012IAUS..284..337N,2019MNRAS.487.4870V}}. At scales of hundreds of kiloparsec hot gas ($10^{6}-10^{7}$\,K) needs to cool in order to accrete into galaxies to fuel their on-going star-formation. The classical explanation for the cooling of gas is through bremstrahlung or line emission in the Xray, although inelastic collisions with dust grains is the most efficient mechanism for gas cooling at these temperatures \citep{1981ApJ...248..138D,2000A&A...354..480P,2004A&A...417..401M,2009AdSpR..44..440P,2010ApJ...725..955N,2018MNRAS.474.2073V}, providing grains exist in the IGM around star-forming galaxies. At kiloparsec scales within the disc of galaxies dust grains heat the ISM via the photoelectric effect and the thermodynamic balance of the gas is maintained through an equality between the photoelectric heating and the FIR cooling lines powered by inellastic collisions with gas particles (e.g. \citealt{2003ApJ...591..258J}). At parsec scales the cooling needed to precipitate the final stages of gravitational collapse in star-forming regions is provided by inelastic collisions of molecules with dust grains {\citep{1969MNRAS.145..271L}}. Dust thus influences the condensation of gas from the IGM into the ISM, then into denser structures within the ISM, and finally into cloud cores and stars.

Dust also has an important effect in providing the low energy seed  photons which are inverse-Compton scattered by the cosmic-ray electron population (CRe) in galaxies, accounting for a substantial fraction of the gamma-ray emission at energies above 0.1GeV from the diffuse ISM in the  disks of star-forming galaxies {\citep{1968PhRv..167.1159J,1970RvMP...42..237B,1981Ap&SS..79..321A,1993A&A...275..325N,2000APh....13..107B,2000ApJ...543...28S}}. Moreover, dust also traces the total gas mass in disks, including the mass of cold molecular Hydrogen, which has no direct spectroscopic tracer {\citep{1997ApJ...480..173S}}. Collisions of
cosmic-ray nucleons (CRp) with this gas give rise to pions whose decay provides the other main component of the diffuse $E > 0.1$\,GeV gamma-ray emission in star-forming galaxies.  A quantitative knowledge of the diffuse ISRF and distribution of dust, is therefore an essential prerequisite for decoding the gamma-ray emission of these systems {\citep{Popescu17}}, and deriving the distributions of CRs over space and energy. In particular, CRp comprise a major energetic component of the ISM {\citep{2007ARNPS..57..285S,2013PhPl...20e5501Z,2015ARA&A..53..199G}}, controlling key processes like powering galactic winds, shaping star and planet formation, promoting chemical reactions in the interstellar space, in turn leading to the formation of complex and ultimately life-critical molecules.

Deriving the distribution of dust in galaxies and its heating sources - stars of different ages and {metallicities} - is thus crucial for understanding almost every aspect of galaxy formation and evolution. The decoding of the UV/optical/far-infrared/submm images of galaxies \citep{2010AIPC.1240...35P} via self-consistent radiative transfer methods 
{\citep{2013ARA&A..51...63S}} is in principle the most reliable translation method between observations and intrinsic physical quantities of galaxies, allowing the distributions of stars and dust to be derived.  Broadly speaking, the SEDs of galaxies are influenced by two major factors: dust properties and geometry of the system.  {In recent years most of the focus has been in the former \citep{2018ARA&A..56..673G}, but the latter is arguably the most important}.  While dust properties are best constrained from extinction measurements, polarization and  dust emission in regions with known radiation fields, the geometry of the system is always a prerequisite of radiative transfer modelling of galaxy SEDs. Indeed, the RT methods are the only ones that allow the geometry of a system to be self-consistently incorporated into calculations, using constraints from imaging in both direct and dust re-radiated stellar light.

The SED modelling with RT methods  was first applied to edge-on galaxies where the vertical distribution of stars and dust can be derived \citep{Kylafis1987,1997A&A...325..135X,1998A&A...331..894X,1999A&A...344..868X}. The first edge-on galaxy that was modelled consistently from the UV to the FIR/submm was NGC891 \citep{PM00}, where the main ingredients of this kind of model were established: parameterisation, constraints from data, formalisms for calculating various quantities of interest, like for example fractions of stellar light emitted by various stellar populations in heating dust as a function of infrared wavelength, formalisms for calculating emission from stochastically heated grains, etc. The overall methodology and formalism from \cite{PM00} has been adopted and followed by the various groups in the field, albeit sometimes different terminology or refinements needed for solving specific problems/cases. Further work on modelling edge-on galaxies include: \cite{2001A&A...372..775M,2008A&A...490..461B,2010A&A...518L..39B,2015MNRAS.451.1728D,2016A&A...592A..71M,Popescu17,2018A&A...616A.120M}.

More recently efforts started to be devoted to modelling face-on galaxies, in particular driven by the recent availability of high resolution panchromatic images, including those in the important FIR/submm regime. First attempts have been done in {\cite{2014A&A...571A..69D}}, \cite{2017A&A...599A..64V}, and \cite{2019MNRAS.487.2753W} using non-axi-symmetric RT models. However, because of their non-axi-symmetric nature and the way they are implemented, they were not used to fit the geometry of the system, due to the prohibited computational time, but only used to fit the spatially integrated SEDs. As such, these models are not implemented to solve the inverse problem\footnote{In mathematics the inverse problem is that of determining the set of
unobserved parameters of a function which uniquely
predicts the recorded or observed data.} {for the geometry of the system}, but instead they assume the spatial distribution of stars and dust to be known. Here we go beyond these attempts and  present the first self-consistent model of a non-edge-on galaxy that explicitly solves the inverse problem, albeit using axi-symmetric models. We present this model for the case of M33. In a further work we will also present the non-axi symmetric model of M33.

{M33, the ``most beautiful spiral known" \citep{1918PLicO..13....9C} and ``a close rival to the nebula of Andromeda" \citep{1918PLicO..13....9C}, was described in the Hubble Atlas of Galaxies \citep{1961hag..book.....S} as the ``nearest Sc to our own galaxy". M33 or the Triangulum Galaxy is the third-largest member in the Local Group, after the Andromeda Galaxy (M31) and the Milky Way. It is a 
metal poor ({$12 + \log ({\rm O/H}) = 8.36 \pm 0.04$; \citealt{2008ApJ...675.1213R}}) spiral galaxy, rich in gas {(total gas mass of $3.2\times 10^9\,{\rm M}_{\odot}$, similar to the stellar mass content; \citealt{1997ApJ...479..244C}, \citealt{2003MNRAS.342..199C})} and dark matter dominated {(dark matter mass within the total gaseous extent of $5\times 10^{10}\,{\rm M}_{\odot}$; \citealt{2003MNRAS.342..199C})}. With stellar mass $\sim$30 times less than that of the Milky Way and a dark halo mass $\sim$20 times less than that of the Local group \citep{2019arXiv191104557C}, M33 is a minor satellite of the group.} At only 0.84-0.86 Mpc \citep{1991ApJ...372..455F,2006AJ...132.1361S,2015MNRAS.449.4048K}, M33 has been {amply} observed throughout the electromagnetic spectrum, and can be considered, together with the Milky Way, the Magellanic Clouds and M31 our nearest laboratories for studying galactic evolution. {Because of the large amount of multiwavelength high-resolution observations available, M33 is ideal for deriving spatial distributions of stellar emissivity and of dust, morphological components and intrinsic physical properties}. M33 has an intermediate inclination of $54-56^{\circ}$ \citep{1989AJ.....97...97Z,2015MNRAS.449.4048K}, which, although not close to face-on orientation, it is still within the range where projection effects due to the vertical distribution of stars and dust do not become dominant.  M33 hosts no significant bulge nor a prominent bar \citep{2016AA...590A..56H,2007ApJ...669..315C}, thus simplifying the radiative transfer analysis to modelling disc-like only morphological components. 

{A puzzling result that emerged from previous modelling of the dust emission SED of M33 is the existence of a so-called ``submm excess" \citep{2016AA...590A..56H,2019MNRAS.487.2753W}, in the sense that models underestimated observations in the submm spectral range. This excess was interpreted as an effect of dust grain properties being different in M33 than those used in the models. A submm excess has also been invoked in other studies of low-metallicity galaxies \citep{2010A&A...523A..20B,2013A&A...557A..95R}. However, none of these studies consider the effect of geometry (and the resulting dust temperature distributions) on the predicted dust emission SEDs. Here we address this problem with our radiative transfer model, whereby the geometry is self-consistently derived from fitting multi-wavelength imaging observations, both in direct and in dust-reradiated stellar light.}

The paper is organised as follows. In Sect.~\ref{sec:data} we present the various multiwavelength imaging observations used for modelling M33 and the photometry analysis performed on the data. In Sect.~\ref{sec:model} we describe the geometrical model of the stellar and dust distributions and in Sect.~\ref{sec:rt} we describe the radiative transfer model used in this paper. The fitting procedure is outlined in Sect.~\ref{sec:fitting}. We present the fits to the surface brightness distribution from the UV to the FIR/submm and the resulting global SED of M33 in Sect.~\ref{sec:results}. The derived intrinsic properties of M33 -  star-formation rate, star-formation surface density, dust optical depth, dust mass, and dust attenuation are discussed in Sect.~\ref{sec:intrin}. In the same section we also present the derived morphological components of M33 and their intrinsic properties, as well as the solutions for the radiation fields of M33. We discuss the predictions of our model in Sect.~\ref{sec:disc}. A comparison of the properties of M33 with those of the Milky Way and other local universe galaxies is also performed in Sect.~\ref{sec:disc}. We give the summary and conclusions of our results in Sect.~\ref{sec:sum}.

\section{Data}\label{sec:data}
In this section we describe the panchromatic data of M33 used in this project. The data were obtained from the archives of recent scientific missions,  as summarised in \autoref{tab:data_sum}. We assume a distance to M33 of 859\,kpc, for which the conversion between angular and linear size is 4.16 parsec per arcsec. We assume an inclination of $56^{\circ}$. Below we describe the data used in this work and a summary of these observations is given in Table~\ref{tab:data_sum}.

\subsection{GALEX}
We use the far-ultraviolet (FUV), $0.15\,\micron$, and near-ultraviolet (NUV), $0.22\,\micron$, observations presented and reduced in \cite{2005ApJ...619L..67T}. These observations were obtained by the Galaxy Evolution Explorer (GALEX, \citealt{2005ApJ...619L...1M}) and distributed by \cite{dePaz07} as part of the GALEX Ultraviolet Nearby Galaxy Survey. We assume a conservative flux calibration error of $10\%$ following \cite{2007ApJS..173..682M}. {The FWHM is 4.3 and $5.34^{\prime\prime}$ for the FUV and NUV images, respectively corresponding to 17.9 and 22.3\,pc.} 

\subsection{Local Galaxy Group Survey}
Images spanning the UBVI wavelength range observed at Kitt Peak National Observatory (KPNO) have been obtained from the Local Galaxy Group Survey (LGGS). A description of the observations and their reduction can be found in \cite{Massey06}. These data, providing {a} uniform coverage for the entire galaxy, have so far only been used for star/stellar cluster photometry. {SDSS data is also available for M33, but we have not used it in this study because it is less deep than the LGGS data.}

In order to calibrate the data we performed photometry on a group of stars within the field of view. The results of this photometry were then compared to the published measurements in the stellar photometry catalogue of \cite{Massey06}. We estimate an uncertainty of $8\%$ on our calibrations of the data. Our total flux measurements, for the entire galaxy, in the U, B, V bands agree within the uncertainties with prior measurements listed on the NASA/IPAC Extragalactic Database\footnote{The NASA/IPAC Extragalactic Database (\href{https://ned.ipac.caltech.edu/}{NED}) is operated by the Jet Propulsion Laboratory, California Institute of Technology, under contract with the National Aeronautics and Space Administration.} {The images in the U, B and V bands have superior angular resolution than that of the GALEX images, with a FWHM of $1.4^{\prime\prime}$ or 5.8\,pc. As expected, in the I band the resolution is courser than  that of the GALEX observations, with a FWHM of $3.2^{\prime\prime}$ or 13.3\,pc}.

\subsection{Moses Holden Telescope}\label{sec:MHT} 
While the LGGS survey provided most of the information regarding total flux densities and surface brightness profiles, it failed to provide accurate measurements within the inner 100\,pc of M33 in the B and V bands, where the observations had bad pixels. Because of this we decided to do our own observations to overcome this problem. For this we used the Moses Holden Telescope (MHT) of the University of Central Lancashire. This is a 0.7-m PlaneWave Instrument CDK700 optical telescope located at Alston Observatory near Preston, Lancashire, UK. In combination with a focal reducer its Apogee Aspen CG16M CCD camera provides imaging of a 40.2 x 40.2 arcminute field of view, sampled with 1.18\arcsec pixels.

Our observations were undertaken during the evening of October 9th 2018.  The median seeing was 2.8\arcsec and all observations were undertaken with airmass $<$1.2. In Table~\ref{tab:data_sum} the FWHM quoted in both the B and V bands is elongated because of wind shift. We obtained five 60s and four 300s exposures in the B and V bands, at a position angle of 113.1$^\circ$ (i.e. aligned with the minor axis). The centre of M33 was offset towards the southeast of the frame by around 4.8 arcminutes to provide additional radial coverage along the minor axis, which was used for background subtraction. Along with the raw science images a series of bias images, dark frames and twilight sky flats for each filter were also obtained.

The Python $\textsc{CCDProc}$ \citep{astropyI,astropyII} package was utilised to produce a master bias, dark and filter dependent flats, and then to apply these to the raw data, to produce reduced science images. Finally the 
$\textsc{SWarp}$ \citep{Swarp} was used to align and combine each individual exposure and sum them to produce the final combined science frames for each filter.

The resulting surface-brightness distribution  and fluxes obtained from the MHT observations were cross-checked with the LGGS corresponding photometry. We obtained consistent results within the quoted errors.

\subsection{2MASS}
We use observations in the J and Ks band from the Two Micron All-Sky Survey (2MASS) with a calibration error of $3\%$. Further information regarding the stacking of these observations can be found in \cite{2003AJ....125..525J}. {The FWHM of both the J and K band images is $3^{\prime\prime}$, corresponding to a linear resolution of 12.5\,pc.} These observations become very noisy beyond a  radial distance of 4\,kpc from the centre of M33, and the galaxy is not detected beyond 5\,kpc. As such, we could not use these observations to constrain the outer disc of M33 at these wavelengths.

\subsection{Spitzer}
Data from the infrared Array Camera (IRAC, \citealt{IRAC04}) and the Multiband Imaging Photometer (MIPS, \citealt{MIPS04}) instruments on \emph{Spitzer} have been obtained from the Spitzer Local Volume Legacy Survey \citep{2009ApJ...703..517D}. We use IRAC $3.4\,\micron$, $4.5\,\micron$, $5.8\,\micron$, $8\,\micron$ along with MIPS $24\,\micron$.{
The resolution of the IRAC images is comparable to that of the optical images, with FWHM ranging from 1.98 to $2.02^{\prime\prime}$, or from 8.2 to 8.4\,pc. Because of this excellent resolution of the IRAC images we use these in preference to the WISE images in the corresponding bands. 
The MIPS 24\,${\mu}$m observations have a FWHM of $6^{\prime\prime}$. We therefore use these in preference to the WISE images at 22\,${\mu}$m which are less sensitive and have a resolution of only $12^{\prime\prime}$.
}

\subsection{Herschel}
We obtain data in the $(70 - 500)\,\micron$ range from the Herschel M33 Extended Survey (HERM33es\footnote{\href{http://www.iram.es/IRAMES/hermesWiki}{http://www.iram.es/IRAMES/hermesWiki}}, \citealt{2010A&A...518L..67K,2011AJ....142..111B,2012AA...543A..74X,2015AA...578A...8B}). Observations were done with the Herschel's Photoconductor Array Camera and
Spectrometer (PACS; \citealt{PACS10}) at 70, 100 and 160\,${\mu}$m and with Spectral and with the Photometric Imaging Receiver (SPIRE; \citealt{SPIRE10}) at 250, 350 and 500\,${\mu}$m. {At the longest wavelength ($500\,{\mu}$m) the resolution is FWHM=$35.2^{\prime\prime}$ or 146.6 pc, making this the lowest resolution image available for this study.}

\begin{table*}
	\centering
	\caption{{Summary of observations. The table lists the name of the telescope used for each observation, the corresponding filter/instrument, the reference wavelength ${\lambda_0}$, the FWHM of the observation, the calibration errors on the total flux densities $\frac{\varepsilon_{\rm cal}}{F_{\nu}}$ in percentages, the RMS noise due to the background fluctuations 
	$\sigma_{\rm  bg}$ (see Eqn.~\ref{eqn:bgRMS}) at the native resolution of the observations, the band name, and the references from which the photometry was taken.}}
	\label{tab:data_sum}
	\begin{tabular}{lccccccl} 
		\hline
		Telescope & Filter/Instrument & $\lambda_0$ [$\micron$] & FWHM [\arcsec] & $\frac{\varepsilon_{\rm cal}}{F_{\nu}}$ [$\%$] & {$\sigma_{\rm  bg}$ [kJy/sr]} & Band name & References\\
		\hline
        GALEX                     & FUV   & 0.1528 & 4.3                & 10 & {0.13} & FUV      & \cite{2005ApJ...619L..67T}\\
                                  & NUV   & 0.2271 & 5.3                & 10 & {0.20} & NUV      & \cite{2005ApJ...619L..67T}\\
		\hline

		Moses Holden Telescope    & B     & 0.4381 & 4.65 $\times$ 3.82 & 10 & {3.5} &   B       & This work\\
		                          & V     & 0.5388 & 4.53 $\times$ 3.62  & 10 & {3.2} &      V   & This work\\

		\hline
		KPNO 4.0-m Mayall         & U     & 0.3552 & 1.4                & 8  & {7.4} &      U    & \cite{Massey06} \\
		                          & B     & 0.4381 & 1.4                & 8  & {19.} & LGGS B    & \cite{Massey06}\\
		                          & V     & 0.5388 & 1.4                & 8  & {25.} & LGGS V    & \cite{Massey06}\\
		                          & I     & 0.8205 & 3.2                & 8  & {48.} &      I    & \cite{Massey06}\\
		\hline
		Whipple Observatory 1.3-m & J     & 1.235  & 3.                 & 3  & {14.} & J         & \cite{2003AJ....125..525J}\\
		                          & Ks    & 2.159  & 3.                 & 3  & {12.} & K         & \cite{2003AJ....125..525J}\\
		\hline
		Spitzer                   & IRAC  & 3.550  & 1.95               & 3  & {14.} & I1    & \cite{2009ApJ...703..517D}\\
                                  & IRAC  & 4.493  & 2.02               & 3  & {1.7} & I2    & \cite{2009ApJ...703..517D}\\
                            	  & IRAC  & 5.731  & 1.88               & 3  & {4.5} & I3    & \cite{2009ApJ...703..517D}\\
                            	  & IRAC  & 7.872  & 1.98               & 3  & {11.} & I4    & \cite{2009ApJ...703..517D}\\
                            	  & MIPS  & 23.68    & 6.                 & 4  & {2.8} & MIPS 24   & \cite{2009ApJ...703..517D}\\

		\hline
		Herschel                  & PACS  & 70.    & 5.2                 & 15 & {19.} & PACS 70   & \cite{2015AA...578A...8B}\\
	                              & PACS  & 100.   & 7.7                 & 15 & {290.} & PACS 100  & \cite{2011AJ....142..111B}\\
		                          & PACS  & 160.   & 12.                 & 15 & {96.} & PACS 160  & \cite{2011AJ....142..111B}\\
		                          & SPIRE & 250.   & 17.6                & 15 & {12.} & SPIRE 250 & \cite{2012AA...543A..74X}\\
		                          & SPIRE & 350.   & 23.9                & 15 & {9.7} &  SPIRE 350 & \cite{2012AA...543A..74X}\\
		                          & SPIRE & 500.   & 35.2                & 15 & {8.3} & SPIRE 500 & \cite{2012AA...543A..74X}\\
		\hline
	\end{tabular}
\end{table*}

\subsection{Masking foreground stars}
Within the UV/optical/NIR wavelength ranges, bright foreground stars can affect the estimates of the background on the observation, and the total integrated flux of the galaxy. Because of this foreground stars were masked independently at each wavelength. For this we produce a median map in which each pixel has a value equal to the median of the $3\times 3$ pixels surrounding each pixel on the observation. Following this we subtract the median map from the observation. By considering the distribution of residual values we take  pixels with values $>15\,\%$ the maximum residual to be pixels coinciding with the centres bright point sources, generally bright foreground stars. Once identified, these bright points are masked using a mask of radius 20 pixels. In order to ensure we do not mask bright sources associated with the galaxy, such as the galactic centre or the bright star formation region NGC\,604, each of the produced masks are checked by eye and any false positives removed. This approach to identifying and masking bright foreground stars has been taken as it makes no assumption of intrinsic colours of the foreground star populations. We find that the masking of foreground stars has a negligible effect on the total integrated flux of M33, however, it does have an effect on the derived background fluctuations and total flux uncertainty, where bright foreground stars dominate the background fluctuations. We also made visual tests to ensure that the number of stars within the extent of the galaxy does not exceed that in the background region.

\subsection{Convolutions}
All the data for wavelengths shorter than  $160\,\micron$ have a higher resolution than that normally employed by our model, which, for the distance of M33, matches the PACS 160 resolution. We have therefore degraded these data to PACS 160 resolution, using the convolution kernels\footnote{\href{https://www.astro.princeton.edu/~ganiano/Kernels.html}{https://www.astro.princeton.edu/\textasciitilde{}ganiano/Kernels.html}} of \cite{2011PASP..123.1218A} where available. The convolutions of LGGS, MHT, and 2MASS data have been performed using a two dimensional Gaussian. After the convolutions all data shortwards of  $160\,\micron$ were thus degraded to a physical resolution of $50\mathrm{pc}$. For the SPIRE bands we have retained the original physical resolution of $73\, \mathrm{pc}$, $100\, \mathrm{pc}$, and $146.6\, \mathrm{pc}$ at $250\,\micron$, $350\,\micron$, and $500\,\micron$ respectively, and instead degraded the resolution of the model images to match the data.

\subsection{Photometry}
\label{sec:phot}
 Each observation was first corrected for foreground dust extinction assuming E(B-V)=0.0356 \citep{2011ApJ...737..103S}, where the extinction coefficients have been derived from the extinction curve parametrization of \cite{1999PASP..111...63F} with $R_V=3.1$. 
 A single correction factor was used at each wavelength and no attempt was made to correct for gradients along the galaxy.
 We performed curve of growth (CoG) photometry for all wavelengths in order to derive the azimuthally averaged surface brightness profiles of M33 and {total integrated fluxes   $F_{\nu}$}. The annuli used were ellipses derived using the position angle and inclination of M33. The size of each annuli has been defined manually in order to sample interesting features on the surface brightness profile, identified from a coarse linearly sampled CoG. A total of 119 annuli were used. {Errors on total fluxes $\varepsilon_{F_\nu}$}, take into account calibration error $\varepsilon_{\rm cal}$, fluctuations in the background $\varepsilon_{\rm bg}$, and Poisson noise $\varepsilon_{\rm poisson}$, the latter being applied for all wavelengths up to and including the I band.{The error calculation method is given in Appendix~\ref{sec:err_calc}}. 
 {In the case of the MHT observations, the background level and its fluctuations have been derived from an offset image that provided additional coverage along the minor axis (see Sect.~\ref{sec:MHT}). }
 We found that the errors on total fluxes are dominated by calibration errors, as seen in Table~\ref{tab:err_contribution}. Subsequently, colour corrections have been applied to IRAC, MIPS, and PACS observations, in an iterative procedure that inputs the best SED shapes from the model. Colour corrected spatially integrated flux densities are shown in Table~\ref{tab:obs_ccorr}. 

\begin{table}
	\centering
	\caption{Colour corrected fluxes}
	\label{tab:obs_ccorr}
	\begin{tabular}{lcc}
		\hline
		Band & {$F_\nu$}[Jy] & {$\varepsilon_{F_\nu}$} [Jy]  \\
		\hline
		 FUV        & 2.1   & 0.1 \\
		 NUV        & 3.04   & 0.09 \\
		 U     & 5.9   & 0.6 \\
		 LGGS B     & 14  & 2 \\ 
		 MHT B      & 14  & 1 \\
		 LGGS V     & 20  & 2 \\
		 MHT V      & 22   & 2  \\
		 I     & 32   & 4  \\
		 J      & 16.8  & 0.6 \\ 
		 K      & 16.4  & 0.5 \\
		 I1     & 16  & 1 \\
		 I2     & 11.2  & 0.3 \\
		 I3     & 20.1  & 0.7 \\
		 I4     & 108  & 3  \\
		 24    & 48  & 2 \\
		 70    & 600  & 90 \\
		 100   & 1400 & 200\\
		 160   & 2200 & 300\\
		 250  & 1300 & 200\\
		 350  & 710  & 100\\
		 500  & 330  & 50 \\
		\hline
	\end{tabular}
\end{table}

\section{Model Description}\label{sec:model}
Our model of M33 is based on the axi-symmetric RT model of \citeauthor{PT11} (\citeyear{PT11}, hereafter \citetalias{PT11}) for the UV to submm emission of spiral galaxies, in which dust opacity and stellar emissivity geometries are described by parameterized analytic functions. {The model incorporates the effect of anisotropic scattering and explicit calculation of the stochastic heating of dust grains of various sizes and chemical composition}.  

We only retain the overall formalism of the \citetalias{PT11} model, and we derive the geometrical parameters of M33 through an optimisation process in which the model is fitted to the detailed imaging observations available for M33, spanning the FUV to submm wavelength range.
From the detailed surface brightness distributions we found it necessary to model M33 with four different morphological components: a nuclear, an inner, a main and an outer disc. Physical quantities related to these will carry the notations {``}n",{``}i",{``}m",{``}o" throughout this paper. Each of the morphological components is further made up of the generic stellar and dust components from \citetalias{PT11} (see also Sect.~\ref{sec:stellcomp} and \ref{sec:dustcomp}) : the old and young stellar discs (the stellar disc and the thin stellar disc respectively) and the dust disc and the thin dust disc respectively. In principle we allow for each of the four morphological components to be made up of two stellar and two dust components. For example we can speak of a main stellar disc, a main thin stellar disc, a main dust disc and a main thin dust disc. In practice we found that not all morphological components require so many stellar or dust components, as is the case for the nuclear disc, which could be fitted with only one thin stellar disc.

All disc components j in the model are described by the following general formula:

\begin{equation}
\label{eq:model}
w_{\rm j} (R,z) = \begin{cases} 
    {\displaystyle
        0  \hspace{4.8cm}{\rm if} \hspace{0.1cm} R < R_{{\rm tin, j}}
    }\\
    {\displaystyle
        A_{0, {\rm j}} \left[ \frac{R}{R_{{\rm in, j}}}\left( 1 - \chi_{\rm j} \right) + \chi_{\rm j} \right] 
        \exp{\left(-\frac{R_{{\rm in, j}}}{h_{\rm j}}\right)}
        \exp{\left(-\frac{z}{z_{\rm j}}\right)}
    }\\
    {\displaystyle
        \hspace{4.1cm}{\rm if} \hspace{0.1cm} R_{{\rm tin, j}} \leq R < R_{{\rm in, j}}
    }\\
    {\displaystyle
        A_{0, {\rm j}} \exp{\left(-\frac{R}{h_{\rm j}}\right)}
        \exp{\left(-\frac{z}{z_{\rm j}}\right)} 
         \hspace{1.1cm} {\rm if} \hspace{0.1cm} R_{{\rm in, j}} \leq R \leq R_{{\rm t, j}}
    }
\end{cases}
\end{equation}
with: 
\begin{equation}
    \label{eq:chi}
    \chi_{\rm j} = \frac{w_{{\rm j}}(0,z)}{w_{{\rm j}}(R_{{\rm in, j}},z)}
\end{equation}

\noindent
where
$R$ and $z$ are the radial and vertical coordinates, $h_{\rm j}$ is the scale-length, $z_{\rm j}$ is the scale-height, $A_{0, {\rm j}}$ is a constant determining the scaling of $w_{\nu, {\rm j}}(R,z)$, $\chi_{\rm j}$ is a parameter describing the linear slope of the radial distributions interior to an inner radius $R_{{\rm in, j}}$, $R_{{\rm tin, j}}$, is the inner truncation radius of the linear slope interior to $R_{{\rm in, j}}$, and $R_{{\rm t, j}}$ is the outer truncation radius of the exponential distribution. Eqns. \ref{eq:model} and \ref{eq:chi} are wavelength dependent.
The spatial integration of these distributions between the inner and the outer truncation radius provides the intrinsic luminosities if they refer to stellar distributions, or the dust mass, if they refer to dust density/dust opacity distributions. The corresponding analytic formulas used in  the calculations are given in Appendix~\ref{sec:formula}.

\begin{figure*}
    \centering
    \includegraphics[width=\linewidth]{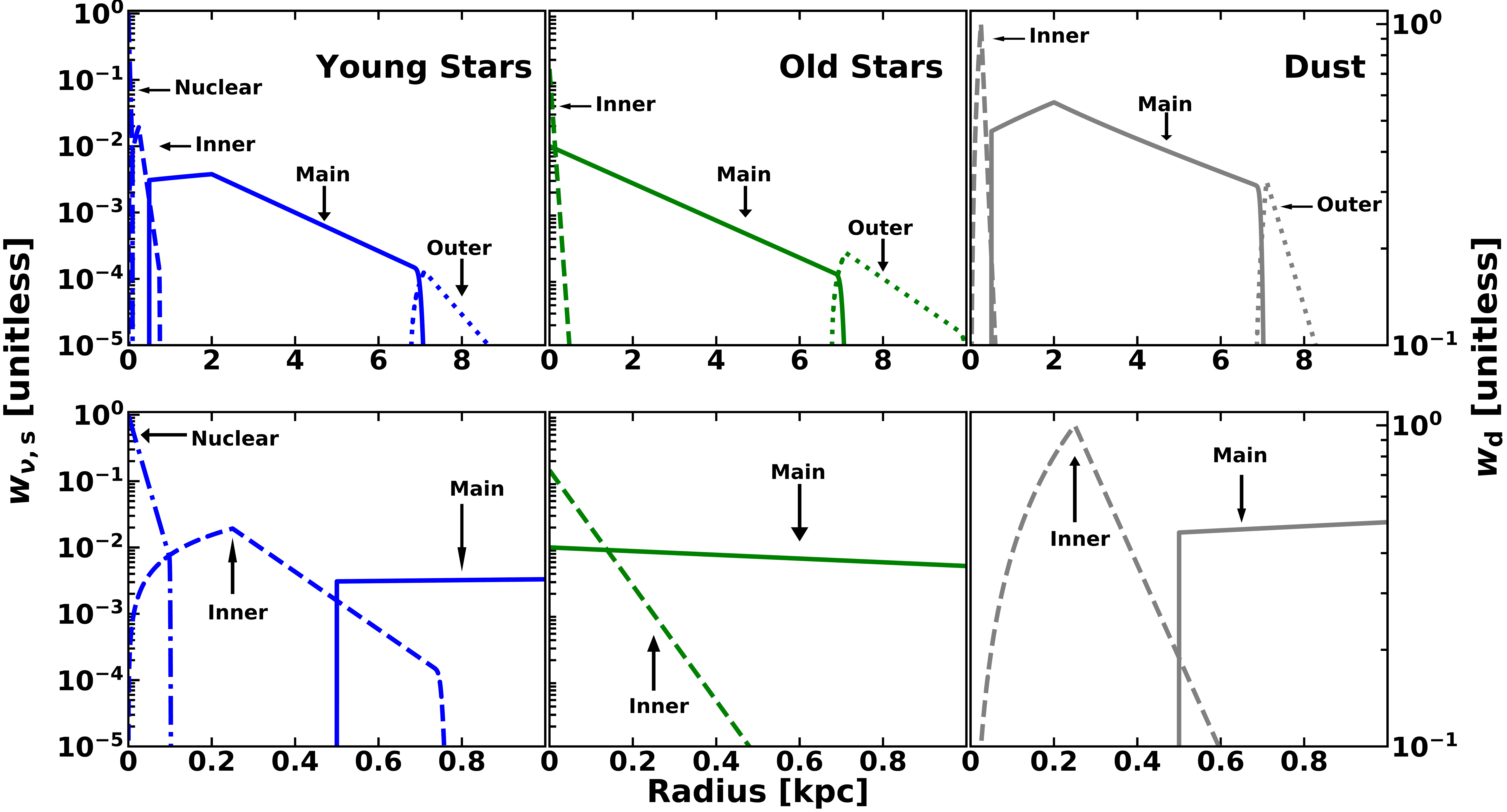}
    \caption{Top: Radial profiles for the stellar emissivity $w_{\nu,s} (R,0)$ of the young (left) and old (middle) stellar populations in the B band, and for the dust density distribution $w_{d} (R,0)$ (right), for the different morphological components: nuclear and inner (dashed-lines), main (solid-line), outer (dotted-lines) discs. The profiles of stellar emissivity have been normalised to the maximum intensity corresponding to the nuclear disc of the young stellar populations. The profiles for the dust density distribution have been normalised to the maximum intensity corresponding to the inner dust disc. It should be noted that the dust distribution for each morphological component is made up of a thin and a thick dust disc having different scale-lengths, and as such the resulting radial profile of the dust distribution does deviate from an exponential form. Bottom: As above, but zoomed into the inner 1\,kpc. } 
    \label{fig:stel_dust}
\end{figure*}

\begin{figure*}
    \centering
    \includegraphics[width=\linewidth]{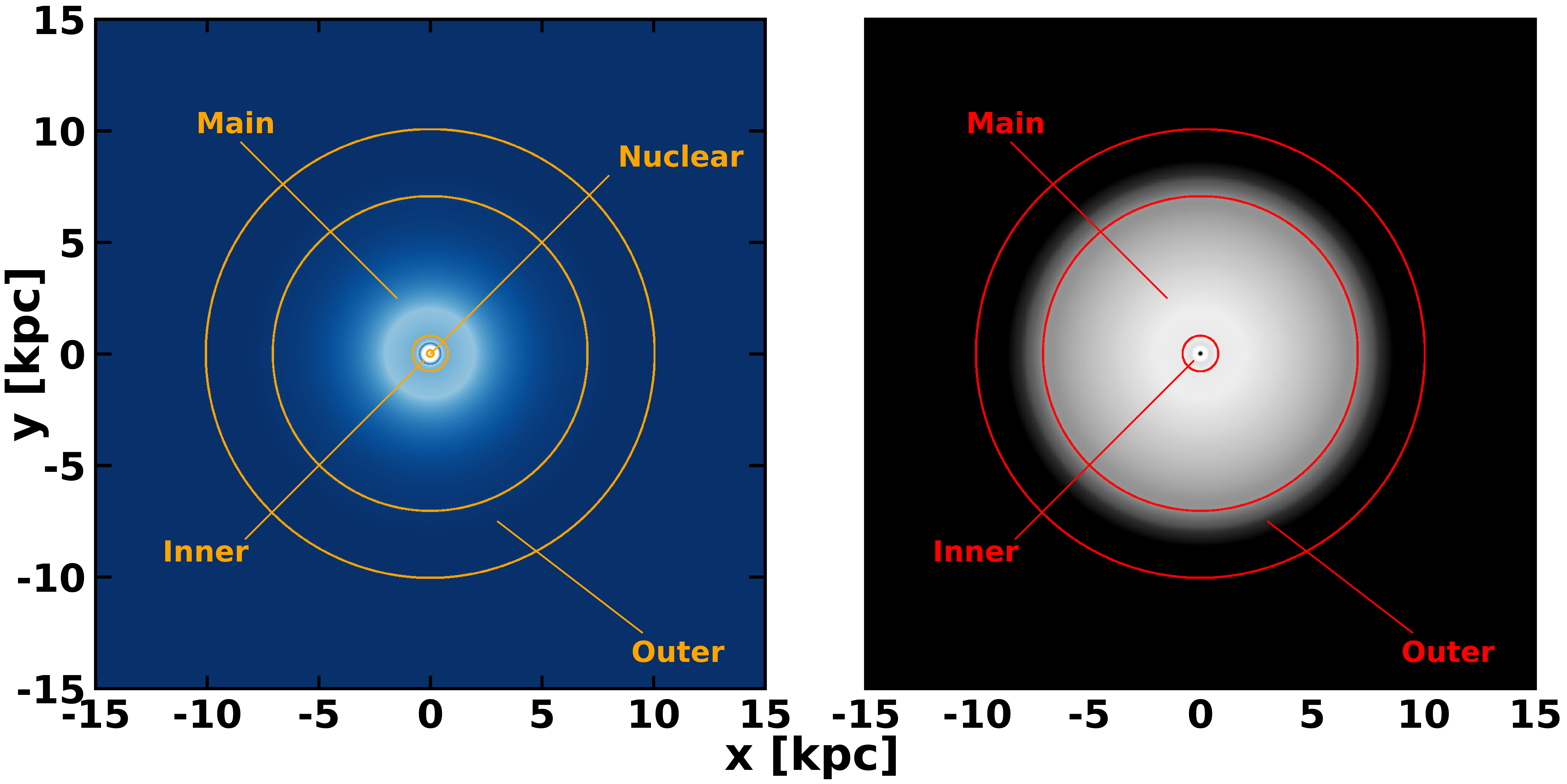}
    \caption{The model image for the stellar emissivity of the young stellar population (left), and for the dust density distribution (right) in M33, as seen face-on. The image shows the different morphological components found for M33.} 
    \label{fig:comp_nuv}
\end{figure*}

\subsection{Stellar components}\label{sec:stellcomp}
\subsubsection{The stellar disc}
Preferentially emitting in optical and NIR wavelengths, the stellar disc is made up of old stellar populations and is described by the geometrical parameters $h^{\rm disc}_{\rm s}$, $z^{\rm disc}_{\rm s}$, $R^{\rm disc}_{\rm in, s}$, and $\chi^{\rm disc}_{\rm s}$, and the amplitude parameters $L^{\rm disc}(\lambda)$.  Most of these parameters were constrained from observations at available wavelengths (in the U, B, V, I, J, K, I1, I2, and I3 bands) as described in Sect.~\ref{sec:fitting}.  It should be noted that the inclusion of  an old stellar population in the U-band represents a departure from the generic model from \citetalias{PT11}, which was based on edge-on systems. In those systems very little observational {constraints} from the disc are available in this band, and as such no modelling in this band was available for inclusion in the model of \citetalias{PT11}. In the case of a non-edge-on galaxy like M33 there is clear evidence that an old stellar populations is required by the imaging data in the U-band.

The parameters  $R^{\rm disc}_{\rm in, s}$ and $\chi^{\rm disc}_{\rm s}$ were found to be wavelength independent. The model of M33 contains an inner stellar disc, a main stellar disc and an outer stellar disc, with radial profiles as depicted in the middle panel of Fig.~\ref{fig:stel_dust}. In principle an old stellar population associated with the nuclear disc (or with a very small bulge) could not be excluded, {however, due to the small extent of this component (relative to the resolution of the measurement), we could not use any geometrical constraint to disentangle such a contribution}, and as such a nuclear (old) stellar disc is not included in the current model. To conclude the stellar disc of each morphological component containing an old stellar population (inner, main and outer disc) is described by 4 geometric parameters and one amplitude parameter.

\subsubsection{The thin stellar disc}
Containing young stellar populations, the thin stellar disc produces the majority of the UV output and is described by the geometrical parameters $h^{\rm tdisc}_{\rm s}$, $z^{\rm tdisc}_{\rm s}$, $R^{\rm tdisc}_{\rm in, s}$, and $\chi^{\rm tdisc}_{\rm s}$, and the amplitude parameters $L^{\rm tdisc}(\lambda$).  All the  parameters except $z^{\rm tdisc}_{\rm s}$ have been mainly constrained from the NUV data under the assumption that $h^{\rm tdisc}_{\rm s}$ and $z^{\rm tdisc}_{\rm s}$ do not vary with wavelength (see \citetalias{PT11}). The model of M33 contains a nuclear thin stellar disc, an inner thin stellar disc, a main thin stellar disc and an outer thin stellar disc, with radial profiles as depicted in the left panel of Fig.~\ref{fig:stel_dust}. An image of the stellar emissivity seen face-on is also shown in Fig.~\ref{fig:comp_nuv}, where the different morphological components are indicated on the map. 

Since, in non-edge-on galaxies, the UV emission, though
strongly attenuated, is still readily measurable, we have included amplitude parameters as free variables
for the emission of the young stellar populations in different UV-optical bands. This is a necessary departure
from the use in \citetalias{PT11} of a fixed emission template SED, calculated for a steady state SFR, for modelling in particular the
UV emission of edge-on galaxies, where the UV emission
is almost completely obscured by the dust lanes. All this opens up the possibility, which we will explore in future works, of investigating the star-formation history on timescales of a few 10s to a few 100s of Myr through
analysis of the derived intrinsic UV colours as a function of radial position. 
Keeping in line with previous modelling, we express the spectral integrated luminosity of the young stellar disc $L^{\rm tdisc}$ in terms of a star-formation rate ${\rm SFR}^{\rm tdisc}$, using Eqns. 16, 17, and 18 from \citetalias{PT11}. Because we use the total bolometric luminosity of the young stellar disc to derive the SFR, our method is less sensitive to assumptions regarding steady-state or IMF used. 

To conclude, the thin stellar disc of each morphological component containing a young stellar population (nuclear, inner, main and outer disc) is described by 4 geometric parameters and one amplitude parameter.

\subsection{Dust components}\label{sec:dustcomp}
\subsubsection{The dust disc}
Describing the large scale distribution of the diffuse dust associated with the majority of the stellar population in a galaxy and with the HI gas, the dust disc is one of the main components of the \citetalias{PT11} model. Mainly characterised by a smaller scale-height $z^{\rm disc}_{\rm d}$ than that of the old stellar population $z^{\rm disc}_{\rm s}$ , while still being larger than that of the young populations $z^{\rm tdisc}_{\rm s}$, the dust disc is usually more radially extended than the old stellar disc (e.g. \citealt{1999A&A...344..868X}).  On a similar vein to the stellar discs the geometrical parameters of the dust disc are $h^{\rm disc}_{\rm d}$, $z^{\rm disc}_{\rm d}$, $R^{\rm disc}_{\rm in, d}$, and $\chi^{\rm disc}_{\rm d}$.  The amplitude parameter is the B-band face-on optical depth at the inner radius $\tau^{\rm f}_{B}\left( R^{\rm disc}_{\rm in,d}\right)$. The model of M33 contains an inner dust disc, a main dust disc and an outer dust disc. When modelling the radial profiles (see Sect.~\ref{sec:fitting}), no necessity was found to include a dust counterpart for the nuclear disc. To {summarise}, the dust disc for each morphological component incorporating such a disc (inner, main and outer) is described by 4 geometric parameters and an amplitude parameter.

\subsubsection{The thin dust disc}
A generic feature of the \citetalias{PT11} model, the thin dust disc represents the diffuse dust associated with the young stellar population. As such this dust component is constrained to have the same scale-length and scale-height as the young stellar disk (see  \citetalias{PT11}). The geometrical parameters of the thin dust disc are $h^{\rm tdisc}_{\rm d}$, $z^{\rm tdisc}_{\rm d}$, $R^{\rm tdisc}_{\rm in, d}$, and $\chi^{\rm tdisc}_{\rm d}$. The amplitude parameter is the B-band face-on optical depth at the inner radius $\tau^{\rm f}_{B}\left( R^{\rm tdisc}_{\rm in,d}\right)$. The model of M33 contains an inner thin dust disc, a main thin dust disc and an outer thin dust disc. In our modelling we found no need to include a thin dust disc within the nuclear disc. As such, the thin dust disc for each of the inner, main and outer disc is described by 4 geometric parameters and one amplitude parameters. 

The total dust distribution from both the dust disk and the thin dust disk is shown in the right panel of Fig.~\ref{fig:comp_nuv}.

\subsubsection{Clumpy component}

The clumpy component is another generic feature of the \citetalias{PT11} model and represents the dust around young star-formation regions. Clumps in our model have a small filling factor and thus, the effect on light propagating at kpc scales is not significantly {affected}. The clumps do however efficiently block the light from young stars inside the clouds. 
The amplitude of the clumpy component is described by the parameter $F$, which was defined in \citetalias{PT11} to represent the fraction of the total luminosity of massive stars locally absorbed in star-forming clouds (see Sect. 2.5.1 from \citetalias{PT11} for a detailed description of the escape fraction of stellar light from the clumpy component).

\vspace{0.5cm}
\noindent
The parameters associated with all these structures are constrained from data as described in Sect.~\ref{sec:fitting}. The model 
geometric parameters are listed in Table~\ref{tab:ogeom}, {while the derivation of the value of these parameters and their errors is described in  Sect.~\ref{sec:fitting}}. The amplitude parameters (luminosity densities and dust optical depth) are listed {in} Table~\ref{tab:lum_dens}.

\begin{table*}
\caption{The geometrical parameters of the model that are constrained from 
data. All the length parameters are in units of kpc.}
\label{tab:ogeom}
\begin{tabular}{ll}
\hline\hline
$h_{\rm s}^{\rm i-disc}(\rm U,B,V,I,J,K,\rm I1,I2,I3)$      & (0.05, 0.05, 0.05, 0.05, 0.05, 0.07, 0.07, 0.07, 0.08)$\pm20\%$\\
$h_{\rm s}^{\rm m-disc}(\rm U,B,V,I,J,K,\rm I1,I2,I3)$         & (1.8, 1.8, 1.7, 1.55, 1., 1.1, 1.7, 1.7, 1.5)$\pm$10\%\\
$h_{\rm s}^{\rm o-disc}(\rm U,B,V,I,\rm I1,I2,I3)$     & (1., 1., 1., 1., 1., 1., 1.)$\pm 20\%$\\
&\\
$h_{\rm s}^{\rm n-tdisc}$                                     & 0.02 \\
$h_{\rm s}^{\rm i-tdisc}$                                      & 0.10$\pm$0.01\\
$h_{\rm s}^{\rm m-tdisc}$                                         & 1.50$\pm 0.15$\\
$h_{\rm s}^{\rm o-tdisc}$                                     & 0.60$\pm$0.06\\
&\\
$h_{\rm d}^{\rm i-disc}$                                       & 0.15$^{+0.05}_{-0.03}$\\
$h_{\rm d}^{\rm m-disc}$                                          & 9.0$^{+2.7}_{-1.8}$\\
$h_{\rm d}^{\rm o-disc}$                                      & 1.0$\pm$0.2  \\
&\\
$\chi_{\rm s}^{(\rm i-disc, m-disc, o-disc)}$               & (1$\pm0.1$, 1$\pm 0.1$, -20$\pm4$)\\
$\chi_{\rm s}^{(\rm i-tdisc, m-tdisc, o-tdisc)}$ & (0., 0.75$\pm 0.08$, -20$\pm 2$)\\
$\chi_{\rm d}^{(\rm i-disc, m-disc, o-disc)}$               & (0., 0.75$^{+0.23}_{-0.15}$, -20$\pm 4$)\\
&\\
$R_{\rm in,s}^{(\rm i-disc, m-disc, o-disc)}$                  & (0.,0.,7.1)\\
$R_{\rm in,s}^{(\rm n-tdisc, i-tdisc, m-tdisc, o-tdisc)}$    & (0.,0.25,2.,7.1)\\
$R_{\rm in,d}^{(\rm i-disc, m-disc, o-disc)}$                  & (0.25,2.,7.1)\\
&\\
$R_{\rm tin,s}^{(\rm i-disc, m-disc, o-disc)}$                  & (0., 0., 6.76)\\
$R_{\rm tin,s}^{(\rm n-tdisc, i-tdisc, m-tdisc, o-tdisc)}$    & (0., 0., 0.5, 6.76)\\
$R_{\rm tin,d}^{(\rm i-disc, m-disc, o-disc)}$                  & (0., 0.5, 6.76)\\

&\\
$R_{\rm t,s}^{(\rm i-disc, m-disc, o-disc)}$                  & (0.5, 7., 10.)\\
$R_{\rm t,s}^{(\rm n-tdisc, i-tdisc, m-tdisc, o-tdisc)}$    & (0.1, 0.75, 7., 10.)\\
$R_{\rm t,d}^{(\rm i-disc, m-disc, o-disc)}$                  & (1., 7., 10.)\\
\hline
\end{tabular}
\end{table*}

\section{The radiative transfer codes}\label{sec:rt}
We used the radiative transfer code of \citetalias{PT11}, a modified version of \cite{Kylafis1987}, which employs a ray-tracing algorithm and the method of scattered intensities. We also used the DART-Ray\footnote{\href{http://www.star.uclan.ac.uk/~gn/dartray_doc/}{http://www.star.uclan.ac.uk/\textasciitilde{}gn/dartray\_doc/}} code \citep{Natale14,Natale15,Natale17}. Optimisation of the model has been made using the \citetalias{PT11} code. The surface brightness maps for the dust emission, as seen by an observer, have been produced with DART-Ray. The radiation fields in the dust emission have also been produced using DART-Ray. Cross-check calculations between the codes 
show agreement in the calculation of radiation fields at a few
percent level \citep{Natale14}. For the absolute and comparative
performance of the codes, we refer the reader to \cite{2013MNRAS.436.1302P} and the above references for DART-Ray.
In order to model the detailed central region of M33, we adopt a minimum
spatial sampling of 50pc.

\section{Fitting the surface brightness photometry from the FUV to FIR}\label{sec:fitting}
The process of fitting the detailed surface brightness profiles of the observations is equivalent to optimising for the detailed geometry and amplitude (luminosity/opacity) of the stellar populations and of the dust. Due to the large number of geometrical parameters needed to model M33, a complete search of the whole parameter space with radiative transfer calculations is computationally prevented. Instead, we used an intelligent algorithm that takes into account the preferential effects of different parameters on the emission at specific wavelengths (previously shown in \citetalias{PT11}), making thus possible to avoid unnecessary parameter combinations. 

As M33 is a non-edge-on galaxy and therefore does not offer a mean for directly determining the vertical distribution,  we fixed  the relative scale-heights of stars and dust to the general trends derived from edge-on galaxies (see \citetalias{PT11}). We thus fixed $z^{\rm disc}_{\rm s}$ from \citetalias{PT11} (their Table E.1), to be the same at all wavelengths and to bear the same ratio to the B-band scale-length of a single exponential, as in \citetalias{PT11}. {This led to a value of 190\,pc for $z^{\rm disc}_{\rm s}$.} The scale-height of the dust disc $z^{\rm disc}_{\rm d}$ was taken to be in the same ratio to that of the old stellar disk as in \citetalias{PT11}, {namely 160\,pc}. The scale-height of the young stellar disk $z^{\rm disc}_{\rm s}$ was fixed to have the same absolute value as in \citetalias{PT11}, namely 90\,pc. Tests made to see how changes to the values adopted for the scale-heights could modify our solution showed minimal effects, as long as we did not change the general characteristic of the solution, with the old stellar disk being thicker than the dust disk, which, in turns, remains thicker than the young stellar disk. 

Azimuthally averaged radial profiles for the model were produced in the same manner as those for the observations, allowing a direct comparison between observed and model profiles. We first started the optimisation by considering single exponential functions for all stellar and dust distributions and some initial guess parameters taken either directly from data without dust effects considered (e.g. running GALFIT to available UV/optical/NIR profiles) or from general trends derived from our previous modelling.

{The use of single exponential functions in the radial direction turned out to produce a poor description of
 the observations. It became immediately apparent that the profiles at all wavelengths do not follow a single increasing exponential towards the centre of the galaxy, but have a series of changes at characteristic radii. These changes are observed as flattening/steepening of the profiles, meaning  changes in the gradient of the exponential function. This gradient clearly changes four times, indicating the need to use four disc components instead of only one. Because of this unambiguous feature of the observations we did not try to optimise for the number of components, but rather fix this to four components.}
\begin{figure*}
    \centering
    \includegraphics[width=\textwidth,trim={4cm 8cm 4cm 13cm},clip]{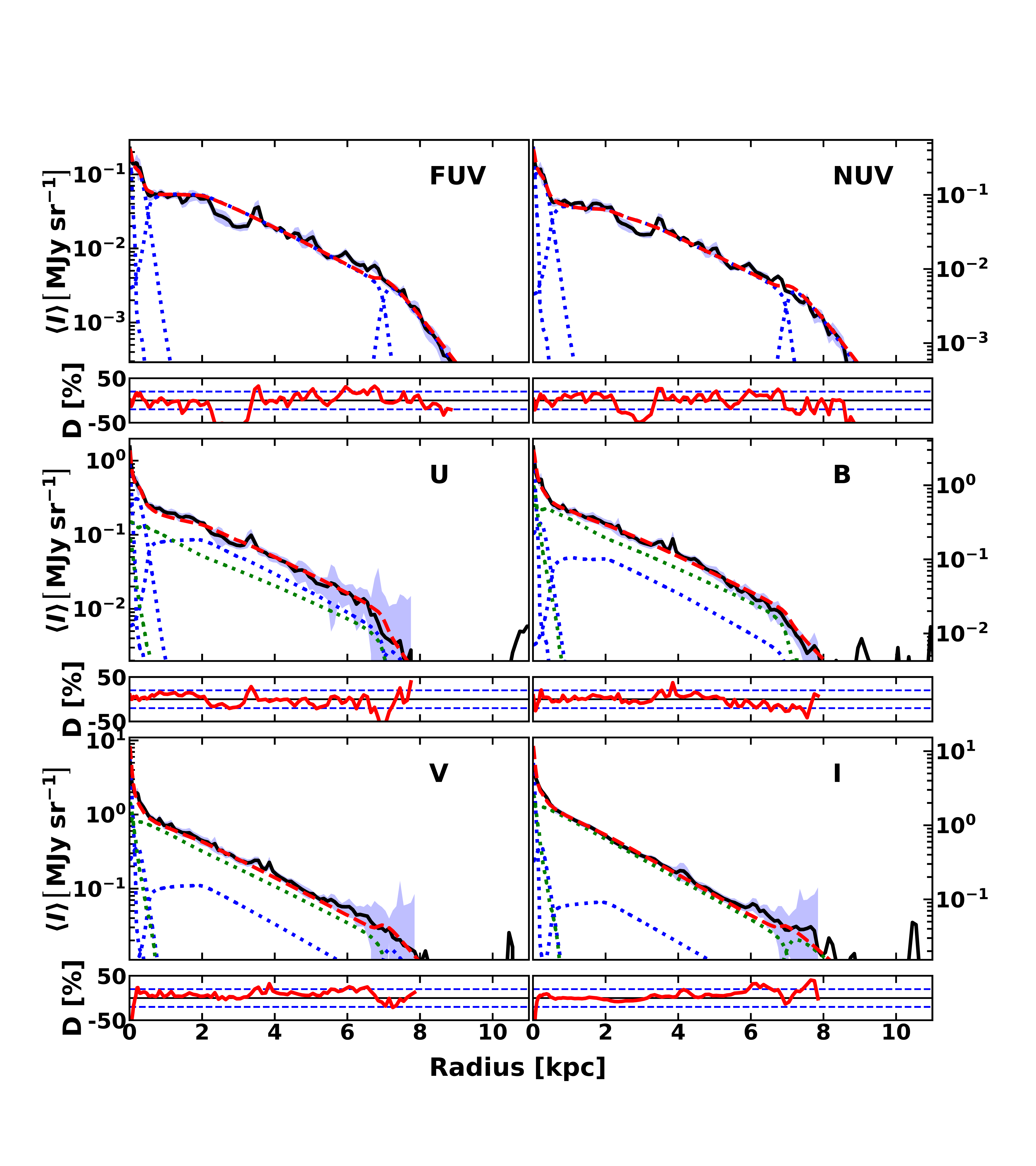}
    \caption{Comparison between azimuthally averaged radial profiles obtained from our model of M33 (red dashed-lines) and from observations (black solid-line). The sampling of both the observed and modelled profiles is as described in Sect.~\ref{sec:phot}. {The blue shaded area around the observed profiles represents the errors, which were calculated taking into account both systematic errors due to photometric calibration, as well as random errors due to background fluctuations and configuration noise, as described in Appendix~\ref{sec:err_calc} and calculated using Eqns.~\ref{eqn:av_SB_gal}-\ref{eqn:epsilon_SB}} The contribution of the nuclear, inner, main and outer thin stellar discs are plotted with blue dotted-lines. The contribution of the inner, main, and outer stellar discs are plotted with green dotted-lines. Small lower panels: The different panels show profiles in the FUV, NUV, U, B, V and I bands. At each band we also show residual profiles (plotted with red solid-line) in the lower panels. To guide the eye, the horizontal blue dashed-lines indicate the $\pm 20\%$ residuals.}
    \label{fig:avintprof1}
\end{figure*}
\begin{figure*}
    \centering
    \includegraphics[width=\linewidth,trim={4cm 8cm 4cm 13cm},clip]{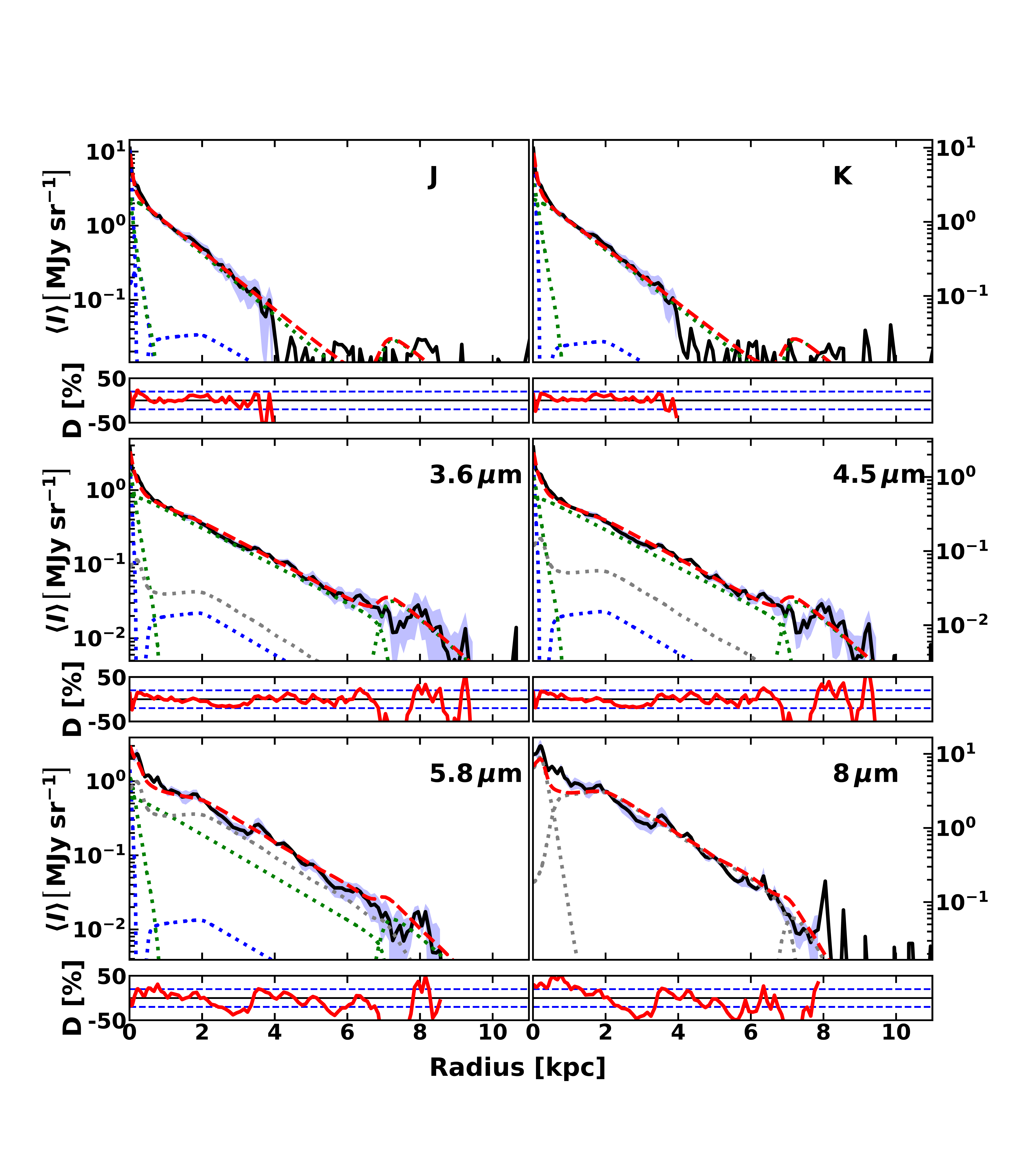}
    \caption{Same as in Fig.~\ref{fig:avintprof1}, but for the J, K, I1, I2, I3 and I4 bands. The dust emission contribution is plotted with grey-dotted lines. In the J and K bands the residuals are only plotted out to 4 kpc, since beyond this radius the galaxy is not detected in these bands, due to the observations being shallow.}
    \label{fig:avintprof2}
\end{figure*}

\begin{figure*}
    \centering
    \includegraphics[width=\linewidth,trim={2cm 13cm 2cm 17.5cm},clip]{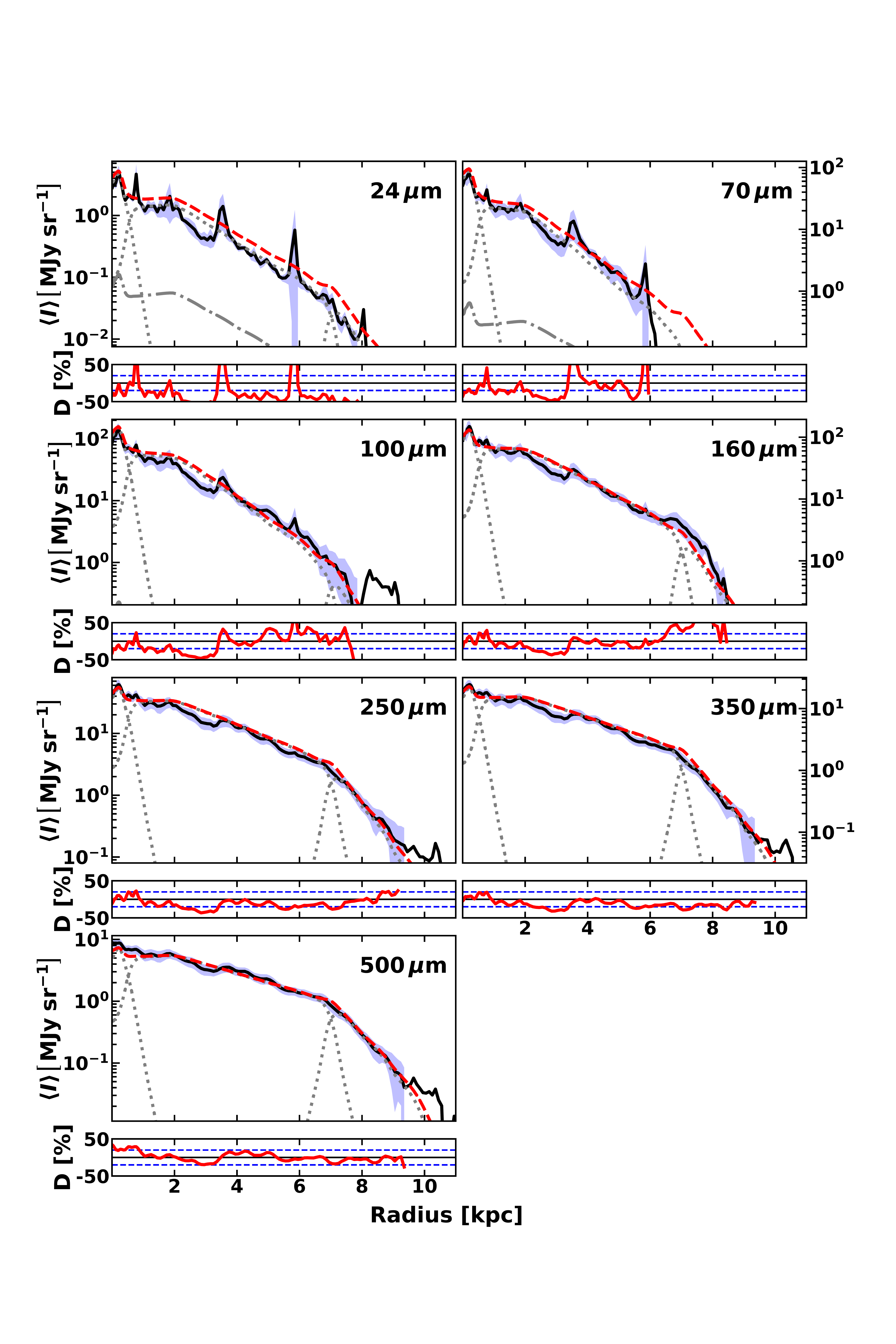}
    \caption{Same as in Fig.~\ref{fig:avintprof2}, but for the 24, 70, 100, 160, 250, 350 and 500\,$\mu$m bands.}
    \label{fig:avintprof3}
\end{figure*}

The following steps were taken in the optimisation process:
\\ \\
\noindent
1. Using the surface brightness profile of the UV data we constrained the geometry of the young stellar populations for an initial guess of the dust opacity. {As mentioned before, the UV profiles do not follow a single increasing exponential towards the centre of the galaxy. In order to fit the observed profiles} we split the stellar emissivity into four distinct components, each with an inner truncation radius $R_{\rm tin}$, an inner radius $R_{\rm in}$, and an outer truncation radius $R_{\rm t}$. In the range $R_{\rm tin}\le R < R_{\rm in}$ the profile is a simple linear function described by the parameter $\chi$ (see Eqn. \ref{eq:chi}). Beyond $R_{\rm in}$, each component follows an exponential as in the generic model, and is truncated at $R_{\rm t}$. These four components of stellar emissivity seen in the UV were taken to reside in  the four different morphological components mentioned in Sect.~\ref{sec:model}:  the nuclear, inner, main and outer discs.  We assumed a constant thin disc scale-height for all of them. The radius $R_{\rm in}$, at which the change in the gradient of the exponential occurs, was unambiguously determined for each morphological component by visual inspection of the radial profiles. The other parameters, such as scale-length 
$h_{\rm s}^{\rm tdisc}$
and spectral luminosity density 
$L^{\rm tdisc}(\lambda)$ were derived iteratively, by searching a grid of RT models for various combinations of these parameters. As 
mentioned before, the inclusion of the amplitude $L^{\rm tdisc}(\lambda)$ as free parameters represents a departure from the use of a fixed spectral template in the generic model from \citetalias{PT11}, and allows us to investigate variations in the star-formation history on timescales of a few 10s to a few 100s of Myr through analysis of the derived intrinsic UV colours. 

Thus, the optimisation of the UV data provided a first guess value for the parameters  $\chi_{\rm s}$,  $h_{\rm s}$ as well as for the amplitude
parameters $L^{\rm tdisk}(\lambda)$, {and a definitive value for 
$R_{\rm in,s}$, $R_{\rm tin,s}$,}
for each of the four young stellar disc components.
\\ \\
\noindent
2. Following \citetalias{PT11}, we fix the geometrical parameters of the thin dust disc to equal  that of the young stellar disk. Thus we set:
\begin{align*}
z^{(\rm n-tdisc, i-tdisc, m-tdisc, o-tdisc)}_{\rm d}&=z^{(\rm n-tdisc, i-tdisc, m-tdisc, o-tdisc)}_{\rm s},\\
h^{(\rm n-tdisc, i-tdisc, m-tdisc, o-tdisc)}_{\rm d}&=h^{(\rm n-tdisc, i-tdisc, m-tdisc, o-tdisc)}_{\rm s},\\ 
\chi^{(\rm n-tdisc, i-tdisc, m-tdisc, o-tdisc)}_{\rm d}&=\chi^{(\rm n-tdisc, i-tdisc, m-tdisc, o-tdisc)}_{\rm s},\\ 
R_{\rm in, d}^{(\rm n-tdisc, i-tdisc, m-tdisc, o-tdisc)}&=R_{\rm in, s}^{(\rm n-tdisc, i-tdisc, m-tdisc, o-tdisc)}, \\
R_{\rm tin, d}^{(\rm n-tdisc, i-tdisc, m-tdisc, o-tdisc)}&=R_{\rm tin, s}^{(\rm n-tdisc, i-tdisc, m-tdisc, o-tdisc)},\\ 
R_{\rm t, d}^{(\rm n-tdisc, i-tdisc, m-tdisc, o-tdisc)}&=R_{\rm t, s}^{(\rm n-tdisc, i-tdisc, m-tdisc, o-tdisc)}.
\end{align*}
\\ \\

\noindent
3. At $500\,\micron$ the emission from a galaxy is dominated by cold dust, coming from the diffuse component, and as such this emission is primarily an indicator of dust opacity. We thus used the SPIRE 500 band to constrain the parameters of the dust distribution.
Using the parameters determined in steps 1-2, we ran a new RT calculation and compared the $500\,\micron$ profile with the corresponding SPIRE 500 profile. As in the UV, we observed clear breaks to the exponential profile. However, we only find three distinct components in the diffuse dust emission rather than the four observed from stellar emission. These components correspond to the inner, main, and outer components of the galaxy seen in the UV, with the same inner radii, and the same inner flattening of the radial profiles. Thus, we constrained the parameters: 
\begin{align*}
&R_{\rm in, d}^{\rm i-tdisc}=R_{\rm in, d}^{\rm i-disc}=R_{\rm in, s}^{\rm i-tdisc}, \\
&R_{\rm in, d}^{\rm m-tdisc}=R_{\rm in, d}^{\rm m-disc}=R_{\rm in, s}^{\rm m-tdisc}, \\
&R_{\rm in, d}^{\rm o-tdisc}=R_{\rm in, d}^{\rm o-disc}=R_{\rm in, s}^{\rm o-tdisc}, \\
\end{align*}
\begin{align*}
&\chi_{\rm d}^{\rm i-tdisc}=\chi_{\rm d}^{\rm i-disc}=\chi_{\rm s}^{\rm i-tdisc}, \\
&\chi_{\rm d}^{\rm m-tdisc}=\chi_{\rm d}^{\rm m-disc}=\chi_{\rm s}^{\rm m-tdisc}, \\
&\chi_{\rm d}^{\rm o-tdisc}=\chi_{\rm d}^{\rm o-disc}=\chi_{\rm s}^{\rm o-tdisc}.
\end{align*}
Using a grid of models we derived a dust disc scale-length $h^{\rm disc}_{\rm d}$ and amplitude parameter, the opacity at the inner radius of the thick and thin dust disc   $\tau^{\rm f,(disc)}_{B}\left( R_{\rm in,d}\right)$ and $\tau^{\rm f,(tdisc)}_{B}\left( R_{\rm in,d}\right)$ respectively, for each morphological component. It should be noted that for each model containing a new trial value ($h_{\rm d}$, $\tau^{\rm f}_{B}$) we had to run step 1 to readjust the luminosity of the young stellar components. In summary, from the $500\,\micron$ data we derived first estimates for  $\chi_{\rm d}$, $\tau^{\rm f}_{B}\left( R_{\rm in,d}\right)$, $h^{\rm disc}_{\rm d}$ and a {definitive value for $R_{\rm in, d}$}, for each of the morphological components.
\\ \\

\noindent
4. With the constraints from steps 1-3, we ran a new RT model and compared the radial profiles in the optical and NIR wavelengths. At these wavelengths we see the dust attenuated emission from the old stellar population as the dominant source of emission. Nevertheless, residual emission from the young stellar populations is also present, in particular in the blue optical range, while at long NIR wavelengths emission from warm dust grains (PAHs) start to affect the bands. Throughout the optical/NIR range we found the same  functional form of the stellar emissivity as in the UV, including a very compact feature in the centre of the galaxy. We therefore modelled the emission with  nuclear, inner, main and outer stellar components, each consisting of an old and young stellar disc, except for the nuclear component for which only a young stellar disc was considered.  We found that for the inner and main components the old stellar discs were best fitted with pure exponential functions throughout $0\le R \le R_{\rm t}$, with no inner radii. 

The outer component was found to follow the same behaviour as in the UV range, with an inner radius $R_{\rm in}^{\rm o-disc}=R_{\rm in}^{\rm o-tdisc}$ and $\chi^{\rm o-disc}=\chi^{\rm o-tdisc}$. 
From these observations we optimised for the scale-length $h_{\rm s}^{(\rm disc)}$  and amplitude parameters  
$L^{(\rm disc)}(\rm U,B,V,I,J,K,\rm I1,I2,I3)$, and  $L^{\rm tdisc}(\rm U,B,V,I,J,K,\rm I1,I2,I3)$ for all the morphological components. Exception to this is the outer disc, for which we were not able to optimise for the scale-length in the J and K band, since these observations were too noisy. Because we found the scale-length of the outer disc to be constant at all the wavelengths we optimised, we simply fixed to this constant value the J and K band scale-length.

Thus, in this step we derive estimates for $h_{\rm s}^{(\rm disc)}$, $L^{(\rm disc)}(\lambda)$, and $L^{\rm tdisc}(\lambda)$, with $\lambda$ in the optical range, while also setting $R_{\rm in}$ and $\chi$ for the stellar emissivity of the old population.\\ \\

\noindent
5. 
Using constraints from steps 1-4, we performed a new RT calculation and looked at all available wavelengths, in particular in the  NUV and $70\,\micron$. This allowed us to fine-tune the parameters of the inner component. Thus we had to slightly increase the scale-length and truncation radius of the inner component's thin dust disc $h^{\rm i-tdisk}$  with respect to that of the young stellar disc. At this stage we also used the 24 micron observations to constrain the localised component.
\\ \\

\noindent
6. Using constraints from steps 1-5, we ran a new RT model and compared the model to observation at all available wavelengths. Various rescalling of the amplitude parameters were needed and several iterations were required before convergence was achieved at all wavelengths.
\\\\

{It should be noted that, despite the complexity of the problem, the optimisation is relatively straightforward.
This is because the optimisation of the different parameters can be done sequentially, one wavelength at a time, which is a  consequence of the fact that it is possible to separate the signature of the dust from that of the stars.
In other words it is possible to identify a spectral range where only one component dominates - e.g. the 500 micron data is mainly shaped by dust opacity and not by heating sources.

In addition, for any wavelength where a fit is performed, and for a given component, the main fitted parameters are only a (radial) exponential scale-length and a corresponding amplitude parameter. This is because 
the scale-height parameters are fixed from generic trends, the inner radius of each morphological component is also fixed from observations,  while the $\chi$ parameter  is  mainly  a  shallow  inner truncation,  which  is  usually  derived  as  to  provide a  smooth  transition  between  various  components.
As such the initial guess of the two main fitted parameters
is achieved by fitting the apparent one-dimensional
profile at the required wavelength, followed by only
a couple of radiative transfer calculation iterations
(going  through  steps  1, 2, 3, 4,  as  described  above).
Because of its simplicity the fit is done by eye, and checked through a calculation of the corresponding chi-squared ${\rm chi}^2$ value. The parameter space of each length and amplitude parameter
is usually sampled within the $\pm 30\%$ around the best
fit model, with a $5\%$
step accuracy.

Another important aspect of our optimisation method is that there are no degeneracies in the parameter space. This was already discussed at length in \citetalias{PT11} for the modelling of the integrated SED, but it applies even more so for a resolved study as the one in this paper. The main point here is again the fact that the 500\,${\mu}$m wavelength is almost entirely shaped by the distribution of dust, with small influence from heating sources, and therefore can be used to derive the parameters of the dust. At the short wavelength end, the UV data are dominated by the young stellar populations, and, for a good initial guess of the dust distribution, they can be used via a radiation transfer calculation to derive the geometry of the young stellar disc in each UV band. In addition there are no degeneracies between reddening and age/metallicity, since we do not fit a stellar population model to the data, but derive the intrinsic UV and optical luminosity density at each (observed) wavelength by directly solving the inverse problem.    
}

The resulting model profiles together with the corresponding observed ones are plotted in Figs. \ref{fig:avintprof1}-\ref{fig:avintprof3}, showing an overall good agreement between the model and observations at all observed wavelengths. 
We also calculated the residuals D between observation and our model
\begin{equation}
    {\rm D} = \frac{{\rm observation}-{\rm model}}{{\rm observation}}
\end{equation}
and plotted these residuals in Figs. \ref{fig:avintprof1}-\ref{fig:avintprof3}. The residual plots show that indeed a good fit was achieved, {with residuals typically within $7\%$}. 

\begin{figure}
    \centering
    \includegraphics[width=\linewidth]{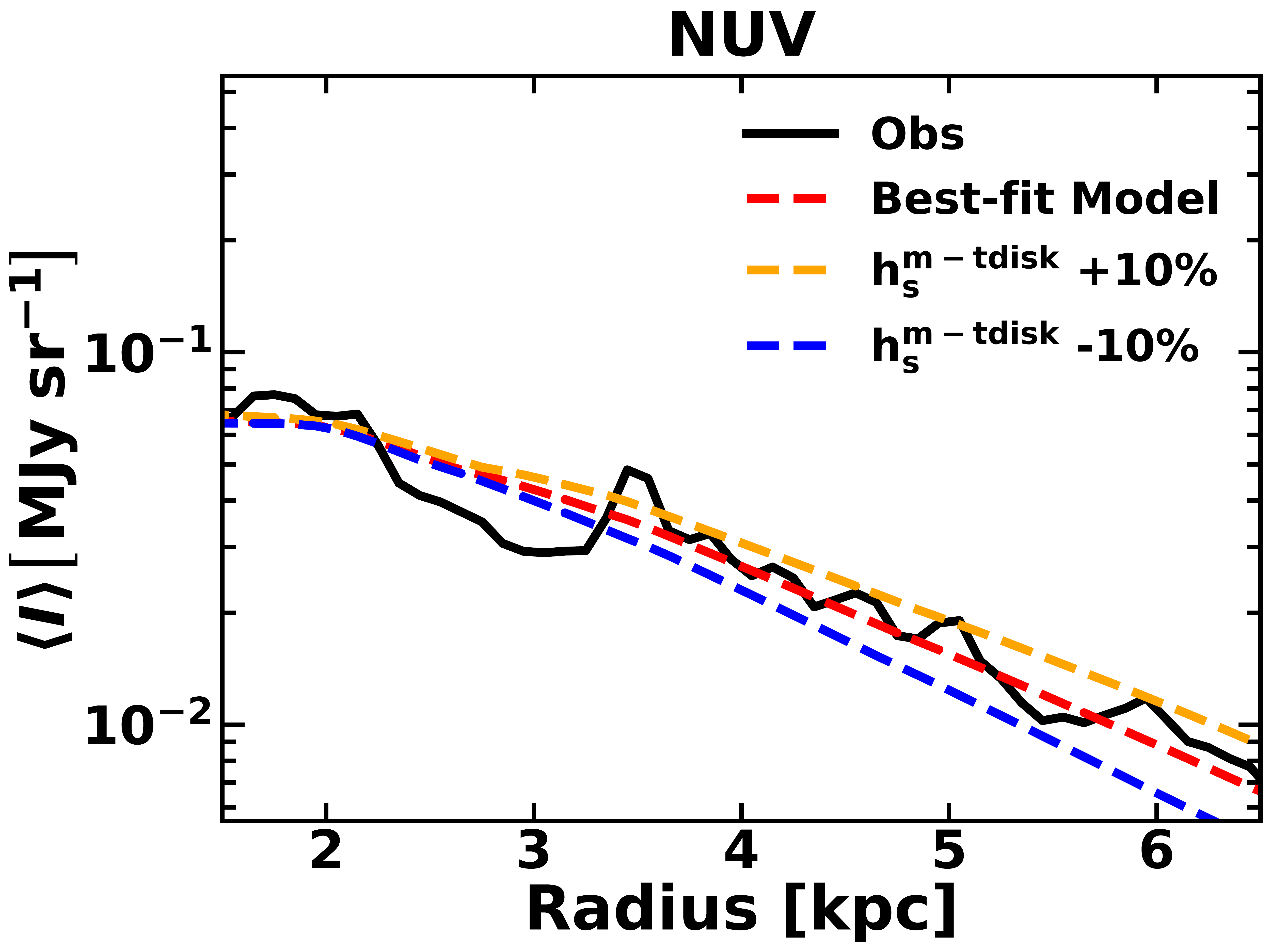}
    \caption{Azimuthally averaged radial surface brightness profiles in the NUV band showing effects of a $\pm10\%$ deviation in the scale-length of the main thin stellar disc and $\left(^{+3}_{-1}\right)\%$ in $L_{\rm s}^{\rm m-tdisc}$, from the best fit model.}
    \label{fig:nuvem}
\end{figure}

Uncertainties in the resulting model  parameters  have been derived by looking at the deviation from the best-fit model, one pair of parameters at a time, where the pair represents a geometrical parameter and a corresponding amplitude parameter, at the wavelength it was optimised. This is because any change in a geometrical parameter is accompanied by a change in the amplitude of the profile, and as such these parameters are not independent to each other. For example the scale-length of the dust disc changes jointly with dust opacity,  the scale-length of the young stellar disc changes jointly with the luminosity of the young stellar disk, and the scale-length of the old stellar disc changes jointly with the luminosity of the old stellar disc. An example of deviations from best fit values is shown in Fig. \ref{fig:nuvem}, for the scale-length of the main thin stellar disc, $h_{\rm s}^{\rm m-tdisc}$ and corresponding  $L_{\rm s}^{\rm m-tdisc}$ in the NUV band. The figure shows the average radial surface brightness profiles for changes of  $\pm 10\%$  in $h_{\rm s}^{\rm m-tdisc}$ and corresponding $+3/-1\%$ in $L_{\rm s}^{\rm m-tdisc}$ , which are taken to be representative errors in these parameters. The resulting best-fit model parameters and their associated errors are given in Table~\ref{tab:ogeom}.

{The errors in the fit can also be quantified through a chi-squared calculation at the specific wavelengths where the model has been optimised. Thus, for a given $\lambda$, the chi-squared was calculated as:}
\begin{align}
    {\rm chi}_{\lambda}^2&=\sum\limits_{n=1}^{N}\frac{({O}_{\rm n} - {M}_{\rm n})^2}{\varepsilon_{{\rm SB,n}}^{2}}\label{eqn:chisq}\\ 
    {\rm chi}^2_{r,\lambda}&=\frac{{\rm chi}_{\lambda}^2}{N}\label{eqn:chisqr}\\
\end{align}
{where $N$ is the number of annuli for which the photometry was performed, $O_{\rm n}$ and $M_{\rm n}$ are the azimuthally averaged surface brightnesses within the annulus $\rm n$ for the observed and modelled radial profiles, respectively, and $\varepsilon_{\rm SB, n}$ is the error within annulus n, as derived using Eqns.~\ref{eqn:av_SB_gal}-\ref{eqn:epsilon_SB}. The variables $N$, $O_{\rm n}$, $M_{\rm n}$ and $\varepsilon_{\rm SB,n}$ are all functions of $\lambda$.} In Table~\ref{tab:chi2} we show the reduced {chi-squared} ${\rm chi}^2_{r,\lambda}$ values for the best-fit model and the upper and lower error models, in the wavelengths range for which various parameters were optimised. Thus for the parameters related to the young stellar populations the ${\rm chi}^2_{\lambda}$ is calculated in the NUV, for the old stellar populations in the NIR range, and for the dust parameters at 500\,${\mu}$m. The values from Table \ref{tab:chi2} show a minimum for the best-fit model in  most cases. 

{The reduced chi-squared value for the model across all wavelengths (for which comparison with observed data has been performed) $l=[1,L]$  is given by:
\begin{align}
    {\rm chi}_{\rm r}^2&=\frac{\sum\limits_{l=1}^{L}{\rm chi}_{\lambda,l}^2}{\sum\limits_{l=1}^{L}N_{\rm l}}\label{eqn:modelchisq} 
\end{align}
where ${\rm chi}^2_{\lambda, \rm l}$ is the chi-squared as defined by Eqn.~\ref{eqn:chisq}. The derived value is ${\rm chi}_{\rm r}^2=2.25$.}

\begin{table}
	\centering
	\caption{The ${\rm chi}^2_{\rm r}$ {(see Eqns.~\ref{eqn:chisq}-\ref{eqn:modelchisq})} values for the best-fit model and the upper and lower error models at the wavelengths where the model was optimised.}
	\label{tab:chi2}
	\begin{tabular}{llll}
		\hline
		Band & Best & e+ & e- \\
		\hline
		 NUV & 2.85 & 4.21 & 3.46\\
		 I1  & 0.95 & 1.26 & 1.09 \\
		 SPIRE 500 & 0.42 & 0.89 & 9.19 \\
		 \hline
	\end{tabular}
\end{table}

\section{Results}\label{sec:results}

\subsection{Fits to the surface brightness distribution}

As mentioned before, the fits to the surface brightness profiles, as depicted in Figs.~\ref{fig:avintprof1}, \ref{fig:avintprof2} and \ref{fig:avintprof3} show an overall good agreement with the data, with relative residuals less than $20\%$ in most cases. Inspection of the observed profiles show that the process of azimuthally averaging produces smooth exponential profiles, making the galaxy suitable for fitting with analytic axi-symmetric functions. Nonetheless, some imperfections to the smooth nature of the curves still occurs, in particular where the young stellar population dominates the output. This is due to some local strong asymmetries, but also to some residual radial features not included in our model. {A particular notable feature is that occurring around 3\,kpc in the UV profiles. There is a strong depression seen throughout the annulus centred at this position, indicating a real radial feature.} {There is also a peak at around 3.8kpc, which is due to the bright star-forming region NGC604}.
These features make the residuals at this location and these wavelengths rather larger (above $20\%$). The worse deviation from smoothness in the profile is, as expected, at 24\,${\mu}$m, where the contrast between the localised emission from SF regions and the diffuse emission is at a maximum. {At this wavelength, the features seen in the NUV, namely the depression at 3\,kpc and the peak at 3.8\,kpc due to NGC604 are even more pronounced}. In addition there are several peaks at around 1, 6  and 8\,kpc, where emission from SF regions, preferentially located in spiral arms, dominates. This situation in particular affects the profiles of M33, since the galaxy is at such close proximity, and therefore we resolve lots of small structure.

Figs ~\ref{fig:avintprof1} and \ref{fig:avintprof2} show that the observed profiles in the B,V,I,J,K bands in the region dominated by the main 
disc continuously steepen with  increasing wavelength. Although the observations in the J and K range become quite noisy beyond 4\,kpc, the slope of their profiles is still clearly defined within the inner 4\,kpc, enabling us to infer the steepening effect mentioned above. In our model this trend is well fitted through a decrease in the intrinsic scale-length of main disc old stellar populations with increasing wavelength. However, at even longer wavelengths, in the IRAC 3.6, 4.5 and 5.8\,${\mu}$m, the slope of the profile reverses again, with the profiles becoming shallower. This is consequently fitted by an increase in scale-length.

The fits to the IRAC profiles (Fig.~\ref{fig:avintprof2}) show how the old stellar population, the dominant component at $3.6\,{\mu}$m, starts to decrease in weight in progressing to $4.5\,{\mu}$m, and is taken over by the dust emission component at $5.8\,{\mu}$m. Only in the nuclear region does the stellar component still dominate the emission at $5.8\,{\mu}$m. At $8.0\,{\mu}$m, which covers a PAH emission feature,  the profiles are completely dominated by the dust emission components. An interesting feature of these profiles is the outer disc, who is well defined at these wavelengths, and is counterpart to the UV emitting outer disc.

At 350 and 500 microns the profiles are well fitted by our model, showing no sign of a so-called {``}submm excess" (\cite{2010A&A...523A..20B,2011A&A...532A..56G,2013ApJ...778...51K,2013A&A...557A..95R}). In particular the outer disc is well defined out to 10\,kpc, and is again well fitted by our models. In the FIR the profiles show steeper profiles (smaller extent) and a very weak outer disc. This behaviour is naturally predicted by our self-consistent model, and can be explained as a result of the low intensity radiation fields in the outer disc producing a low temperature heating of the dust grains, with the peak of their emission SEDs shifted towards longer submm wavelengths. It is interesting to note that in fact all the images in the 70 to 350 micron range are in fact not fitted, but predicted by our model. Accordingly, having fitted the 500 micron images to constrain the bulk of the dust distribution, and the UV/optical images to constrain the heating sources, the FIR regime, which sees the convolution between dust opacity and heating sources, needs to be {exactly predicted by the model}, if the geometry is correctly derived. And indeed, it is not only the overall amplitude, but  also the increase in scale-length of the FIR emission with increasing wavelength,  that is well accounted for and predicted by our model. 

\subsection{The global SED of M33}

\begin{figure}
    \centering
    \includegraphics[width=\linewidth]{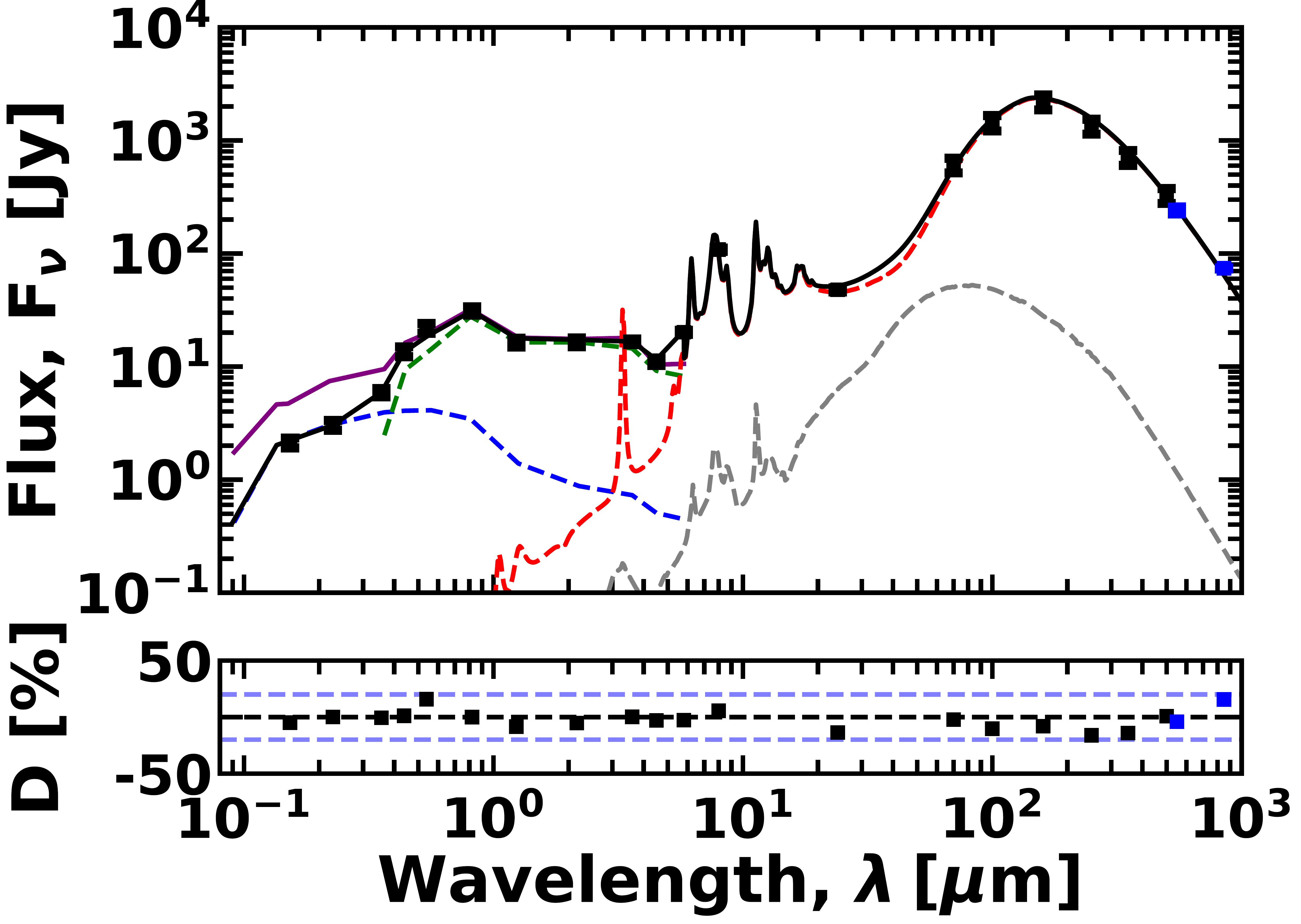}
   \caption{Comparison between the model and the observed global SED of M33. The data used in the optimisation procedure are shown as black square symbols. The error bars on the data (mainly contained within the square symbols in the UV/optical range) represent the one-sigma uncertainty in the observed flux densities, as listed in Table~\ref{tab:obs_ccorr}. The model is represented by the black solid line. The blue squares represent the data from \citeauthor{2018MNRAS.477.4968T} (\citeyear{2018MNRAS.477.4968T}), which were not used in the optimisation, but only shown for comparison. The different components of the model SEDs are plotted with dashed-lines, as follows:  old stellar disks in green, young stellar disks in blue, diffuse dust emission in red, clumpy component in grey. The intrinsic stellar SED is also plotted with solid purple line. Relative residuals D between data and model are plotted in the lower panel, with the $\pm20\%$  residuals indicated with light-blue dashed-lines (for the purpose of guiding the eye).} 
    \label{fig:sed_stell}
\end{figure}

The fit to the UV-FIR/submm azimuthally average profiles resulted in a model that can successfully account for the global emission SED 
of M33. Indeed, in Fig. \ref{fig:sed_stell} one can see that the spatially integrated model SED resembles very well the observed SED of M33, with the average relative residuals between model and data $\left<|{\rm D}|\right>$ of only $5.9\%$ and a maximum residual |D| of $16.2\%$. In particular the balance between energy absorbed in the UV/optical and energy emitted in the infrared is well matched, with no predicted flux density outside the expected error range. This is consistent with axi-symmetric RT models being well suited to fit SEDs of star-forming non-edge-on galaxies. 

Fig.~\ref{fig:sed_stell}  also shows the intrinsic stellar SED of M33 (as it would be observed in the absence of dust), with the difference between the apparent (black) and intrinsic (purple) model SED representing the absorption by dust. We have calculated that $35\pm3\%$ of the stellar light is absorbed by dust and re-emitted in the far-infrared, which is a typical value for late-type spiral galaxies \citep{2002MNRAS.335L..41P,2016A&A...586A..13V,2018A&A...620A.112B}. Another aspect of interest shown by Fig.~\ref{fig:sed_stell} is that the global intrinsic FUV/NUV, is quite red compared to the unity ratio expected for a constant SFR, showing that at the present epoch, the global SFR of M33 is rapidly declining, on a timescale of order 100\,Myr. {Despite this M33 shows signs of recent star-formation activity, at least in selected regions. For example \cite{2009ApJ...699.1125R} found that they could model a set of luminous HII regions in M33 assuming a recent burst of age 4 Myr. Taken together with our results, this would imply that star-formation in M33, although ongoing at the present epoch, was higher 100 Myr ago. This decrease in SFR may also account in part for the relative paucity of localised emission from star formation regions that we derive in our modelling compared to other galaxies we have analysed.}

\begin{figure}
    \centering
    \includegraphics[width=\linewidth]{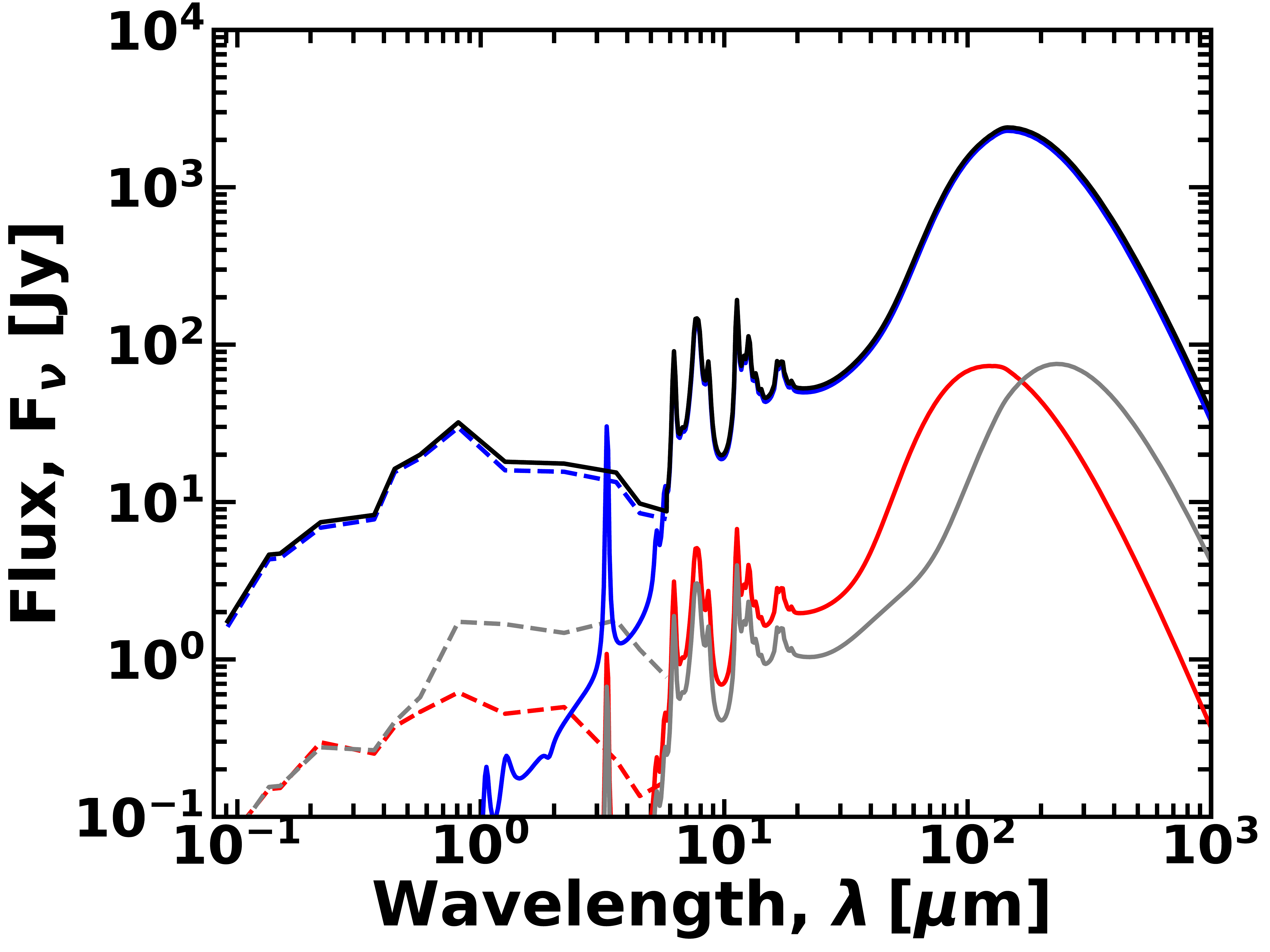}
    \caption{The predicted intrinsic SED of M33, together with the contribution from the different morphological components: main disc, inner+nuclear disc, and outer disc. The model is represented by the black solid line. The individual stellar and dust components are plotted with dashed-lines and solid lines, respectively, as follows: the inner in red, the main in blue and outer in grey.}
    \label{fig:sed_comp}
\end{figure}

When optimising for the geometry of M33 we found several morphological components in addition to the main disc. It is therefore of interest to see what is the contribution of these components to the global SED. For this purpose we plotted in Fig.~\ref{fig:sed_comp} the predicted intrinsic SED of M33, together with the contribution of the nuclear+ the inner, the main and the outer discs. It should be emphasised that the dust emission from the different morphological components do not necessary correspond to the same stellar emission component, in the sense that for example the heating of the dust can come from photons emitted by all morphological components. As expected from its spatial extent, it is the main disc that dominates the emission SED. The main disc thus contributes $92.6\%$ to the stellar light, and $94.2\%$  to the dust emission. By contrast, the inner disc contributes only $2.5\%$ to the stellar emission and $3.7\%$ to the dust emission. The outer disc contribution is similar to that of the inner disc, making $4.6\%$ and $2.1\%$ to the stellar and dust emission output of M33. There is also a nuclear disc, but, due to its small spatial extent, it has a negligible contribution to the bolometric output of M33 (only $0.3\%$ to the stellar emission). 

The dust emission SED of the inner disc peaks at around $100\,\micron$, characteristic of $\sim29{\rm K}$ dust, being much warmer than the SED of the main disc peaking at around $160\,\micron$ and having $\sim 18{\rm K}$. The warm infrared SED of the inner disc is due to an increased surface density of SFR within its confines, as we will see in Sect. \ref{sec:SFR}. Conversely, the outer disc contains cold dust, around $\sim12{\rm K}$ , and peaks at around $250\,\micron$. We find that  dust emission in M33 is mainly powered by emission from the young stellar disks, which account for
{$80\pm{8}\%$}  of the dust heating.

The predicted intrinsic flux (spectral) densities (integrated out to the truncation radius) in several UV-optical photometric bands of interest are listed in Table~\ref{tab:intrinsic_flux}. The fluxes are given both for the whole of M33, but also for the individual morphological components. In addition, we also give in Table~\ref{tab:intrinsic_flux_effective} the corresponding fluxes out to the effective radius in I-band. 
The fluxes from Table~\ref{tab:intrinsic_flux_effective} may be used when comparing the properties of M33 to those of other galaxies, in particular in statistical surveys of distant galaxies with tabulated values of $R_{\rm eff}$. The I-band effective radii in both intrinsic and dust-attenuated light are tabulated in Table~\ref{tab:intrinsic_radius_effective} for the global emission as well as for the individual morphological components.

\section{Intrinsic properties of M33}
\label{sec:intrin}

\subsection{Star-formation rate}\label{sec:SFR}
We derive a ${\rm SFR}=0.28^{+0.02}_{-0.01}\, {\rm M_{\sun}yr^{-1}}$, where the quoted errors only include a random component due to uncertainties in the model parameters within the axi-symmetric formalism. Here we did not include any systematic sources of error, like for example errors due to departure from axi-symmetry. The derived SFR is within the range of recent values found in the literature for M33. \cite{2018MNRAS.479..297W} derived a ${\rm SFR}$ of $ 0.25^{+0.01}_{-0.07}\, {\rm M_{\sun}yr^{-1}}$  using the FUV+$24\,\micron$ method \citep{Leroy_2008} and $ 0.33^{+0.05}_{-0.06}\, {\rm M_{\sun}yr^{-1}}$ from a global fit and the MAGPHYS code \citep{2008MNRAS.388.1595D}.  \cite{2019MNRAS.483..931E} derive a SFR of  $0.34^{+0.08}_{-0.07}\, {\rm M_{\sun}yr^{-1}}$  using $12\,\micron$ estimates and the calibrated \emph{global WISE} W3 luminosities to SFRs presented in \cite{Cluver_2017}, and $ \left(0.44\pm 0.10\right)\, {\rm M_{\sun}yr^{-1}}$, based on the linear relation between $100\,\micron$ emission and total SFR presented by \cite{2010A&A...518L..70B}.

{As expected}, we find that the majority of star formation occurs within the main disc ($ 0.5 \le {\rm R} \le 7\,{\rm kpc}$),  {since it is this component that dominates the galaxy both in terms of size and bolometrics}. We find ${\rm SFR}^{\rm m} = 0.26^{+0.02}_{-0.01}\, {\rm M_{\sun}yr^{-1}}$. The remaining $0.02\, {\rm M_{\sun}yr^{-1}}$  is distributed throughout the nuclear, inner, and outer thin stellar discs.
Taking into account the physical extent of each component, we find  
a monotonically decreasing ${\rm SFR}$ surface density, $\Sigma_{\rm SFR}$, from the inner to the outer disc. Looking across the entire galaxy we derive a $\Sigma_{\rm SFR}=16.3^{+1.}_{-0.7}\times 10^{-4}{\rm M_{\sun}yr^{-1}kpc^{-2}}$ {within the truncation radius of the model}.

The SFR and $\Sigma_{\rm SFR}$ out {to} the truncation radius of M33, as well as of its morphological components are listed in Table \ref{tab:SFR_Rt}. In addition, we give in Table \ref{tab:SFR_Reff} the corresponding numbers calculated out to the intrinsic I-band effective radius.

\subsection{Dust optical depth and dust mass}

The dust face-on optical depth has a maximum value at the inner radius of the inner disc, with $\tau^{f}_{\rm B}\left(R_{\rm in, d}^{\rm i-disc}\right)=1.3\pm 0.1$. The face-on optical depth at the inner radius of the main and outer disc are $0.89^{+0.02}_{-0.04}$ and $0.40\pm 0.02$ respectively. This is consistent with M33 being optically thin in the B band throughout the main and outer disc, and becoming moderately optically thick in the inner disc. {The average face-on optical depth of the galaxy, weighted by the surface area, is $\tau^{f}_{\rm B, area}=0.35$, indicating again that over much of the extent of the galaxy, M33 is optically thin in the B-band when observed face-on. However, when weighting by flux we get $\tau^{f}_{\rm B, flux}=1.24$, showing that most of the luminosity is emitted where dust opacity is higher.

Although M33 is optically thin over most of its extent, the relatively larger optical depth ($\tau^f_B=1.3$) at the inner radius of the inner disk, where the luminosity of the galaxy is higher, means that corrections for total luminosity densities due to dust attenuations will be significant in the B-band and in the UV range. This can be seen both in the difference between the observed and intrinsic flux densities of M33, as listed in Table~\ref{tab:obs_ccorr} and Table~\ref{tab:intrinsic_flux}, but also in the fraction of stellar light absorbed by dust in M33. In addition, dust attenuation not only affects the integrated luminosities of the system, but also the appearance of the images, in particular in the UV. This will be discussed in Sect.~\ref{sec:EoA}.}

We have calculated a dust mass of $M_{\rm d}=14.1^{+0.3}_{-0.5}\times10^6{\rm M_{\sun}}$ for M33. The main disc contains the majority of the dust mass $M^{\rm m}_{\rm d}=11.3^{+0.3}_{-0.5}\times10^6{\rm M_{\sun}}$, with the remainder  $2.8\times10^6{\rm M_{\sun}}$ distributed between the inner and the outer disc. We list the dust masses of the individual morphological components and of M33 in Table~\ref{tab:dust_mass}. {Comparison with other estimates of dust mass only make sense if the same optical constants are used in the calculations. Because of this we compare our results with those of \cite{2016AA...590A..56H} who used our generic RT model (\citetalias{PT11}) to fit the integrated SED of M33. They derived a dust mass of $M_{\rm d}=13.8^{+17.5}_{-10.0}\times10^6{\rm M_{\sun}}$ for M33, which is consistent with our results within errors. Our dust mass estimates seem to be larger than the $M_{\rm d}=9.75\times10^6{\rm M_{\sun}}$ derived in \cite{2018MNRAS.479..297W} from a pixel-to-pixel analysis of M33 and same optical constants for the dust. However, since no errors have been given in \citeauthor{2018MNRAS.479..297W} it is difficult to assess whether the difference is significant or not.} 

Assuming a total gas mass ($ M_{\rm G}={ M_{\rm H_I} + M_{\rm H_2} + M_{\rm He}}$) $M_{\rm G}\sim 3.2 \times 10^9{\rm M_{\sun}}$ \citep{2003MNRAS.342..199C} with a $20\%$ uncertainty, we derive a gas-to-dust ratio ${\rm GDR} = 230\pm50$. This is in agreement with  \cite{Gratier17} who finds a GDR of ~200-400 from the central to the outer regions of M33. We are also in agreement with  \cite{2014A&A...563A..31R} for the expected GDR of a half solar metallicity galaxy, as is the case of M33.  
When comparing our GDR derived for M33 with the GDR derived from our previous RT modelling of the Milky Way \citep{Popescu17} we find that the GDR nicely scales with the metallicity as $Z^{-1}$, as expected if both galaxies have the same fixed proportion of metals in the solid state.

\subsection{The attenuation curve of  M33}
\begin{figure}
    \centering
    \includegraphics[width=\linewidth]{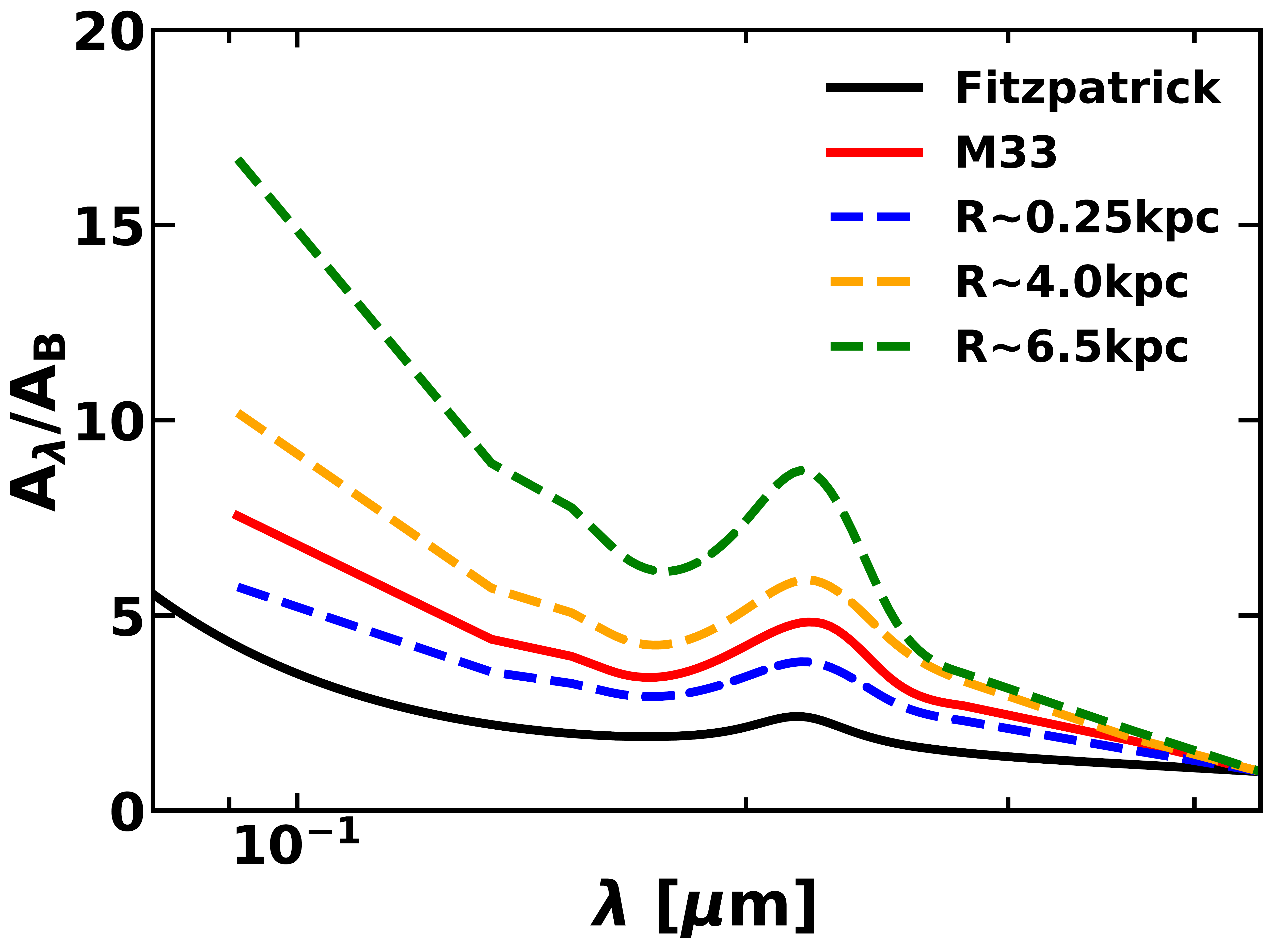}
    \caption{Comparison of the global attenuation curve of M33 (at the observed inclination of the disc; red solid line) with the average extinction curve of the Milky Way (black solid line) from \citeauthor{1999PASP..111...63F} (\citeyear{1999PASP..111...63F}),  normalised in the B band. Also over-plotted with dashed-lines are the attenuation curves of M33 at various radii.}
    \label{fig:atten-rad}
\end{figure}

The attenuation curve of a galaxy is an important, yet usually unknown function, as it incorporates not only the effect of dust extinction (depending on the optical properties of dust grains), but also the effect of geometry \citep{1989MNRAS.239..939D,2004A&A...419..821T}. Since the output of our model is the geometry of stars and dust in M33, we can predict the effect the geometry has on the overall attenuation and exactly calculate the attenuation curve, for the fixed dust model used in this paper. In Fig.~\ref{fig:atten-rad} we show the predicted UV attenuation curve of M33 derived from our model, as compared with the extinction curve of the Milky Way. The latter is computed from \cite{1999PASP..111...63F}, which is a principal empirical constraint for the grain model from \cite{2001ApJ...548..296W}, that we use as input to our RT calculations. Therefore, Fig~\ref{fig:atten-rad}  is to be interpreted entirely in terms of the effects of geometry. It is immediately apparent that M33's attenuation curve is much steeper in the UV than the Milky Way extinction curve (if both curves are normalised in the B-band). Similar results have been obtained in the RT modelling of M51 \citep{2014A&A...571A..69D} and of M31 \citep{2017A&A...599A..64V}. Another interesting feature of the attenuation curve is the $2200$\,\AA\ bump, which for M33 is deeper than in the MW extinction curve. We find the width of the bump to be only marginally larger than the MW bump. In the modelling of M51 and M31 \cite{2014A&A...571A..69D} and \cite{2017A&A...599A..64V} find a significant broader width for the bump, although \cite{2017A&A...599A..64V} does not consider this to be a real effect, but attribute this to the less good fit their model has in the NUV. 

\begin{figure}
    \centering
    \includegraphics[width=\linewidth]{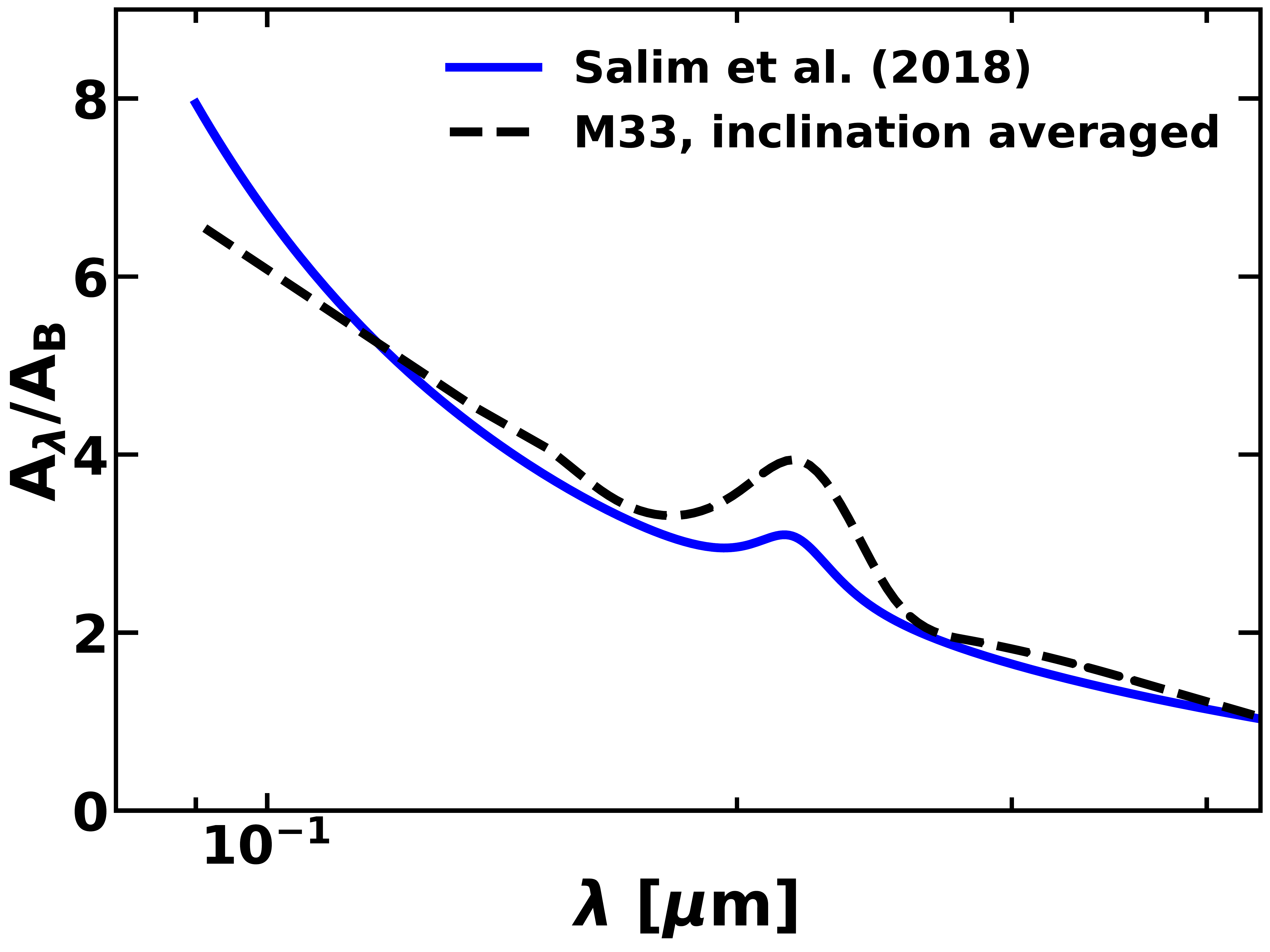}
    \caption{{Comparison between the inclination averaged attenuation curve of M33 and the average attenuation curve from \citeauthor{2018ApJ...859...11S} (\citeyear{2018ApJ...859...11S}), for the mass range of M33.}}
    \label{fig:atten-rad-salim}
\end{figure}

{In Fig.~\ref{fig:atten-rad-salim} we also compare our results for the inclination-average attenuation curve of M33 with the average attenuation curve from \cite{2018ApJ...859...11S}, for the mass range of M33. The inclination-average is needed since the curve from \cite{2018ApJ...859...11S} is also an average over a population of galaxies seen at various random orientations. The overall curves do not look dissimilar, at least down to the FUV filter, where observations do constrain the model of M33. The range between 912\AA\ and the FUV filter is more uncertain, since the attenuation of M33 is only predicted at these wavelengths, and not determined from data. The 2200\AA\ bump is more pronounced for M33. This may again be due to the effect of the detail geometry being taken into account in the modelling of M33 with respect to the energy-balance approach in the \citeauthor{2018ApJ...859...11S} curve.}

Because we derive the radial dependence of the stellar emissivity and dust distribution, we can also predict the attenuation curve of M33 as a function of radial position. This is shown in Fig. \ref{fig:atten-rad} with dashed lines. We find a monotonic trend of steepening the slope of the attenuation in progressing from the inner disc to the outer disc, and a more pronounced 2200 \AA\ bump in the outer regions than in the inner ones. This trend can be understood in terms of the galaxy being optically thick throughout the UV range in the inner disc, but transiting from optically thick in the FUV to optically thin in the B band within the outer disc.

The monotonic radial dependence of the attenuation curve in M33 is quite remarkable, and shows that local measurements of SFRs and other related quantities would present systematic errors, dependent on galactocentric radius,  if only a fixed attenuation law would be used to correct for dust effects. 

We conclude that the data is consistent with the dust in M33 having the optical properties of Milky Way dust, with the large 
scale variations in attenuation curve controlled by geometric effects.

\subsection{The morphological components of M33}

One of the main findings of our modelling of M33 is the existence of several morphological structures, each characterised by different geometrical parameters for stars and dust. In the following {section} we describe the characteristics of these structural components.

\subsubsection{The nuclear disc}

The nuclear disc is dominated by young stars concentrated within a small structure of 100\,pc radial extent and scale-length $h_s^{\rm n-tdisc}$ of 20\,pc, as seen in the lower panel of Fig~\ref{fig:stel_dust} and also in Fig.~\ref{fig:comp_nuv}. In the optimisation analysis, we did not include a counterpart of this structure in the distribution of old stars (as it was not possible to spatially separate such a component), although the nuclear disc spatially overlaps with the smooth distribution of old stars coming from the underlying inner and main discs, the latter not being truncated  at an inner radius. The data didn't require a dust counterpart to the young stellar emission from the nuclear disc, neither is there much dust from the underlying larger scale structure; the nuclear disc resides in a central {``}hole" for the distribution of dust, in which the dust attenuation due to the inner disc steeply decreases with decreasing offset from the centre. The nuclear disc hosts the bright blue compact nuclear cluster of M33, which is known to have young populations of stars (of age $\sim10^7-10^8$ yr) and mass $10^6{\rm M}_{\odot}$ \citep{1993AJ....105.1793K}.

The nuclear disc contributes to the star-formation rate of M33 with only $0.0018\pm0.0001\, {\rm M_{\sun}yr^{-1}}$. However,  since its spatial extent is very small, it has the highest surface density of star-formation ($\Sigma_{\rm SFR}^{\rm n}=(1030\pm70)\times 10^{-4}{\rm M_{\sun}yr^{-1}kpc^{-2}}$).

Despite being the morphological component with the highest surface density of SFR, the intrinsic stellar SED of the nucleus appears rather red, when comparing with the SEDs of other morphological components (see Fig.~\ref{fig:int_FUV}). In fact it is the reddest SED in M33. Since our methodology is to fit geometrical shapes (e.g. thin disk for the nuclear component) rather than stellar populations, it is open to the possibility that there may be an older stellar population inhabiting this region, in form of either a disc or a small hidden classical bulge,
that would be difficult to infer from the data. M33 is considered to be a bulgeless galaxy \citep{1992AJ....103..104B}, but this has been a subject of controversy \citep{1993ApJ...410L..79M,1993AJ....105.1793K,1994ApJ...434..536R,2001AJ....122.2469G,2007ApJ...669..315C}.

\subsubsection{The inner disc}

The inner disc appears more as a ring structure, when looking at the distribution of young stars and dust, extending between 250\,pc to 2\,kpc, and having a scale-length of about $100$ parsec (see Figs.~\ref{fig:stel_dust} and\ref{fig:comp_nuv}). The old stars instead are distributed down to the centre of M33, so no ring structure is defined. {Past photometric studies \citep{1992AJ....103..104B,1993ApJ...410L..79M,1994ApJ...434..536R} claimed an excess emission in the inner region with respect to the inward extrapolation of a disk exponential law, which was attributed to either a bulge component with an effective radius of 0.5\,kpc \citep{1993ApJ...410L..79M}, to a pseudobulge, since it was modelled to have  underwent a star formation episode less than 1\,Gyr ago, or to a bar \citep{1994ApJ...434..536R}. Later dynamic studies inferred the existence of an oval bar \cite{2007ApJ...669..315C} within the confines of the inner disc. We thus identify the inner disc with the morphological component hosting the bar of M33.}

The inner disc has a ${\rm SFR}^{\rm i}$ of $0.011\pm0.001\,{\rm M_{\sun}yr}^{-1}$, which is an order of magnitude higher than that of the nuclear disc, but still small in terms of the total star-formation rate of M33. Likewise, $\Sigma_{\rm SFR}^{\rm i}$ is $(100\pm10)\times 10^{-4}{\rm M_{\sun}yr^{-1}kpc^{-2}}$, making the inner disc the morphological component with the second highest surface density of SFR after the nuclear component. 

The intrinsic stellar SED of the inner disc is the bluest of all the other morphological components (see Fig.~\ref{fig:int_FUV}), although, as mentioned above, the nuclear component may also have a very blue SED, but may be contaminated by a small bulge emitting preferentially in the optical/NIR. The blue SED of M33 is in line with this component having the second highest surface density of SFR rate in M33.

The inner disc reaches at its inner radius the highest dust opacity in the galaxy, of $\tau^{f}_{B}\left(R_{\rm in, d}^{\rm i-disc}\right)=1.3\pm 0.1$. Thus, in the inner disc the galaxy starts to be moderately optically thick in the B-band and is optically thick throughout the UV spectral range. This explains why the attenuation curve at the position of the  inner disc, as depicted in Fig.~\ref{fig:atten-rad} with blue dashed-line, is rather flat in the UV, very similar to the extinction curve of the Milky Way, and definitively flatter than the global attenuation curve of M33.

The dust mass contained within the inner disc is $M^{\rm i}_{\rm d}=8.2^{+0.8}_{-0.7}\times10^4{\rm M_{\sun}}$. This dust is strongly heated by the high density of star-formation, and because of this it reaches a high average temperature of 29\,K.

\subsubsection{The main disc}
The main disc extends from about 2\,kpc out to 7\,kpc, in particular when viewed in the distribution of young stars and dust (see Fig.~\ref{fig:comp_nuv}). The distribution of old stars continues exponentially to the centre, and overlaps with that from the inner disc. The scale-length of the young stellar population is 1.5\,kpc. The dust disc has instead a very flat distribution, with a scale-length of 9.0 kpc, but truncated at 7\,kpc.

The main disc contains the majority of recent star-formation in M33, with ${\rm SFR}^{\rm m} = 0.26^{+0.02}_{-0.01}\, {\rm M_{\sun}yr^{-1}}$. This is because the main disc extends over a large area. Nonetheless the surface density of SFR is small, with  $\Sigma_{\rm SFR}^{\rm m}=30^{+2}_{-1}\times 10^{-4}{\rm M_{\sun}yr^{-1}kpc^{-2}}$.

\begin{figure}
    \centering
    \includegraphics[width=\linewidth]{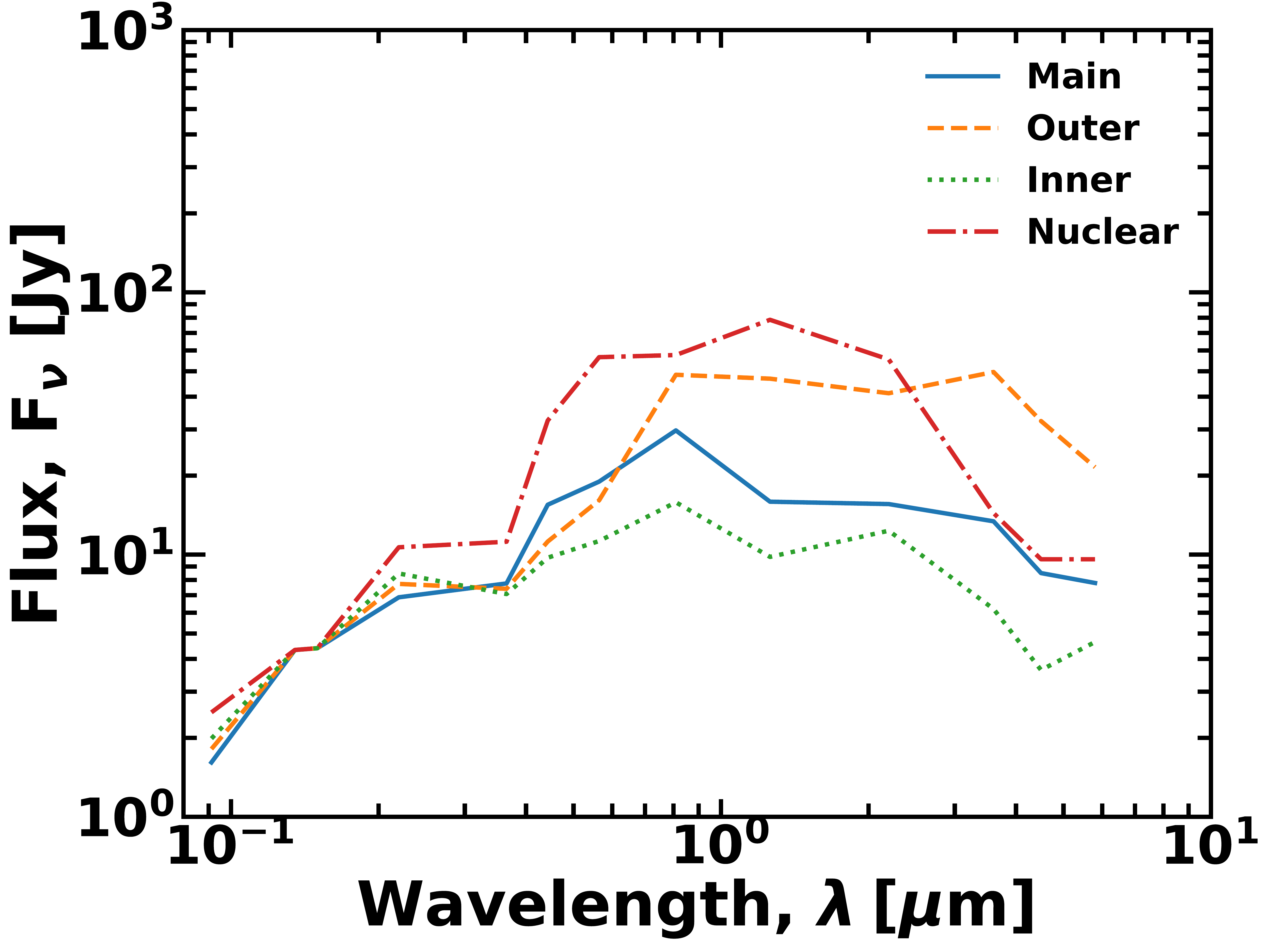}
    \caption{The intrinsic stellar SEDs of the morphological components of M33. The SEDs are normalised to overlap in the FUV.}
    \label{fig:int_FUV}
\end{figure}

The intrinsic SED of the main disc (see Fig.~\ref{fig:int_FUV}) is representative for the galaxy as a whole, since it dominates the bolometric output of M33. Apart from the red intrinsic FUV/NUV colour previously noted, the colours appear qualitatively typical for a late-type spiral galaxy containing no significant bulge.

The dust opacity of the main disc at its inner radius is $0.89^{+0.02}_{-0.04}$. This means that the main disc is optically thin in the B-band and transitions from being optically thick in the FUV to more optically thin regime towards longer UV wavelengths. This is why the attenuation curve at the position of the main disc (see orange dashed-line in Fig.~\ref{fig:atten-rad}) is steeper than the global attenuation curve of M33, is definitively steeper than the attenuation curve at the position of the inner disc and becomes even steeper towards the outer disc (see green dashed-line in Fig.~\ref{fig:atten-rad}).

Like with the SFR, due to its large extent, the main disc contains the majority of the dust mass $M^{\rm m}_{\rm d}=11.3^{+0.3}_{-0.5}\times10^6{\rm M_{\sun}}$. This dust is heated to an average temperature of 18\,K, which is a typical temperature for grains in the diffuse interstellar medium of spiral galaxies \citep{2002ApJ...567..221P,2005SSRv..119..313S,2005MNRAS.364.1253V,2009AJ....138..146W,2010A&A...518L..65B,2010A&A...518L..88B,2010A&A...518L..61B,2010A&A...518L..67K,2011A&A...536A..16P}.

\subsubsection{The outer disc}

The outer disc extends from about 7\,kpc to 10\,kpc (see Fig.~\ref{fig:comp_nuv}), and has a scale-length of 1\,kpc for the distribution of old stars and dust, and 0.6\,kpc for the young stars. The outer disc produces the same amount of recent star-formation as the inner disc, with 
${\rm SFR}^{\rm o}=0.011\pm0.001\,{\rm M_{\sun}yr^{-1}}$. However, because its spatial extent is larger than that of the inner disc, it is much more quiescent, with $\Sigma^{\rm o}_{\rm SFR}=(1.1\pm0.1)\times 10^{-4}{\rm M_{\sun}yr^{-1}kpc^{-2}}$. This makes it the morphological component with the lowest SFR surface density. 

The intrinsic stellar SED of the outer disc (Fig.~\ref{fig:int_FUV}) is very red, having a rather flat distribution in the optical/NIR, and a strong emission component at long wavelengths. The flat SED in the optical may be an artefact of the observations being rather noisy in the outer disc at these wavelengths. However, in the NIR, the IRAC observations clearly show a well define outer disc, having the highest level of emission relative to the FUV band from all the other morphological components.

Although the inner radius of the outer disc lies at 6.76\,kpc, the face-on dust optical-depth of the outer disk at this point is significant, with a value  in B-band  of $0.40\pm 0.02$. However, the opacity decreases very steeply radially outwards from this point, making this outer region overall very optically thin. The dust mass of the outer disc is $M^{\rm o}_{\rm d}=(2.7\pm0.1)\times10^6{\rm M_{\sun}}$ and its average dust temperature is very low,  of only 12\,K.

\subsection{The radiation fields of M33}

An important by-product derived from our decoding analysis is the calculation of the radiation fields energy density {(RFED)} inside M33.
As previously outlined, these can be used as input to calculations of diverse physical phenomena such as the inverse-Compton scattering of the diffuse MIR/FIR radiation field by cosmic ray electrons to produce gamma-rays, and the photoelectric heating of the diffuse ISM by
the non-ionising UV radiation fields. Generic solutions for the radiation fields within spiral galaxies have been given in \cite{2013MNRAS.436.1302P}, although these were only derived for single exponential disc profiles of stellar emissivity and dust, plus bulge components. As already demonstrated for the case of the Milky Way in \cite{Popescu17}, a more complex radial distribution, given by several morphological components including inner discs (bars), means that the radiation fields will also exhibit a more complex spatial distribution than in a generic model. 

\begin{figure*}
    \centering
    \includegraphics[width=\linewidth]{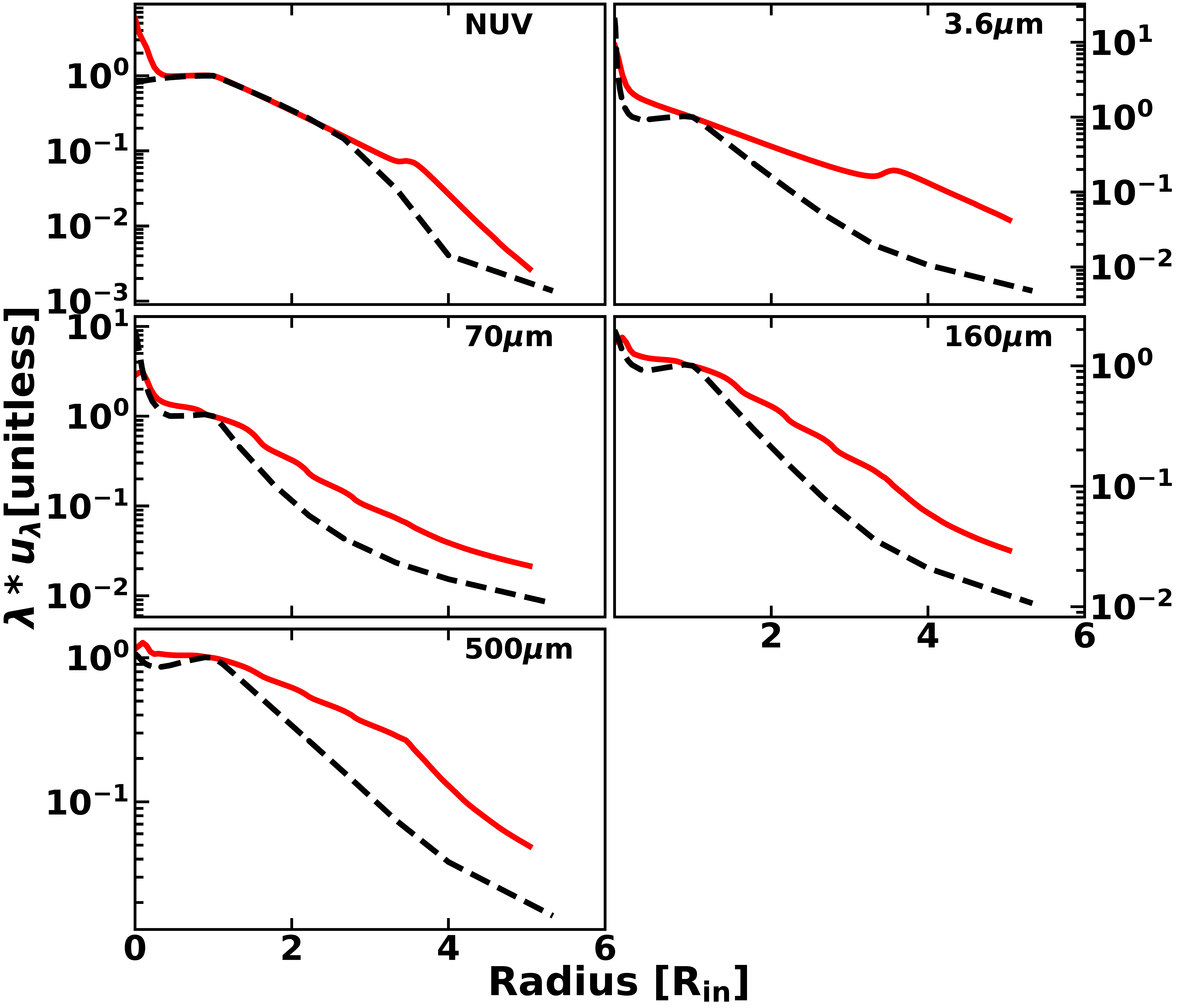}
    \caption{Radial profiles at $z=0$ of the radiation fields energy density (RFED) of M33 (solid red line) at selected wavelengths. The profiles are normalised to the value of RFED at the inner radius of the main disc, $R_{\rm in}^{\rm m-tdisk}=2\,{\rm kpc}$, and plotted against the radial distance given in units of $R_{\rm in}^{\rm m-tdisk}$. For comparison we show the corresponding RFED of the Milky-Way (dashed black line) taken from \citeauthor{Popescu17} (\citeyear{Popescu17}), normalised again to the value of the RFED at the inner radius of the Milky Way of $R_{\rm in}^{\rm MW}=4.5$\,kpc, and plotted against the radial distance in units of $R_{\rm in}^{\rm MW}$.
    {The normalisation factors used in these plots for both M33 and the MW are listed in Table~\ref{tab:RFED}.}}
    \label{fig:RFED}
\end{figure*}

The radiation fields of M33 are calculated both in direct stellar light and in the dust emission, and are made publicly available from the CDS data base. 
Examples of radial profiles of RFED at selected wavelengths are given in Fig.~\ref{fig:RFED}. 
{In order to assess their spatial variation, also with respect to the Milky Way RFED, the profiles are normalised to their value at the inner radius of the main disk, with the normalisation factors given in Table~\ref{tab:RFED}. }
The profiles show the imprint of the various morphological components of M33, although their shape is strongly modulated by the complex radiative transfer effects. Thus, in the panel showing the NUV profile (top left of Fig.~\ref{fig:RFED}), the contribution of the nuclear/inner, main and outer discs are clearly reflected in the shape of the profile. At other wavelengths these differences are less pronounced, although still detectable. This behaviour is to be expected, both for the RFED in direct stellar light and in dust emission. Thus, in the NUV the galaxy is more optically thick than at other optical/NIR wavelengths, and as such the radiation fields follow more closely the stellar emissivity. 
By contrast, at $3.6\,{\mu}$m (top right of Fig.~\ref{fig:RFED}), where the galaxy is optically thin, the profile has less changes in the slope, due to the larger horizon seen at any radial position, with photons arriving from different depths within the galaxy, and thus smoothing the shape of the profile.

At 500\,${\mu}$m (bottom left of Fig.~\ref{fig:RFED}) the radiation fields follow more closely the dust distribution, at least in the main and outer disc. The shallow slope of the profile within the main disc confines is, at least in part, due to the shallow distribution of the dust disc. At shorter infrared wavelengths, which are strongly affected by the heating effects on the dust, the profiles of radiation fields do not resemble either stellar emissivity or dust distribution, showing that only an explicit RT calculation can derive the spatial and spectral distribution of the radiation fields.

Of particular interest is to compare the radiation fields of M33 with those of the Milky Way, for which detailed solution of RFED have been obtained in \citep{Popescu17}.  To allow for a meaningful comparison, in Fig.~\ref{fig:RFED} we overplot the radiation fields of the Milky Way, normalised in the same way as for the M33, whereby the inner radius $R_{\rm in}^{\rm MW}$ of the Milky Way is 4.5\,kpc \citep{Popescu17}.

Interestingly, in the NUV the slope of the radiation fields within the main disc seems to be identical for the MW and M33. This is quite remarkable (from a technical point of view), taken into account that the MW is completely obscured in the UV from our solar position, and that the UV radiation fields have been derived in \cite{Popescu17} without direct observational constraints, and mainly from the FIR all-sky maps in conjunction with the RT modelling. The drop in the profile towards the centre is also similar, probably coming from the inner disc in both M33 and MW. However the very centre is dominated by the contribution from the strong nuclear disc in M33, while in the MW we have a slight drop in the profile, in the absence of such a prominent nuclear component. At large radii the profile of M33 is enhanced over that of the MW. This is due to the outer disc of M33. It is possible that the MW may also have a faint outer disc emitting in the UV, however this is difficult to infer in the absence of direct observational constraints. At 3.6\,${\mu}$m the profiles of radiation fields are different, with M33 exhibiting a more shallower decrease with radial distance. 

In the dust emission the radiation fields of M33 exhibit shallower profiles than those of the MW, due to the combination of a flatter dust distribution in M33 plus a strong outer dusty disc.

Overall the radiation fields of M33 have more power at large radial distances with respect to the inner part than in the MW.

\begin{figure}
    \centering
    \includegraphics[width=\linewidth]{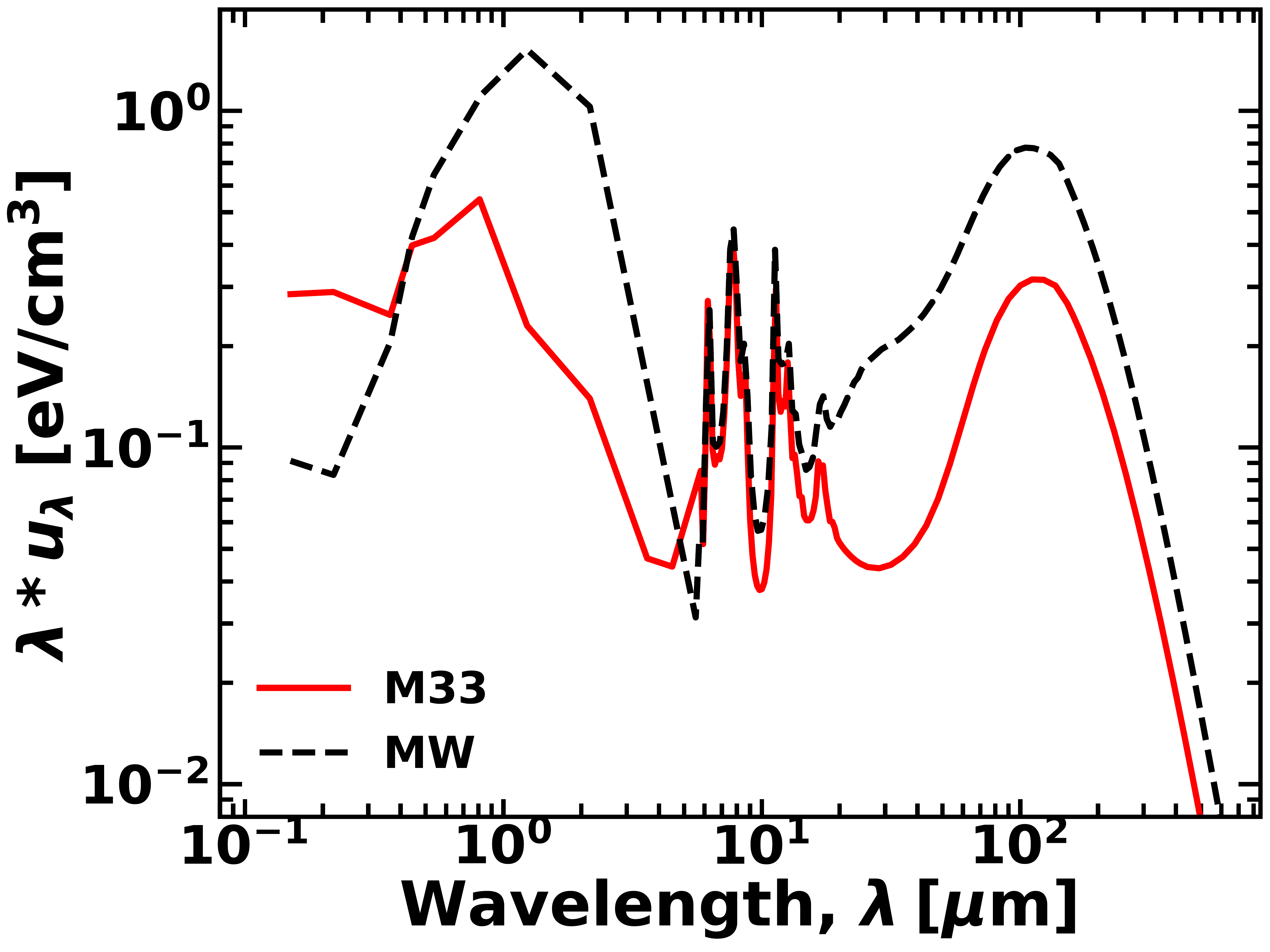}
    \caption{{The SED of the radiation fields of M33 (red line) at $R_{\rm in}=1$, with $R_{\rm in}$ as defined in Fig.~\ref{fig:RFED}. For comparison we plot (dashed black line) the SED of the Milky Way, at a similar characteristic radius $R_{\rm in}=1$.}}
    \label{fig:RFED_sed}
\end{figure}

{While the differences in the spatial distribution of the radiation fields between M33 and the Milky Way are to be expected, due to the differences in the different morphological components of these galaxies, of particular interest is to also compare the absolute values of the energy densities of these fields.  At the characteristic distance of 1 $R_{\rm in}$, as defined in Fig.~\ref{fig:RFED}, we plot in Fig.~\ref{fig:RFED_sed} the SED of the RFED of M33 and of the MW. One can immediately see that the UV radiation fields of M33 are higher (by a factor of 
$\sim2.9$
- see Table~\ref{tab:RFED}) than those of the MW. However, for the same characteristic radius, the FIR radiation fields of M33 are lower by a factor of
$\sim 2-3$ 
than those of the MW. The largest discrepancy is in the optical-NIR, where the RFED of M33 can reach a factor of 10 with respect to those of the MW. These findings are in broad agreement with M33 having a higher star-formation rate surface density than the Milky Way, but lower surface density of old stars and lower dust masses. However, the quantitative differences in the strength and colour of the radiation fields between the two galaxies, also, as a function of position, arise from a combination of differences in the geometry of stars and dust, dust mass, SFR and star formation  history. The complexity of all these factors points towards the need for a self-consistent radiative transfer calculation, as done in this paper.}

\begin{table}
    \caption{{Radiation field energy density scaling factors as applied in Fig.~\ref{fig:RFED}}}
    \centering
    \begin{tabular}{ccc}
        \hline
         & {M33}  & {MW} \\
       {$\lambda$ $[\mu\mathrm{m}]$} & {$\lambda * u_{\lambda}$ [eV/cm$^3$]} & {$\lambda * u_{\lambda}$ [eV/cm$^3$]} \\
        \hline
        {0.22} & {0.290 } & {0.083 } \\
        {3.6}  & {0.047 } & {0.250 } \\
        {70}   & {0.177 } & {0.546 } \\
        {160}  & {0.247 } & {0.523 } \\
        {500}  & {0.008 } & {0.015 } \\
        \hline
    \end{tabular}
    \label{tab:RFED}
\end{table}

\section{Discussion}
\label{sec:disc}

One of the most important features of our model is that, because it explicitly solves the inverse problem with radiative transfer calculations, it has a powerful predictive power. In this section we shall discuss some of these predictions, and their implications for our understanding of how M33 came to be as observed now, and, beyond this, what this implies for our understanding of galaxy formation. For this we shall also compare our results with corresponding findings from other local galaxies.

\subsection{The assembly of stellar discs in M33}\label{sec:assembly}

The fits to the surface brightness profiles of M33 resulted in the derivation of a trend of decreasing the scale-length of the main stellar disc, from the B band to the K band.
This behaviour is predicted by semi-analytical hierarchical models for galaxy formation (e.g. {\citealt{1998MNRAS.295..319M}}), whereby discs grow with cosmic time through continuous accretion from the IGM. If galaxies grow from inside out, one would predict the stellar populations to be younger and have a lower metallicity at larger radii than at low radii, such that local universe galaxies would be intrinsically larger at the shorter wavelengths where light from the younger stellar populations is more prominent. However, we find the situation in M33 to be more complex than presented in the inside out scenario of disc growth, since the scale-length of the main stellar disc was found to increase again at the longer NIR IRAC bands. This suggest that, in addition to the disc growth through  accretion, there must have been an underlying extended older stellar disc, nowadays mainly detectable in the form of the outer disc and the outer extent of the main disc. Its low star-formation rate surface density and its very red 3.6-FUV colours support this supposition. We speculate that the inverted colours in the outer disk compared to the expectations of the inside-out growth scenario may be a consequence of M33 being a satellite galaxy in a compact group hosting the Milky Way and M31 as dominant central galaxies. In such an environment, star formation in the outer disc may periodically be suppressed by ram-pressure stripping of the gas in the outer disk. By contrast, gas fuelling and star formation in the main body of the galaxy may be relatively unaffected by the group environment, with the consequence that, even in the group environment, the main disc grows according to the inside-out scenario. This would be in accordance with recent statistical studies showing that the group environment has surprisingly little effect on the global SFR of satellite galaxies in groups in the local Universe (see {\citealt{2017AJ....153..111G}}), even for relatively low mass systems like M33. A further factor affecting the switch in intrinsic colour gradient with radius in M33 may be the removal of stars from the inner to the outer disk  through the accumulative effects of tidal interactions with the Milky Way and M31 between the present epoch and the epoch at which M33 was accreted into the local group.

\subsection{The mysterious {``}submm excess" gone}

Studies of low-metallicity galaxies (e.g. {\citealt{2010A&A...523A..20B,2011A&A...532A..56G,2013ApJ...778...51K,2013A&A...557A..95R})} have shown that, when modelling their SEDs with empirical or semi-empirical models, they were not able to reproduce the observed submm SEDs, in the sense that models underestimated observations. This was called a {``}submm excess". M33 has been previously modelled by \cite{2016AA...590A..56H} with our generic RT models from \citetalias{PT11}, which is designed to only  fit the spatially integrated SED, and considers single exponential functions for the spatial distributions of stars and dust. \cite{2016AA...590A..56H} found a submm excess for M33, which they ascribed to possible different grain properties in the low metallicity environments. \cite{2019MNRAS.487.2753W} also modelled M33 with RT codes and again found a submm excess, which they also ascribed to different dust properties. While \cite{2018MNRAS.477.4968T} in their careful spatially resolved modelling of Planck observations of M33 using a superposition of modified Planck functions found no evidence for a submm excess, this result came at the expense of allowing the effective dust emissivity index to vary with position, without having an underlying physical model of why the grain emissivity should change like this.

\begin{figure*}
    \centering
    \includegraphics[width=\linewidth,trim={0cm 0cm 0cm 0cm},clip]{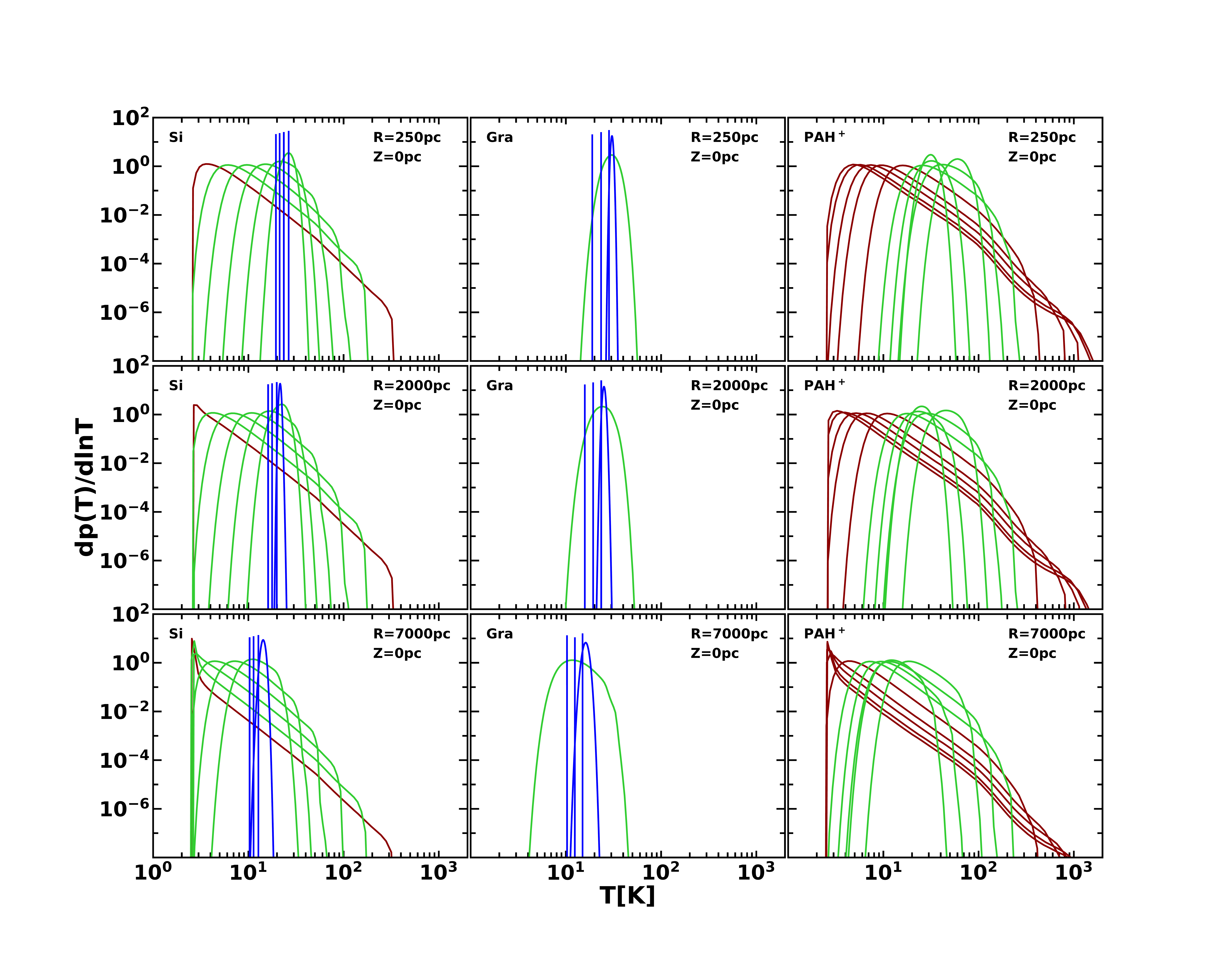}
    \caption{{Temperature distributions for dust grains with varying sizes (plotted as different curves in each panel) and various compositions: Si (left panels), Gra (middle panels) and PAH$^+$ (right panels), heated by the diffuse radiation fields calculated for our model of M33. Temperature distributions for PAH$^0$ are not plotted in this figure. The colour coding is as follows: red is for grains with radius $a<0.001\,\micron$ (0.00035, 0.00040, 0.00050, 0.00063 and 0.00100$\,\micron$), green is for grains with $0.001<a \le 0.01\,\micron$ (0.00158, 0.00251, 0.00398, 0.00631, 0.01000$\,\micron$) and blue is for grains with $a>0.01\,\micron$ (0.0316, 0.10000, 0.31623, 0.7943$\,\micron$). The biggest grains have delta function distibutions as they emit at equilibrium temperature. Going the top to bottom panels the calculations are done for different radial positions in the model: R=250pc; R=2000pc; and R=7000pc.}}
    \label{fig:dust_temp_distro}
\end{figure*}

In our modelling we found that we can fit the SED of M33 with no need to modify dust properties. As such, we find no {``}submm excess" for M33, or, in other words, the so called {``}submm excess" is well predicted by our model. The main difference from the study of both \cite{2016AA...590A..56H} and \cite{2019MNRAS.487.2753W} is that we not only fit the integrated SED, but derive the geometry of M33 by fitting images as well. 
Thus, we find that we need several morphological components to fit the surface-brightness distribution of M33, with a nuclear, an inner, a main and an outer disc, each one represented by different exponential distributions, and each with a different range of dust temperatures, 
determined self-consistently with the UV/optical via the RT analysis. 
In light of this we conclude that the submm excess may well be only an artefact of previous models not being able to self-consistently incorporate the geometry of the system in the SED modelling.

{It should be noted that our success in fitting  the panchromatic images of M33 without the need to invoke modified grain properties does not constitute a proof that the dust in M33 has Milky-Way type properties. However, it does provide a consistency check that existing grain models can account for the observed dust emission and attenuation in M33. In other words, if dust grains with enhanced submm emissivity were to exist in M33, they would not be inferred from a submm excess.
}

\subsection{The temperature distribution of dust grains in M33}\label{sec:tempdistro}

{Since our model calculates the temperature of dust grains throughout the volume of M33, it is of interest to examine their distributions, both temporal (in terms of temperature fluctuations), and spatial. For this we show in Fig.~\ref{fig:dust_temp_distro} an example of probability distributions of dust temperatures for different grains sizes and composition, at three radial locations in the plane of M33. As expected, for the same grain sizes, the stochastic heating effects become more prominent at larger radii. In addition, for the grains close to  thermal equilibrium (those exhibiting close to delta function distributions), there is a visible decrease in temperature with radial distance. This trend of decreasing dust temperature from the inner to the outer regions of M33 was also inferred by \cite{2014A&A...561A..95T} and \cite{2018MNRAS.477.4968T}, who performed modified back-body fits to the dust emission data, either on a pixel-to-pixel analysis (\citeauthor{2014A&A...561A..95T}) or on an azimuthally averaged analysis where the galaxy was divided between 3 concentric ellipses (\citeauthor{2018MNRAS.477.4968T}). A direct comparison between temperatures resulting from modified black-body fits and those derived from an explicit calculation of dust temperatures of various grain sizes and composition is impossible, in particular because it is not possible to define the same concept for what cold and warm dust is, but also because an average over the grain size distribution and composition does not have physical meaning, since heating of the dust is a non-linear effect. However, within the limited scope of such comparison we note that within the inner 4\,kpc of M33 \citeauthor{2014A&A...561A..95T} finds a cold dust temperature between 20 and 24\,K while our big grain equilibrium temperatures vary between 12 and 30\,K. Within the 4-5\,kpc annulus, which is the outermost region reached in the analysis of \citeauthor{2014A&A...561A..95T}, they find a cold dust temperature between 14-19\,K, while we find a variation between 13-20\,K. This would mean that we are broadly consistent with previous studies of dust temperature in M33 out to 5 kpc radius.

 At around 7\,kpc our analysis shows that grains can reach equilibrium temperatures as low as 10\,K (see Fig.~\ref{fig:dust_temp_distro}), which is to be expected in an environment with very low energy densities of the radiation fields. \cite{2018MNRAS.479..297W} also finds cold dust temperatures as low as 10\,K in their pixel-to-pixel analysis of M33 (see their Fig. 7). At even larger radii the dust grains are tendentially heated more and more by long range photons coming from the inner disk, since very little stellar emissivity is to be found at these distances. At a certain far enough radius the heating may even start to be dominated by the intergalactic radiation fields or the CMB. This is consistent with our results derived from the dust emission counterpart detected in the  extended HI disk of NGC891 \citep{2003A&A...410L..21P}, beyond the optical emitting disc of this galaxy. 

Independent of our model calculation, a simple inspection of the dust emission SED of the outer disc of M33 (Fig.~\ref{fig:sed_comp}) shows its peak at around $250\,\micron$, which is consistent with a Wien average temperature of 12\,K. The cold dust in the outer disc of M33 is prominently detected by both Herschel and Planck and has a minor but still significant contribution ($19\%$) to the total dust mass of M33 .}

\subsection{Effects of attenuation}\label{sec:EoA}

\begin{figure}
    \centering
    \includegraphics[width=\linewidth]{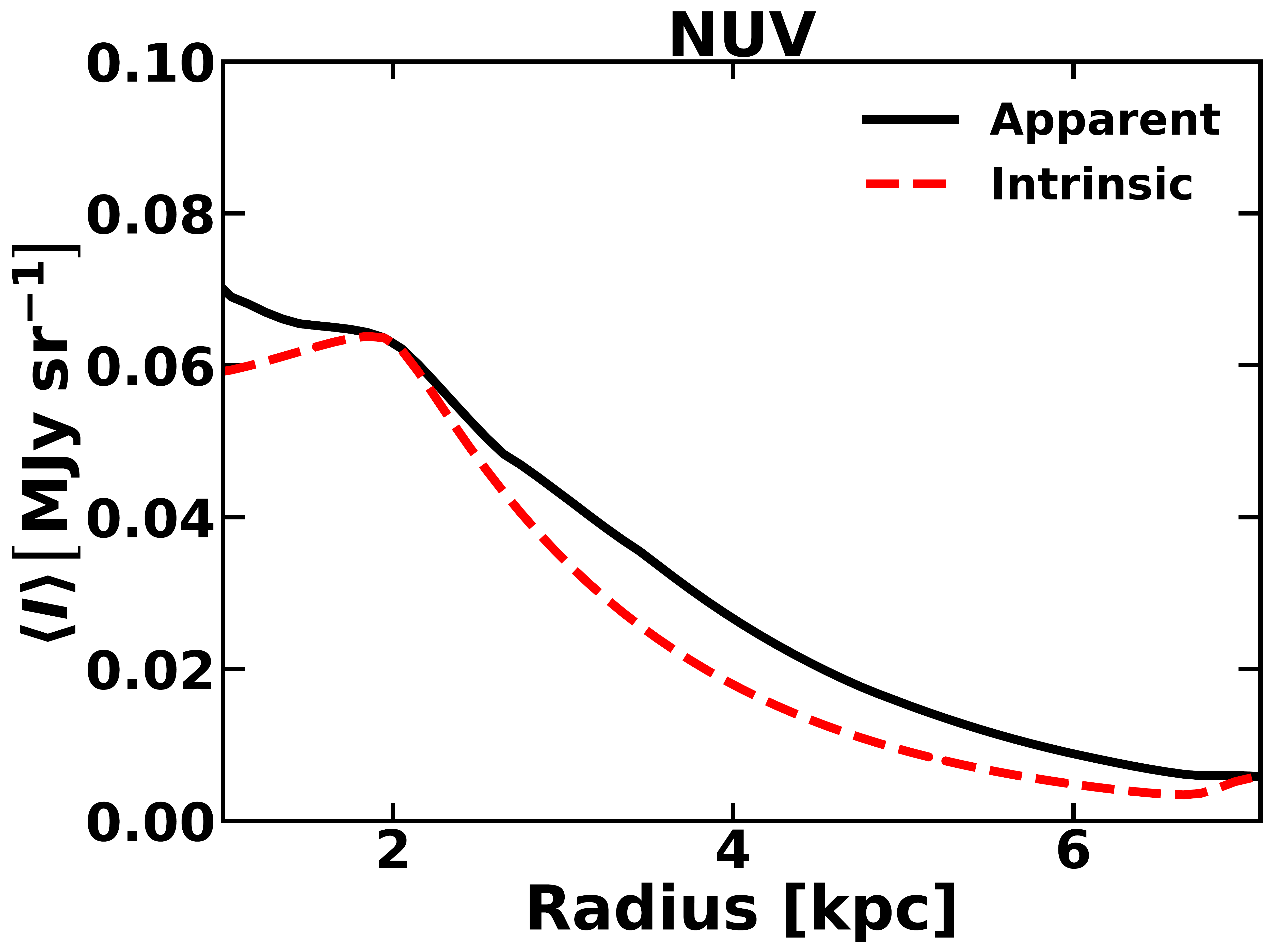}
    \caption{Comparison of our best-fit model of M33 in the NUV to the corresponding model without dust, the latter having been scaled {by a factor of 0.39,} to match the best-fit model at the inner radius of the main component.}
    \label{fig:nuv_atten}
\end{figure}

The effect of dust attenuation in shaping the perceived images of direct stellar light has been demonstrated on generic models \citep{MPT_2006,2013A&A...553A..80P,2013A&A...557A.137P}, but not on individual galaxy cases. Since we now have a {detailed} model of M33, we can illustrate how dust changes the appearance of the surface brightness distributions, in particular in the UV, where the effect is largest. For this we show in Fig. \ref{fig:nuv_atten}  a zoom (between $1-7$\,kpc) into our model for the azimuthally averaged SB profile of M33 in the NUV and the corresponding profile that would be observed in the absence of dust. The dustless model has been scaled to overlap with the attenuated model at the inner radius of the main disc. The comparison reveals important changes in the shape of the profiles. This proves that using observed UV images as proxies for the radial distribution of stellar emissivity of the young stellar population, without taking into account radiative transfer effects, 
introduces systematic errors in the geometry of the model.

\begin{figure}
    \centering
    \includegraphics[width=\linewidth]{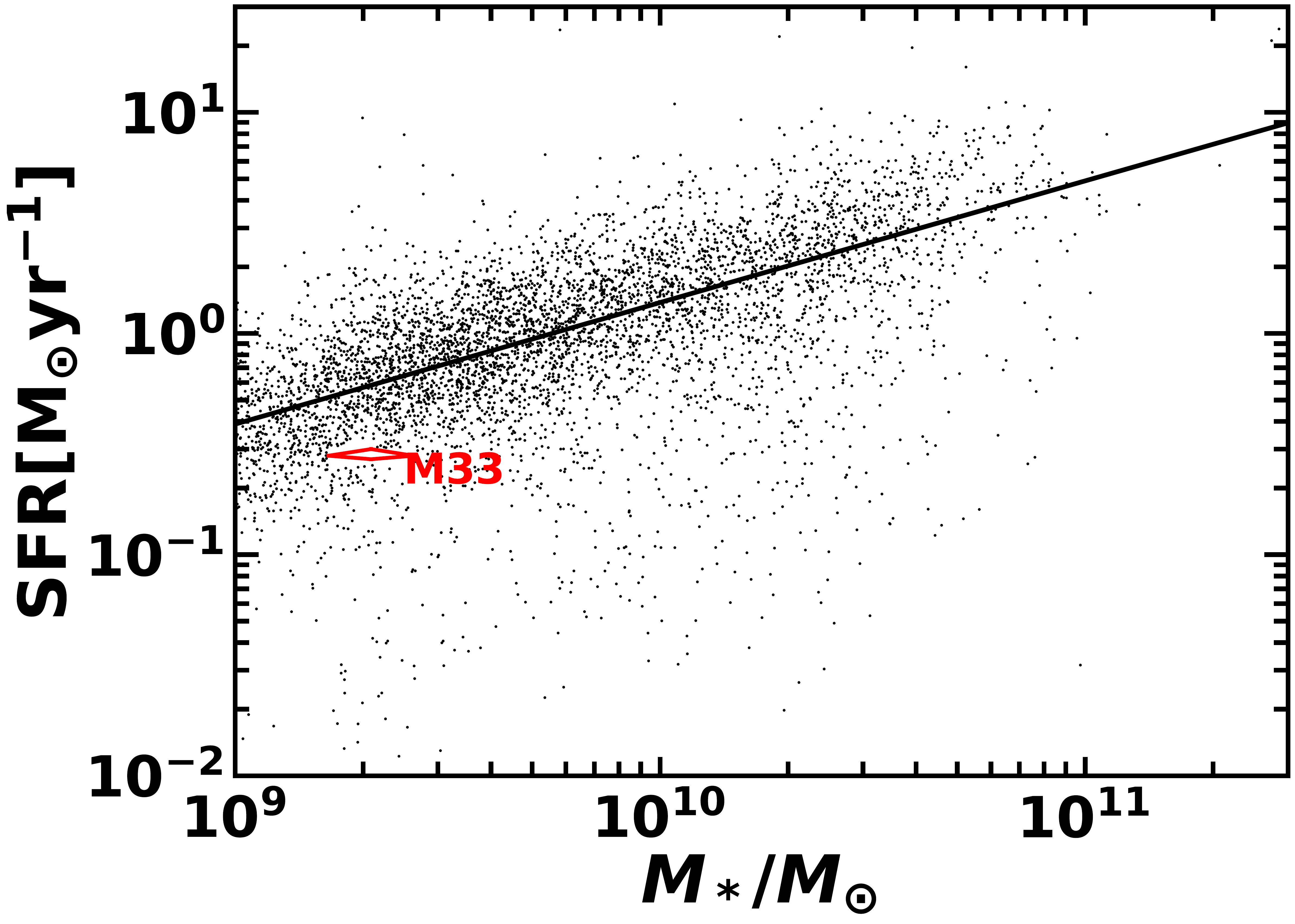}
    \caption{SFR versus stellar mass for a field reference sample of local universe disc-dominated galaxies (black points)
and the global emission from M33 (red rhomboid). The field reference sample is comprised of of 5202 morphologically 
selected, disk-dominated galaxies drawn from the GAMA survey by \citeauthor{2017AJ....153..111G} (\citeyear{2017AJ....153..111G}), which are not members of 
groups of galaxies. The vertical and horizontal apexes of the rhomboid containing the M33 point are placed at the 
1-sigma bounds in SFR and $M_*$, respectively. SFR for the GAMA galaxies were derived following \citeauthor{2017AJ....153..111G} (\citeyear{2017AJ....153..111G}),
while stellar masses for both the GAMA and M33 data were derived from intrinsic i and g-band photometry following 
\citeauthor{2011MNRAS.418.1587T} (\citeyear{2011MNRAS.418.1587T}). The solid line is the regression fit to a single power law model given in Table 2 of \citeauthor{2018MNRAS.477.1015G} (\citeyear{2018MNRAS.477.1015G}).}
    \label{fig:ms_m33}
\end{figure}

\begin{figure}
    \centering
    \includegraphics[width=\linewidth]{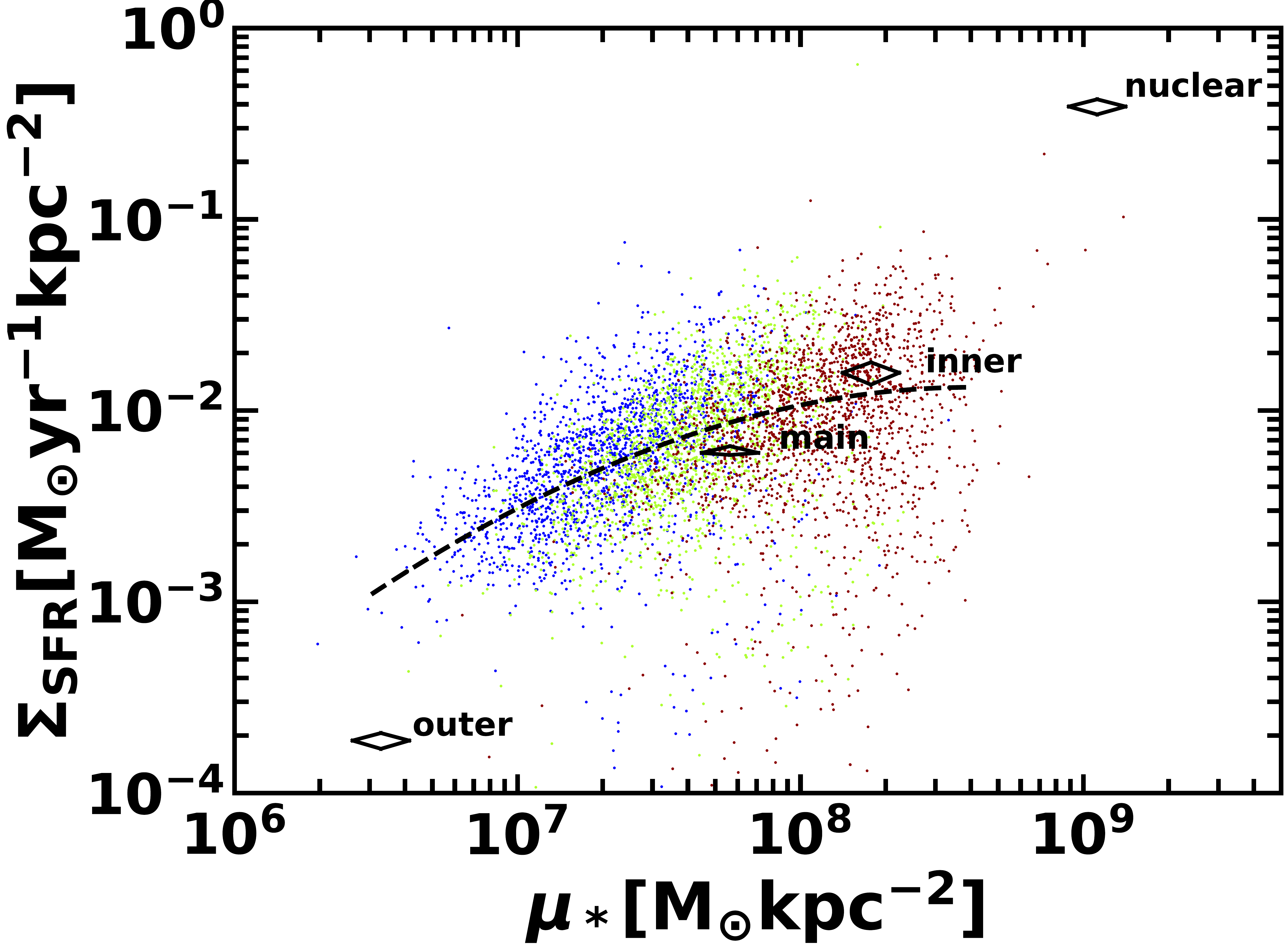}
    \caption{Surface density in SFR versus surface density in $M_*$ for the field reference sample of \citeauthor{2017AJ....153..111G} (\citeyear{2017AJ....153..111G})
(shown in blue, green and red points for low, intermediate, and high global stellar masses, respectively) and the four 
morphological components of M33 (black rhomboids). {The position of M33 in this plot largely overlaps with that of the main disc of M33, and because of this is not shown.} Surface densities were calculated within the $R_{\rm eff}$ in intrinsic i-band, 
with $R_{\rm eff}$ either derived from Sersic surface photometry, for the galaxies in the field reference  sample, or, for the M33 
components, derived from the shapes fitted in the RT  analysis.}
    \label{fig:ms_sigma_m33}
\end{figure}

\subsection{Comparison of M33 with the population of disk-dominated galaxies in the local Universe} \label{sec:galcomp}

Here we compare the integrated SFR and surface density of SFR in M33 and its individual components with
corresponding quantities for the sample of morphologically selected disc-dominated galaxies from the 
Galaxy And Mass Assembly Survey (GAMA: {\citealt{2011MNRAS.413..971D,2015MNRAS.452.2087L}}), as analysed by \cite{2017AJ....153..111G,2018MNRAS.477.1015G}. 
Fig.~\ref{fig:ms_m33} shows the ``main sequence'' relation between integrated quantities SFR and $M_*$ for a flux- limited 
subset (complete to 19.4 mag in the SDSS r-band) of 5202 disc-dominated GAMA galaxies located within redshift 0.13, 
and which are not classified as belonging to a galaxy group (see  Table 1 and Sect. 4.3 of {\citealt{2017AJ....153..111G}}). We refer 
to this subsample, plotted as dots on the figure, as the ``field'' reference sample of disc-dominated galaxies.

For comparison, a red rhomboid denotes the location and uncertainties of the global emission of M33 in the SFR-$M_*$  
plane, as derived from the present analysis. This show that M33 is significantly displaced from the star-forming main 
sequence, delineated by the overplotted regression fit to a single power law relation between SFR and $M_*$ for the field  
sample. This is qualitatively in accordance with our independent conclusion from the red intrinsic FUV/NUV colours that 
M33 has been undergoing a decline in SFR on a timescale of order 100\,Myr.

We can also use the results of the RT analysis of \cite{2017AJ....153..111G}, which were based on applying the \citetalias{PT11} model to 
Sersic surface photometry of the galaxy discs, to plot the surface densities of SFR ($\Sigma_{\rm SFR}$) versus the surface 
densities of stellar mass ($\mu_{\rm *}$), and use this as a benchmark for comparing the surface densities of SFR of the 
individual geometric components of M33. This is done in Fig.~\ref{fig:ms_sigma_m33}. The points for the field reference sample 
of disc-dominated galaxies are colour coded according to the integrated stellar mass of the GAMA galaxies 
(divided equally between low, intermediate and high mass, respectively plotted as blue, green and red points).
M33 {analogues} appear as blue points in this scheme. $\Sigma_{\rm SFR}$ is calculated as the total SFR contained within
the $R_{\rm eff}$ in intrinsic i-band, while $\mu_{\rm *}$ is calculated as the stellar mass contained with the same $R_{\rm eff}$. We  
note that the locus of the points for the field reference sample is similar to that published using data from integrated field 
imaging by \cite{2016A&A...590A..44G}, \cite{2018MNRAS.474.2039E}, and \cite{2018ApJ...865..154H}. In particular, taking into account the 
strong dependence on Hubble type, as quantified by \cite{2016A&A...590A..44G}, the locus of the points corresponds well 
with that found by those authors for Sb to Sd galaxies, as expected for our disk-dominated sample.

For comparison we have overplotted in black on Fig.~\ref{fig:ms_sigma_m33} the corresponding surface densities in SFR 
and $M_*$ (ie calculated from the integrated quantities contained within the effective radii in intrinsic i-band) of the various 
morphological components of the M33 RT model. One sees that the dominant main disk component of M33 is not  
untypical in its $\Sigma_{\rm SFR}$ in relation to $\mu_{\rm *}$ compared to massive and intermediate mass galaxies in the GAMA 
sample. By contrast, the surface density of SFR in the outer disk of M33 is highly suppressed with respect to the GAMA sample (albeit 
that there are only a few GAMA galaxies with comparable $\mu_{\rm *}$ as measured within the intrinsic  $R_{\rm eff}$).
The opposite is true for the nuclear component of M33. These results are consistent with the reason for the relative 
quiescence of M33 in its global SFR being due to a suppression in SFR beyond $R_{\rm eff}$, as compared with
the field reference sample.

Overall, the sequence of points nuclear->inner->main->outer for the morphological
components of M33 lies on a steeper relation than that exhibited by the surface densities within $R_{\rm eff}$ for the field reference sample. {We call this new relation a ``structurally resolved main sequence". In this sequence, the nuclear, inner, main and outer discs of M33 range from high-up in the starburst region to deep in
the ``green valley". The results for the outer disc in the green valley could be interpreted as an environmental effect, due to M33 being a minor satellite of the Local Group. This would be consistent with our other findings related to the wavelength dependence of the scale-lengths of the stars (see our previous discussion in Sect.~\ref{sec:assembly}). In an environment based scenario,
SF in the outer disk may be suppressed by ram-pressure stripping, even in a relatively low mass
group. By contrast, gas fuelling and SF in the main body of the galaxy may be relatively unaffected
by the group environment, explaining the position of the outer disc in the structurally resolved main sequence.

Although the environmental effect is plausible, part or perhaps even all of this trend, may alternatively
be linked to stellar feedback processes regulating the growth of galaxy disks in any environment.
Stellar feedback can operate on a variety of spatial and temporal scales: local feedback from SF
regions into the ISM can heat ISM slowing its continuous conversion into stars on relatively small
time and length scales, while on large scales feedback can drive gas completely out of galaxies
through galactic winds. This in turn induces a variation in SF activity on the corresponding spatial and
temporal scales. This may lead the SFR cycle to be out of phase between the outer and inner disc,
thus offering an alternative explanation of the behaviour we see in M33. In addition stellar migration \citep{2017ASSL..434...77D}
may also play a role (albeit likely subdominant) in changing the slope of the relation by redistributing
stellar mass in the disk on Gyr timescales. At present we cannot formally distinguish between these
possibilities, as M33 is the first, and as yet only galaxy for which such analysis has been done.}

\subsection{Advantages and limitations}

{In this paper we showed that the axi-symmetric models can provide a good solution for the overall energy balance between dust absorption and emission and provide a good match to the observed UV-optical-FIR/submm SED of M33, including the submm regime. In addition the model provides a good match to the UV-optical-submm azimuthally averaged radial profiles of M33. The implementation of the model is such that it makes it possible to derive the geometry of the emitters and absorbers through the inversion of the data, rather than by assuming a fixed geometry. Because of this the main advantage of this method is that it has a strong predictive power. The model makes predictions for the large-scale spatial distributions of stars and dust and for the intrinsic colour of the different stellar populations, since it inverts the data independently at each UV-optical wavelength . 

The main limitation of the model is the fact that it cannot make predictions for the arm-interarm structure of the galaxy. To do this would require a direct inversion of the data to predict non-axisymmetric structure. This, however, would require a solution for the stellar emissivity SED and dust opacity in each independent spatial element sampled by the observations, resulting in a non-linear optimisation problem in of order a million independent variables.
This is a very challenging problem, considering the resources needed
to perform a large number of radiation transfer calculations in any iterative convergence scheme, which to date has not been solved. While  there have been attempts to model observations of face-on galaxies using  non-axisymmetric RT models \citep{2014A&A...571A..69D,2017A&A...599A..64V,2019MNRAS.487.2753W}, these coadd a small
number of fixed spatial-spectral templates to predict the spatially
integrated SED, and do not attempt to solve the inverse problem for the imaging data.
}

\section{Summary and Conclusions}\label{sec:sum}

We modelled M33 using the generic RT model of \cite{PT11} in conjunction with the imaging observations available for this galaxy from the UV to the FIR-submm.  While retaining the generic formalism from \citetalias{PT11}, we {have designed} a new procedure that allowed the detailed geometry of stars and dust of M33 to be fitted jointly with their luminosity output. As such this is the first radiative transfer modelling of a non-edge-on galaxy, where the geometry is self-consistently derived.

The fits to the surface brightness profiles resulted in an overall good agreement with the data, with relative residuals less than $20\%$ in most cases. At submm wavelengths the profiles are well fitted, showing no sign of a {``}submm excess". We also predicted very well the radial temperature gradient of the dust, which is seen  in the change with wavelength of the slope and extent of the 70-350 micron profiles. 

When looking at the global properties of M33 we found that our model predicts quite well the energy balance between absorption and emission and successfully accounts for the global emission SED of M33, with an average relative residual between model and data of $5.9\%$. 
The global properties derived for M33 are as follows:
\begin{itemize}
\item The global star-formation rate is ${\rm SFR}=0.28^{+0.02}_{-0.01}\rm{M}_{\sun} {\rm yr}^{-1}$
\item The star-formation surface density is 
$\Sigma_{\rm SFR}=16.3^{+1}_{-0.7} \times 10^{-4}\rm{M}_{\sun} {\rm yr}^{-1} {\rm kpc}^{-2}$.
\item The dust optical depth has a maximum value at the inner radius of the inner disc of $\tau_B^f=1.3\pm 0.1$.
\item The dust mass is $M_{\rm d}=14.1^{+0.3}_{-0.5}\times 10^6\rm{M}_{\sun}$.
\item The gas-to-dust ratio is ${\rm GDR}=230\pm50$.
\item The percentage of stellar light reprocessed by dust is $35\pm3\%$.
\item The young stellar population accounts for {$80\pm8\%$} of the dust heating
\item The attenuation curve  is much steeper in the UV than the Milky Way extinction curve.
\item M33 lies below the {``}blue sequence" in the SFR versus stellar mass space
\end{itemize}

One of the main findings of our modelling of M33 is the existence of several morphological components: a nuclear, an inner, a main and an outer disc, as follows:
\begin{itemize}
\item The nuclear disc is a small structure of 100\,pc  and scale-length 20\,pc, residing in a {``}hole" of the dust distribution.
\item The inner disc appears more like a ring structure in the distribution of young stars, extending from 250 pc to 2\,kpc and having a scale-length of 100\,pc.
\item The main disc extends from about 2\,kpc out to 7\,kpc, with the scale-length of the young stellar population of 1.5\,kpc and the scale-length of the main dust disc of 9\,kpc, but truncated at 7\,kpc.
\item The outer disc extents from 7\,kpc  out to 10\,kpc and has a scale-length of 0.6\,kpc.
\end{itemize}

\noindent
The morphological components present some interesting radial trends:
\begin{itemize}
\item The dust associated with the inner disc is heated to around $\sim29{\rm K}$, being much warmer than the dust in the main disc which has a temperature of only $\sim 18{\rm K}$. Conversely, the outer disc contain cold dust, at around $\sim12{\rm K}$.
\item The star-formation surface density decreases from the nuclear,  to the inner, to the main, and to the outer disc.
\item In the $\Sigma_{\rm SFR}$ vs stellar mass density space the sequence of points corresponding to the nuclear>inner>main>outer disc defines a much steeper relation than the {``}blue sequence" of local Universe field star-forming galaxies.
\item  There is a  monotonic trend in steepening the slope of the UV attenuation curve when progressing from the inner to the outer disc. 
\end{itemize}

While we find evidence for an inside out growth of the main disc component through accretion, we also find evidence for the existence of an older disc component, mainly associated with the outer disc. We speculate that these findings may be a consequence of M33 being a satellite galaxy in a compact group hosting the Milky Way and M31 as dominant central galaxies. In such an environment, star formation in the outer disc may be suppressed by ram-pressure stripping of the gas in the outer disc. By contrast, gas fuelling and star formation in the main body of the galaxy may be relatively unaffected by the group environment, with the consequence that, even in the group environment, the main disc grows according to the inside-out scenario.  A further factor affecting the switch in intrinsic colour gradient with radius in M33 may be the removal of stars from the inner to the outer disk  through the accumulative effects of tidal interactions with the Milky Way and M31. These scenarios are consistent with our findings that the morphological components of M33 form a much steeper relation in the $\Sigma_{\rm SFR}$ vs stellar mass space than the main sequence of star-forming galaxies. {We called this new relation a ``structurally resolved main sequence". In this sequence, the nuclear, inner, main and outer discs of M33 range from high-up in the starburst region to deep in
the ``green valley". Although the environmental effect is plausible, we discussed that the trend exhibit by the structurally resolved main sequence of M33 may alternatively
be linked to stellar feedback processes regulating the growth of galaxy disks in any environment. Stellar feedback operating on a variety of spatial and temporal scales could lead the SFR cycle to be out of phase between the outer and inner disc, thus offering an alternative explanation of the behaviour we see in M33.

 The finding of a structurally resolved main sequence indicates that the global properties are borne out of an average of (occasionally significantly different) local properties and serve to highlight the rich diversity of galactic properties that we miss out with integrated-scale studies.
}

We derived the spectral and spatial distribution of the radiation fields of M33 and made them available at the CDS database. The radiation fields show the imprint of the different morphological components, although modulated by the complex radiative transfer effects.  Overall the radiation fields of M33 were found to be enhanced at larger radii relatively to the inner regions when compared to the radiation fields of the Milky Way.

\section*{Acknowledgements}
{We would like to thank an anonymous referee for very useful and insightful comments that improved this manuscript. }
Jordan J Thirlwall acknowledges support from a Science and Technology Facilities Council studentship grant (grant number ST/N504014/1). Cristina C Popescu and Giovanni Natale acknowledge support from a past Leverhulme Trust Research Project Grant RPG-2013-418, during which part of this work has been achieved.
Ben Carroll acknowledges support from the University of Central Lancashire's Undergraduate Research Internship Programme 2018 and 2019.
{The authors acknowledge discussions with Prof. Jay Gallagher and Dr. David Murphy}. 
\noindent
This work is based in part on observations made with the NASA Galaxy Evolution Explorer. GALEX is operated for NASA by the California Institute of Technology under NASA contract NAS5-98034.  
This research has made use of the NASA/IPAC Infrared Science Archive, which is operated by the Jet Propulsion Laboratory, California Institute of Technology, under contract with the National Aeronautics and Space Administration. 
This work has also made use of data products from the Two Micron All Sky Survey, which is a joint project of the University of Massachusetts and the Infrared Processing and Analysis Center/California Institute of Technology, funded by the National Aeronautics and Space Administration and the National Science Foundation. 
This work is based in part on observations made with the Spitzer Space Telescope, which is operated by the Jet Propulsion Laboratory, California Institute of Technology under a contract with NASA.
We also utilise observations performed with the ESA Herschel Space Observatory \citep{Herschel10}, in particular
to do photometry using the PACS \citep{PACS10} and SPIRE \citep{SPIRE10} instruments.
This research made use of Astropy,\footnote{http://www.astropy.org} a community-developed core Python package for Astronomy \citep{astropyI, astropyII} and of Matplotlib,\footnote{https://matplotlib.org} \citep{matplotlib}.



\bibliographystyle{mnras}
\bibliography{bibtex} 



\appendix
\newpage
\onecolumn

\section{Error calculation for observed flux densities and surface brightnesses}\label{sec:err_calc}

{The errors in the derived global (spatially integrated) flux densities have been calculated by taking into account the calibration errors, the background fluctuations, and the Poisson noise. 

The level of the background and the errors due to background fluctuations have been derived in most cases by considering $M=6$ annuli determined by eye to be beyond the extent of the galaxy.
 The average surface brightness  $\bar{F}_{{\rm bg,i}}$ 
 within each $i=[1,M]$ annuli is:
 \begin{align}
    \bar{F}_{{\rm bg,i}}=\frac{1}{{N_i}}{\sum\limits_{n=1}^{N_i}F_{{\rm  n}}}
\end{align}
where $N_i$ is the total number of pixels within the annulus $i$. The pixel-to-pixel variation within each annulus   $\sigma_{{\rm bg,i}}$ is then:
\begin{align}
    \sigma_{{\rm bg,i}}=\sqrt{\frac{1}{{N_{i}-1}}\sum\limits^{N_i}_{n=1}(F_{n}-\bar{F}_{{\rm bg,i}})^2}
\end{align}
and the error in the pixel-to-pixel variation is
\begin{align}
\epsilon_{\rm bg,i}=\frac{\sigma_{\rm bg,i}}{\sqrt{N_{\rm i}}}.
\end{align}
We then determine the background level as the average surface brightness over all M background annuli:
\begin{align}
    \bar{F}_{{\rm bg}}=\frac{1}{{M}}{\sum\limits_{i=1}^{M}\bar{F}_{{\rm bg,i}}}
\end{align}
 with the background RMS (annulus-to-annulus variation) $\sigma_{{\rm bg}}$ given by:
\begin{align}
  \sigma_{{\rm bg}}=\sqrt{\frac{1}{M-1}\sum\limits^{M}_{i=1}(\bar{F}_{{\rm bg}}- \bar{F}_{{\rm bg,i}}})^2\label{eqn:bgRMS}
\end{align}
and the error in the background brightness $\varepsilon_{\rm bg}$:
\begin{align}
    \varepsilon_{\rm bg}=\frac{\sigma_{{\rm bg}}}{\sqrt{M}}.
\end{align}
The error in the total flux density due to the background fluctuations $\varepsilon_{\rm bg}$ is then:
\begin{align}
    \varepsilon_{\rm total, bg} = \Omega_{\rm gal}\varepsilon_{\rm bg}
\end{align}
 where $\Omega_{gal}$ is the solid angle subtended by the galaxy and  $\varepsilon_{\rm bg}$ is expressed in flux density per steradian.
 In the case of the MHT observations, the background level and its fluctuations have been derived from an offset image that provided additional coverage along the minor axis (see Sect.~\ref{sec:MHT}). The Poisson noise has been calculated as $\varepsilon_{\rm poisson}=\sqrt{C}$,  where $C$ represents the sum of all counts within  all annuli used to derive the total flux of the galaxy at each wavelength.
The total errors in the flux densities have been derived using:
\begin{align}
    \varepsilon_{F_{\nu}}=\sqrt{\varepsilon_{\rm cal}^2+\varepsilon_{\rm total, bg}^2+\varepsilon_{Poisson}^2}.
\end{align}
In most cases $\varepsilon_{F_{\nu}}$ are dominated by the calibration errors, with the Poisson noise being negligible. Examples of  contributions to total errors are given in table 
\ref{tab:err_contribution}.

The errors in the derived azimuthally averaged surface brightnesses have been calculated by taking into account the  calibration errors, the background fluctuations, and the so called ``configuration noise" (arising from deviations of the observed brightness from an axi-symmetric distribution).
 The errors in the surface brightness due to the background fluctuations $\varepsilon_{\rm SB, bg}$, for each annulus within the galaxy, have been calculated as:
 \begin{align}
     \varepsilon_{\rm SB,bg}=\varepsilon_{\rm bg}\times\sqrt{\frac{\Omega_{\rm bg}}{\Omega_{\rm annulus}}}\label{eqn:av_SB_gal}
 \end{align}
 where $\Omega_{\rm bg}$ is the total solid angle used to determine the background (over M annuli) and $\Omega_{\rm annulus}$ is the solid angle subtended by an individual annulus within the galaxy.
 For the azimuthally averaged surface-brightnesses we also need to account for the noise due to the non-axi symmetric nature of the observations. This is what we call ``configuration noise". The errors due to this noise have been calculated by the following procedure. We divide each annulus used to derive the azimuthally averaged profile into $Q=4$ quadrants. The average surface brightness $\bar{F}_{{\rm gal,q}}$ within each quadrant $q=[1,Q]$ is given by:
  \begin{align}
    \bar{F}_{{\rm gal,q}}=\frac{1}{{N_q}}{\sum\limits_{n=1}^{N_q}F_{{\rm  n}}}
\end{align}
 where $N_q$ is the total number of pixels within the quadrant $q$. The average surface brightness over all quadrants within an annulus is then:
 \begin{align}
    \bar{F}_{{\rm gal}}=\frac{1}{{Q}}{\sum\limits_{q=1}^{Q}\bar{F}_{{\rm gal,q}}}
\end{align}
 The configuration noise RMS (quadrant-to-quadrant variation) $\sigma_{\rm SB,conf}$ is given by:
 \begin{align}
  \sigma_{{\rm SB,conf}}=\sqrt{\frac{1}{Q-1}\sum\limits^{Q}_{q=1}(\bar{F}_{{\rm gal}}- \bar{F}_{{\rm gal,q}}})^2
\end{align}
and the configuration error:
\begin{align}
    \varepsilon_{\rm SB, conf}=\frac{\sigma_{{\rm SB, conf}}}{\sqrt{Q}}.
\end{align}

The total errors in the azimuthally averaged surface brightness profiles have been derived using:}
\begin{align}
    \varepsilon_{SB_{\nu}}=\sqrt{\varepsilon_{\rm cal}^2+\varepsilon_{\rm SB,bg}^2+\varepsilon_{\rm SB, conf}^2}.\label{eqn:epsilon_SB}
\end{align}
{whereby the first term is independent of radius, while the second and third terms are radius dependent.
The total errors derived using Eqn.~\ref{eqn:epsilon_SB} are plotted as blue-shaded areas around the (azimuthally averaged) observed radial profiles in Figs.~\ref{fig:avintprof1}, \ref{fig:avintprof2}, and \ref{fig:avintprof3}. We find that overall the calibration errors  dominate over most radii and wavelengths. At larger radii the errors due to background fluctuations start to dominate, while the configuration noise dominates at radii and wavelengths where strong deviations from axi-symmetries exist (in particular at 24 and 70\,${\mu}$m).)}

 \begin{table}
	\centering
	\caption{{The contribution from calibration errors ($\varepsilon_{\rm bg}$), background fluctuations errors ($\varepsilon_{\rm bg}$) and Poisson noise errors (${\varepsilon}_{\rm Poisson}$) to the total 
	errors in the observed flux densities, at a few selected wavelengths.}}
	\label{tab:err_contribution}
	\begin{tabular}{lcccc}
		\hline
		 Band & {$\varepsilon_{\rm cal}$ $[{\rm Jy}]$}& {$\varepsilon_{\rm total, bg}$ $[{\rm Jy}]$} & {${\varepsilon}_{\rm Poisson}$ $[{\rm Jy}]$}\\
		\hline
		 FUV      & 0.11 & 0.011 & $6.1\times 10^{-6}$\\
		 LGGS B   & 0.95 & 1.3 & $3.0\times 10^{-4}$\\
		 PACS 160 & 320. & 7.0 & - \\
		\hline
	\end{tabular}
\end{table}
 
\newpage
\section{The stellar luminosity and the dust mass}
\label{sec:formula}

The spatial integration of the disc emissivity (Eqn.~\ref{eq:model}) up to the truncation radius  $R_{{\rm t, j}}$ and the truncation height $z_{\rm t}$, where $z_{\rm t} \gg z_{\rm j}$, is given by:

\begin{equation}
\label{eq:sint}
I= 4 \pi A_{0,{\rm j}}z_{\rm j} T_{{\rm z, j}}
    \begin{cases}
        {\displaystyle h_{\rm j}^2 T_{{\rm R, j}} \hspace{9.cm} {\rm if} \hspace{0.1cm} R_{{\rm in, j}} = 0 } \\ \\
        
        {\displaystyle \frac{1}{3} \left[ \left( 1+\frac{\chi_{\rm j}}{2} \right)R_{{\rm in, j}}^2 - \left(1 - \chi_{\rm j} \right)\frac{R_{{\rm tin, j}}^3}{R_{{\rm in, j}}} - \frac{3}{2} \chi_{\rm j} R_{{\rm tin, j}}^2 \right] \exp{\left(-\frac{R_{{\rm in, j}}}{h_{\rm j}}\right)}+h_{\rm j}^2 T_{{\rm R, j}} \hspace{1cm} {\rm if} \hspace{0.1cm} R_{{\rm in, j}} > 0 } \\
    \end{cases}
\end{equation}
where
\begin{equation}
\label{eq:Tr}
    T_{{\rm R, j}}= \exp{\left( -\frac{R_{{\rm in, j}}}{h_{\rm j}} \right)}
        -\exp{\left(-\frac{R_{{\rm t, j}}}{h_{\rm j}} \right)}
        +\frac{R_{{\rm in, j}}}{h_{\rm j}} \exp{\left(-\frac{R_{{\rm in, j}}}{h_{\rm j}} \right)}
        -\frac{R_{{\rm t, j}}}{h_{\rm j}} \exp{\left(-\frac{R_{{\rm t, j}}}{h_{\rm j}} \right)}
\end{equation}
and
\begin{equation}
\label{eq:Tz}
    T_{{\rm z, j}}=1 - \exp{\left(-\frac{z_{\rm t}}{z_{\rm j}}\right)}.
\end{equation}
Eqns.~\ref{eq:sint},~\ref{eq:Tr}, and~\ref{eq:Tz} can be used to calculate both the spatially integrated stellar disc luminosity and dust mass, for the stellar disc and dust disc components respectively. 

In order to calculate the spatially integrated stellar luminosity, one takes $A_0$ to be the central volume luminosity density $L_0$, and $i=$\lq s\rq. Thus the stellar disc luminosity is 
\begin{equation}
    L=I\left(L_0,h_{\rm s},z_{\rm s}\right),
\end{equation}
where $h_{\rm s}$, $z_{\rm s}$ are the scale-length and height of that stellar disc.

In the case of the dust mass, $A_0$ is assumed to be the central volume density of dust
\begin{equation}
    \rho_{\rm c}=\frac{\tau^{\rm f}_{B}\left( R_{\rm in,d}\right)\exp{\left(\frac{R_{\rm in,d}}{h_{\rm d}}\right)}}{2\kappa(B) z_{\rm d}}
\end{equation}
and $i=$\lq d\rq, where $\tau^{\rm f}_{B}\left( R_{\rm in,d}\right)$ is the face-on B-band dust opacity at radius $R=R_{\rm in,d}$ and $\kappa(B)$ is the mass extinction coefficient in the B-band. Thus the dust mass is
\begin{align}
    M_{\rm d}=I\left(\rho_{\rm c},h_{\rm d},z_{\rm d}\right),
\end{align}
where $h_{\rm d}$, $z_{\rm d}$ are the scale-length and height of the dust disc in question.

\newpage

\section{Model amplitude parameters of M33}\label{sec:mod_param}

The amplitude parameters of our model are expressed in terms of spectral luminosity densities for the case of stellar discs, and in terms of B-band face-on dust optical depth for the case of dust discs. In this appendix we tabulate these amplitude parameters as follows. In Table~\ref{tab:lum_dens} we list the (intrinsic) luminosity densities for the stellar disc and for the thin stellar disc, for each morphological component, at the UV/optical/NIR wavelengths were imaging observations were available. In Table~\ref{tab:tau_rin} we list the B-band face-on dust optical depth at the inner radius of each morphological component. In our model we found no dust disc associated with the nuclear component, therefore the nuclear disc is not included in Table~\ref{tab:tau_rin}.

\begin{table}
	\centering
	\caption{Intrinsic spectral luminosity densities of the stellar disc and thin stellar disc for each morphological component. }
	\label{tab:lum_dens}
	\begin{tabular}{cccccccc}
		\hline
		  $\lambda $ & $L^{\rm disc,i}_{\nu}$& $L^{\rm disc,m}_{\nu}$& $L^{\rm disc,o}_{\nu}$& $L^{\rm tdisc,n}_{\nu}$ & $L^{\rm tdisc,i}_{\nu}$&$L^{\rm tdisc,m}_{\nu}$ & $L^{\rm tdisc,o}_{\nu}$\\
		  $[\micron]$&[W/Hz]&[W/Hz]&[W/Hz]&[W/Hz]&[W/Hz]&[W/Hz]&[W/Hz]\\
		\hline
		  0.150 & -  & - & - & $0.080\times10^{19}$ & $1.266\times10^{19}$ & $3.621\times10^{20}$ & $1.387\times10^{19}$ \\
		  0.220 & -  & - & - & $0.195\times10^{19}$ & $2.441\times10^{19}$ & $5.662\times10^{20}$ & $2.441\times10^{19}$ \\
		  0.365 & $0.119\times10^{19}$  & $0.286\times10^{21}$ & $0.010\times10^{20}$ & $0.307\times10^{19}$ & $1.922\times10^{19}$ & $3.715\times10^{20}$ & $2.242\times10^{19}$ \\
		  0.443 & $1.193\times10^{19}$  & $0.954\times10^{21}$ & $0.239\times10^{20}$ & $0.594\times10^{19}$ & $1.650\times10^{19}$ & $3.829\times10^{20}$ & $1.155\times10^{19}$ \\
          0.564 & $1.571\times10^{19}$  & $1.313\times10^{21}$ & $0.422\times10^{20}$ & $1.036\times10^{19}$ & $1.727\times10^{19}$ & $3.339\times10^{20}$ & $5.037\times10^{19}$\\
		  0.809 & $2.443\times10^{19}$  & $2.345\times10^{21}$ & $1.466\times10^{20}$ & $1.056\times10^{19}$ & $2.200\times10^{19}$ & $2.552\times10^{20}$ & $0.660\times10^{19}$ \\
		  1.259 & $1.805\times10^{19}$  & $1.300\times10^{21}$ & $1.444\times10^{20}$ & $1.439\times10^{19}$ & $1.079\times10^{19}$ & $0.918\times10^{20}$ & $0.360\times10^{19}$ \\
		  2.200 & $3.736\times10^{19}$  & $1.299\times10^{21}$ & $1.299\times10^{20}$ & $1.016\times10^{19}$ & $0.028\times10^{19}$ & $0.655\times10^{20}$ & $0.028\times10^{19}$ \\
		  3.600 & $2.264\times10^{19}$  & $0.603\times10^{21}$ & $1.567\times10^{20}$ & $0.264\times10^{19}$ & $0.007\times10^{19}$ & $0.586\times10^{20}$ & $0.024\times10^{19}$ \\
		  4.500 & $1.573\times10^{19}$  & $0.754\times10^{21}$ & $1.019\times10^{20}$ & $0.176\times10^{19}$ & $0.007\times10^{19}$ & $0.408\times10^{20}$ & $0.012\times10^{19}$ \\
		  5.800 & $1.258\times10^{19}$  & $1.006\times10^{21}$ & $0.679\times10^{20}$ & $0.176\times10^{19}$ & $0.005\times10^{19}$ & $0.357\times10^{20}$ & $0.012\times10^{19}$ \\

		\hline
	\end{tabular}
\end{table}

\begin{table}
	\centering
	\caption{Predicted face-on B-band optical-depth at $R_{\rm in}$ for each morphological component of M33.}
	\label{tab:tau_rin}
	\begin{tabular}{lc}
		\hline
		 Component & $\tau^{\rm f}_{\rm B}(R_{\rm in,d})$ \\
		\hline
		  inner & $1.3\pm0.1$\\
		  main  & $0.89^{+0.02}_{-0.04}$ \\
		  outer & $0.40\pm0.02$ \\
		\hline
	\end{tabular}
\end{table}

\newpage

\section{Model flux densities of M33}
In this appendix we present a few tables with the predictions of our model for the intrinsic flux densities of M33 and of its morphological components. The tables list either the fluxes integrated out to the truncation radius (Table~\ref{tab:intrinsic_flux}), or the fluxes integrated out to the I-band effective radius (Table~\ref{tab:intrinsic_flux_effective}). The former are used in all the plots showing the SEDs of M33, while the latter are used in the comparisons we do with local Universe galaxies detected in large wide-field surveys. The corresponding I-band effective radius of M33 and its components are also tabulated here (Table~\ref{tab:intrinsic_radius_effective}). 

\begin{table}
	\centering
	\caption{Model flux densities out to the truncation radius, $F_{\rm model}(R_{\rm t})$ [Jy]. All the fluxes are intrinsic (corrected for the effect of dust attenuation).}
	\label{tab:intrinsic_flux}
	\begin{tabular}{lcccccccc}
		\hline
		 Component & FUV&NUV&U&B&V&I&J&K\\
		\hline
		 Global  &  4.70 & 7.44 & 8.28 & 16.25 & 20.01 & 32.09 & 18.03 & 17.54\\
		 Nuclear & 0.0071 & 0.017 & 0.027 & 0.053 & 0.092 & 0.094 & 0.13 & 0.09  \\
		 Inner   & 0.15 & 0.28 & 0.23 & 0.32 & 0.37 & 0.052 & 0.32 & 0.41  \\
		 Main    & 4.39 & 6.87 & 7.76 & 15.48 & 18.97 & 29.74 & 15.90 & 15.58   \\ 
		 Outer   & 0.16 & 0.28 & 0.26 & 0.40 & 0.58 & 1.73 & 1.67 & 1.47   \\
		\hline
	\end{tabular}
\end{table}

\begin{table}
	\centering
	\caption{Model flux densities out to the I-band effective radius, $F_{\rm model}(R_{\rm eff})$ [Jy]. All the fluxes are intrinsic (corrected for the effect of dust attenuation).}
	\label{tab:intrinsic_flux_effective}
	\begin{tabular}{lcccccccc} 
		\hline
		 Component & FUV&NUV&U&B&V&I&J&K\\
		\hline
		 Global  &  2.17 & 3.45 & 3.89 & 7.80 & 9.93 & 16.71 & 12.05 & 11.32   \\
		 Nuclear & 0.0046 & 0.011 & 0.017 & 0.031 & 0.055 & 0.056 & 0.076 & 0.053  \\
		 Inner   & 0.059 & 0.11 & 0.098 &  0.14& 0.18& 0.27& 0.19& 0.30    \\ 
		 Main    & 1.91 & 2.99 & 3.47 & 7.13& 9.13& 15.54& 11.42& 10.60   \\ 
		 Outer   & 0.11 & 0.19 & 0.18 & 0.23& 0.40& 0.90& 0.87& 0.76    \\
		\hline
	\end{tabular}
\end{table}

\begin{table}
	\centering
	\caption{Model I-band effective radii, $R_{\rm eff}$ [kpc].}
	\label{tab:intrinsic_radius_effective}
	\begin{tabular}{lcccccccc}
		\hline
		 Component & $R_{\rm eff}$ \\
		\hline
		 Global    & 2.65  \\
		 Nuclear   & 0.03  \\
		 Inner     & 0.17  \\
		 Main      & 2.55  \\
		 Outer     & 7.65  \\
		\hline
	\end{tabular}
\end{table}

\newpage
\section{The star-formation rate and dust mass across M33}

In this appendix we present a few tables with the calculated values of the SFR and $\Sigma_{SFR}$ for M33 and its morphological components. The tables list either these quantities integrated out to the truncation radius (Table~\ref{tab:SFR_Rt}) or integrated out to the I-band effective radius (Table~\ref{tab:SFR_Reff}). The latter are used in the scaling relations relating SFR with stellar masses. We also list the dust masses of the individual morphological components of M33 and of the galaxy as a whole in Table~\ref{tab:dust_mass}.

\begin{table}
    \centering
    \caption{SFR and $\Sigma_{\rm SFR}$ out to $R_{\rm t}$.}
    \begin{tabular}{ccc}
        \hline
        Component & SFR $[\rm{M}_{\sun} {\rm yr}^{-1}]$ & $\Sigma_{\rm SFR}$ $[\times 10^{-4}\rm{M}_{\sun} {\rm yr}^{-1} {\rm kpc}^{-2}]$ \\
        \hline
        Global & $0.28^{+0.02}_{-0.01}$ &$16.3^{+1.}_{-0.7}$\\        
        Nuclear & $0.0018\pm0.0001$ & $1030\pm70$\\
        Inner & $0.011\pm0.001$ & $100\pm10$\\
        Main & $0.26^{+0.02}_{-0.01}$&$30^{+2}_{-1}$\\
        Outer & $0.011\pm0.001$&$1.1\pm0.1$\\
         \hline
    \end{tabular}
    \label{tab:SFR_Rt}
\end{table}

\begin{table}
    \centering
    \caption{SFR and $\Sigma_{\rm SFR}$ out to $R_{\rm eff}$.}
    \begin{tabular}{ccc}
        \hline
        Component & SFR $[\rm{M}_{\sun} {\rm yr}^{-1}]$ & $\Sigma_{\rm SFR}$ $[\times 10^{-4}\rm{M}_{\sun} {\rm yr}^{-1} {\rm kpc}^{-2}]$ \\
        \hline
        Global & $0.137^{+0.01}_{-0.005}$ &$33^{+2}_{-1}$\\
        Nuclear & $0.0011\pm0.0001$ & $11400\pm600$\\
        Inner & $0.0015\pm0.0002$ & $2165\pm200$\\
        Main & $0.120^{+0.01}_{-0.003}$&$240^{+20}_{-7}$\\
        Outer & $0.0074\pm0.0007$&$4.8\pm0.5$\\
         \hline
    \end{tabular}
    \label{tab:SFR_Reff}
\end{table}

\begin{table}
	\centering
	\caption{Dust masses.}
	\label{tab:dust_mass}
	\begin{tabular}{lcccccccc}
		\hline
		 Component & $M_{\rm d} [\rm{M}_{\sun}]$ \\
		\hline
		 Global    & $14.1^{+0.3}_{-0.5}\times 10^6$  \\
		 Inner     & $8.2^{+0.8}_{-0.7}\times 10^4$  \\
		 Main      & $11.3^{+0.3}_{-0.5}\times 10^6$  \\
		 Outer     & $(2.7\pm0.1)\times 10^6$ \\
		\hline
	\end{tabular}
\end{table}



\bsp	
\label{lastpage}
\end{document}